\documentclass[twocolumn,showpacs,amsmath,amssymb,pra]{revtex4}

\usepackage{graphicx}
\usepackage{dcolumn}
\usepackage{bm}

\begin{document}
\title{Geometrical Aspects in Optical Wavepacket Dynamics}
%
\author{Masaru Onoda$^{1,2}$}
\email{m.onoda@aist.go.jp}
\author{Shuichi Murakami$^{2,3}$}
\email{murakami@appi.t.u-tokyo.ac.jp}
\author{Naoto Nagaosa$^{1,2,3}$}
\email{nagaosa@appi.t.u-tokyo.ac.jp}
\affiliation{
$^1$Correlated Electron Research Center (CERC),
National Institute of Advanced Industrial Science and Technology (AIST),
Tsukuba Central 4, Tsukuba 305-8562, Japan\\
$^2$CREST, Japan Science and Technology Corporation (JST), Saitama, 332-0012, Japan\\
$^3$Department of Applied Physics, University of Tokyo, 
Bunkyo-ku, Tokyo 113-8656, Japan
}
%
%
\date{\today}
%
%
\begin{abstract}
We construct a semiclassical theory for propagation of an optical wavepacket
in non-conducting media with periodic structures of
dielectric permittivity and magnetic permeability,
i.e., non-conducting photonic crystals.
We employ a quantum-mechanical formalism in order to 
clarify its link to those of electronic systems. 
It involves the geometrical phase, i.e., Berry phase, in a natural way, 
and describes an interplay between orbital motion and the internal rotation.
Based on the above theory, we discuss the geometrical aspects of the optical Hall effect. 
We also consider a reduction of the theory to a system without periodic structure
and apply it to the transverse shift at an interface reflection/refraction.
For generic incident beams with elliptic polarizations,
an identical result for the transverse shift of each reflected/transmitted beam
is given by the following different approaches;
(i) analytic evaluation of wavepacket dynamics,
(ii) total angular momentum (TAM) conservation {\it for individual photons},
and (iii) numerical simulation of wavepacket dynamics.
It is consistent with a result by classical electrodynamics.
This means that the TAM conservation
for individual photons is already taken into account
in wave optics, i.e, classical electrodynamics.
Finally, we show an application of our theory
to a two-dimensional photonic crystal,
and propose an optimal design for the enhancement of the optical Hall effect 
in photonic crystals.
\end{abstract}

\maketitle

\section{\label{sec:Introduction}Introduction}
The geometrical phase known as the Berry phase
\cite{Berry}
has been attracting extensive interests in various fields,
e.g., optics, molecular physics, nuclear physics, 
and condensed matter physics
\cite{GPP,GPQS}.
In particular, in condensed matter physics,
important roles of the geometrical phase in electronic 
transport phenomena have been intensively studied 
in the past several years,
and great strides has been made
both in theoretical and experimental researches.
Although a hint of the Berry phase was recognized 
long time ago as the anomalous velocity
in ferromagnets which leads to the anomalous Hall effect
\cite{Karplus-Luttinger,Luttinger}, it was only 
after the discovery of the quantum Hall effect that 
the role of the Berry phase in electron transport began 
to be recognized.
In the quantum Hall system under strong magnetic field, 
the Hall conductance was related to 
the topological integer, i.e., Chern number
\cite{TKNN,Kohmoto,Aoki-Ando}.
A recent development is the finding that the 
Berry phase structure is a fundamental characteristic
of the Bloch wavefunctions even in ordinary systems. From this viewpoint,
the similarity between the anomalous and quantum Hall effects 
has been revealed \cite{MN,Jungwirth,Fang}.
The spin Hall effect based on the geometrical mechanism
are also proposed recently and opened a new stage of spintronics
\cite{MNZ,Sinova}.
All of these effects are understood from the concept 
of the generalized anomalous velocity due to the Berry phase.
In other words, a trajectory of an electron is affected by the Berry phase.

In optics, one can also find some phenomena related to the Berry phase.
A change of polarization of light during propagation such as in helically
wound optical fibers found in the early days
\cite{Rytov,Vladimirski,Pancharatnam} 
has been related with the Berry phase \cite{Chiao-Wu,Tomita-Chiao, Berry-II}. 
It's influence on the trajectory of light has been studied recently by deriving 
a set of semiclassical equations of motion \cite{MSN}.
From this viewpoint, there are several optical phenomena which are now interpreted as
a change of light trajectory due to the Berry phase. 
One is a transverse shift in reflection/refraction at an 
interface between two homogeneous media 
\cite{Fedorov,Imbert,Boulware,Ashby-Miller,Schilling,Fedoseev-I,Fedoseev-II,Pillon},  
(This effect in the case of internal total reflection
is called as Imbert-Fedorov shift.)
The other is a rotation of the 
beam inside an optical fiber, which is sometimes called as an optical
Magnus effect \cite{Dooghin, Liberman-Zeldovich,Bliokh}. 
These phenomena can be coined as 
the optical Hall effect, because of the similarity to 
the topological Hall effects 
\cite{TKNN,Kohmoto,Aoki-Ando,MN,Jungwirth,Fang,MNZ,Sinova}
in electronic systems. 

Attribution of these optical phenomena to the Berry phase is not
merely a re-interpretation, but also can open a new frontier for 
novel phenomena. 
The present authors \cite{MSN} proposed that in 
photonic crystals the optical Hall effect is enhanced by an 
order of magnitude than the above-mentioned examples. 
It is inspired by its electronic counterpart; the topological Hall effects 
are known to be enhanced by periodic potentials, 
particularly when two bands come close in energy.
Thus by designing a photonic crystal to have near-degenerate 
bands, the predicted shift of a light beam is large enough to be
observable in experiments \cite{MSN}.
To calculate and design such photonic crystals in a quantitative way,
a semiclassical Berry-phase theory of optics 
in such photonic crystals is called for.
For this purpose, the approach by the variational principle 
\cite{Jackiw, Pattanayak, Chang-Niu, Sundaram-Niu}, 
which we take in our previous \cite{MSN} and present papers, 
works better than other approaches~\cite{Liberman-Zeldovich,Bliokh};
it is because the approaches in Refs.~\cite{Liberman-Zeldovich,Bliokh}
use an eikonal approximation,
by which it is rather difficult to fully incorporate the vectorial 
nature of the electromagnetic waves for generic cases like photonic crystals.

In the previous work \cite{MSN},
we have briefly reported the essence of the optical Hall effect
and the mechanism of its enhancement 
in artificial crystals called photonic crystals,
i.e., systems with periodic structures of 
dielectric permittivity/magnetic permeability \cite{JMW}.
In the present paper, we construct a semiclassical theory 
of an optical wavepacket (or a photon wavepacket) 
in full detail by keeping its close connection to 
a theory of an electron wavepacket.
It incorporates the Berry phase in a natural way.
The main focus of the present paper is 
to present basics to the extended geometrical optics
applicable to photonic crystals.
This class of artificial crystals are
attracting great interests as new optical materials.
Photonic crystals can be designed to have a desired band structure,
with an aid of first-principle numerical calculations, which enables
the control of many novel properties of lights \cite{JMW,Sakoda}. 
To serve for such purposes, our theory is presented in a transparent
way suitable for such numerical calculations.
As is briefly presented in our previous work \cite{MSN}, 
the effect of the geometrical phase on an optical wavepacket
can be incorporated in the same manner as that in electronic systems.
These generalized equations of motion correctly describe the interplay between
the orbital motion and the internal rotation, 
e.g., polarization, of wavepackets.
The effect similar to the electrical Hall effects in condensed matter
is derived in photonic systems with periodic structures.
Indeed, our equations of motion are analogous to
the semiclassical equations of motion for electron wavepackets 
in solids \cite{Chang-Niu,Sundaram-Niu}.
However, the latter basically considered spinless electrons, 
and in the case of optics, the polarization
degrees of freedom has to be taken into account, 
where the Berry connection is non-Abelian in general.
In this sense, an optical wavepacket is more similar to 
a spinful electron wavepacket.

Below, for simplicity, we focus on a light propagating 
in a non-conducting medium in which there is neither 
electric nor magnetic order,
i.e., the dielectric permittivity and the magnetic permeability
are symmetric tensors. Also their frequency dependences are neglected
for simplicity.
These conditions ensure 
the equation of continuity of electromagnetic energy \cite{Born}
and we can construct the unitary theory of electromagnetic field.
Based on this theory, the semiclassical equations of motion
can be derived on an equal footing with electronic systems
in which the semiclassical equations of motion
are derived from quantum mechanics.
In order to stress the analogy between electronic and photonic systems,
we formulate a theory for Bloch states of electromagnetic field
in a quantum-mechanical formalism.
Although we focus on the unitary theory in this paper,
its extension to a non-unitary version for systems with
electric/magnetic order and conducting systems
would give some insights to the interesting phenomena and proposals, e.g.
the photonic Hall effect
in a scattering media subject to an external magnetic field
\cite{photonic-Hall-th, photonic-Hall-ex},
the magnetically induced deflection due to the Pitaefskii magnetization
\cite{Landau, Rikken-Tiggelen, comment-I, reply-I, comment-II, reply-II}, 
the one-way waveguide of edge states in magnetic 
photonic crystals \cite{Haldane-Raghu, Raghu-Haldane}, and 
Lorentz force on the light due to the toroidal moment \cite{Sawada-Nagaosa}.

The reduction to a system without periodic structure is straightforward. 
Indeed, in the previous work \cite{MSN}, we have presented
a simple application of our theory to the transverse shift in 
the reflection/refraction at an interface,
and found that this shift is governed by
the conservation of total angular momentum (TAM)
{\it for reflected and refracted photons individually}.
We have also numerically demonstrated the validity of our theory
for the case of an incident wavepacket with circular polarization.
However very recently,
the transverse shift evaluated by the conservation of TAM
has been questioned \cite{Bliokh-PRL}
for the cases of incident wavepackets with elliptic polarizations.
In this paper, we present 
an additional way of estimating the transverse shift
from the asymptotic form of the wavepacket,
and also the numerical calculations for the generic polarized states.
As an important consequence from the study on this issue,
we find that an identical result for the shift of each beam is given by
(i) analytic evaluation of wavepacket dynamics,
(ii) TAM conservation for individual photons in Ref.~\cite{MSN}
and (iii) numerically exact simulation of wavepacket dynamics.
This agreement in different approaches 
supports the validity of the present theory
claiming that the transverse shift is governed 
by the conservation of TAM for individual photons.
These results are also consistent with the more conventional approach 
based on classical electrodynamics \cite{Fedoseev-I,Fedoseev-II}.
In other words, the TAM conservation
for individual photons is already taken into account
in wave optics, i.e., classical electrodynamics.

For a broad readership, we divide the main contents 
into two sections; Sec.~\ref{sec:formalism} is 
devoted to formalisms, explaining in full details the 
derivations of the theory and the resulting formulae,
while in Sec.~\ref{sec:applications} we focus on 
two applications of the theory: the transverse shift in interface
reflection/refraction, and 
the optical Hall effect in
a two-dimensional photonic crystal. 
Readers who are mainly interested in the applications can skip 
Sec.~\ref{sec:formalism} and jump to Sec.~\ref{sec:applications}.
For this purpose we make Sec.~\ref{sec:applications}
to be self-consistent.

The plan of this paper is as follows.
In Sec.~\ref{sec:non-conducting}, 
an electromagnetic field in a non-conducting medium
is quantized in the Hamilton-Jacobi formalism 
by introducing the Dirac bracket
for the constrained system.
Some quantum operators for physical observables are
also presented.
Eigen states in a periodic system are discussed 
in Sec.~\ref{sec:eigenfunctions}
for the application to a photonic crystal.
In Sec.~\ref{sec:EOM}, 
we consider a perturbed modulation superimposed
on a background periodic structure
and discuss corrections for the eigen states
and the expectation values
of physical observables for an optical wavepacket.
The equations of motion are derived
taking into account the Berry phase and
the perturbed modulation.
An application of our theory 
to reflection/refraction problem at a flat interface is discussed
in Sec.~\ref{sec:Imbert}
by reducing the theory to the case without periodic structure.
Recently some criticisms are raised against our approach to 
this reflection/refraction problem \cite{Bliokh-PRL}.
Remarks on these criticisms are presented in Sec.~\ref{sec:remarks}.
In Sec.~\ref{sec:photonic-crystal}, we apply our theory
to a modulated two-dimensional photonic crystal
and present some examples of Berry curvatures
and internal rotations in a periodic system.

Section~\ref{sec:Discussion} is devoted to the discussion 
on the implications of the present work to wider range of phenomena
in physics. Related previous works are mentioned here.

\section{\label{sec:formalism}Formalisms}
\subsection{\label{sec:non-conducting}Electromagnetic field in a non-conducting medium}
We consider an electromagnetic field in a non-conducting medium
with a generic modulation but without electric nor magnetic orders,
and begin with the following Lagrangian,
\begin{eqnarray}
L & = & 
\frac{1}{2}\int d\bm{r}\left[
\bm{E}(\bm{r},t)\cdot\bm{D}(\bm{r},t)
-\bm{H}(\bm{r},t)\cdot\bm{B}(\bm{r},t)
\right],
\end{eqnarray}
where 
\begin{subequations}
\begin{eqnarray}
\bm{D}(\bm{r},t) &=& 
\tensor{\epsilon}(\bm{r})\bm{E}(\bm{r},t)
= \tensor{\epsilon}(\bm{r})[-\partial_{t}\bm{A}(\bm{r},t)-\bm{\nabla}_{\bm{r}}\phi(\bm{r},t)]
,\nonumber\\
&&\\
\bm{B}(\bm{r},t) &=& 
\tensor{\mu}(\bm{r})\bm{H}(\bm{r},t) = \bm{\nabla}_{\bm{r}}\times\bm{A}(\bm{r},t)
.
\end{eqnarray}
\end{subequations}
We take the unit in which $\hbar=c=1$ where
$c$ is the speed of light in vacuum.
The medium where light propagates is treated as an insulating material,
which is characterized by the dielectric 
permittivity $\tensor{\epsilon}(\bm{r})$ and 
the magnetic permeability $\tensor{\mu}(\bm{r})$. 
These are assumed to be locally symmetric and real-valued tensors, 
and their frequency dependences are neglected.
As mentioned in Sec.~\ref{sec:Introduction},
the equation of continuity
for the electromagnetic energy holds
under these conditions \cite{Born}.
It should be noted that they are the sufficient conditions 
but not the necessary conditions.
The functional derivatives with respect to $\partial_{t}\bm{A}(\bm{r},t)$
and  $\partial_{t}\phi(\bm{r},t)$ determine 
the canonical momenta $\bm{\pi}(\bm{r},t)$ 
for $\bm{A}(\bm{r},t)$ 
and $\pi_{\phi}(\bm{r},t)$ for $\phi(\bm{r},t)$
respectively.
The former gives the canonical definition of $\bm{\pi}(\bm{r},t)$  as
$\bm{\pi}(\bm{r},t)=-\bm{D}(\bm{r},t)$,
while the latter gives the constraint $\pi_{\phi}(\bm{r},t) = 0$.
Here we introduce the Lagrange multiplier $\lambda_{\phi}(\bm{r},t)$
for this constraint, and the Hamiltonian is given by
\begin{subequations}
\begin{eqnarray}
H & = & 
\int d\bm{r}\bm{\pi}(\bm{r},t)\cdot\partial_{t}\bm{A}(\bm{r},t)-L
+\int d\bm{r}\lambda_{\phi}(\bm{r},t)\pi_{\phi}(\bm{r},t)
\nonumber\\
 & = & H_{0}+\int d\bm{r}\left[-\bm{\pi}(\bm{r})\cdot\bm{\nabla}_{\bm{r}}\phi(\bm{r})
+\lambda_{\phi}(\bm{r})\pi_{\phi}(\bm{r})\right],\\
H_{0} & = & 
\frac{1}{2}\int d\bm{r}\bigl[
\bm{\pi}(\bm{r})\tensor{\epsilon}^{-1}(\bm{r})\bm{\pi}(\bm{r})
\nonumber\\
&&\quad
+[\bm{\nabla}_{\bm{r}}\times\bm{A}(\bm{r})]
\tensor{\mu}^{-1}(\bm{r})[\bm{\nabla}_{\bm{r}}\times\bm{A}(\bm{r})]
\bigr].
\end{eqnarray}
\end{subequations}
In order for the constraint $\pi_{\phi}\approx 0$ to be consistently satisfied,
the following additional constraint is required,
\begin{eqnarray}
\{\pi_{\phi}(\bm{r}),H\}_{\mathrm{P}} 
& = & -\bm{\nabla}_{\bm{r}}\cdot\bm{\pi}(\bm{r})\approx 0,
\end{eqnarray}
where $\{\cdots\}_{\mathrm{P}}$ is the Poisson bracket.
The symbol ``$\approx$'' means "weak equality"
which is satisfied when all constraints are imposed \cite{Dirac}.
When the Poisson brackets among a set of constraints and a Hamiltonian
vanish on a constrained subspace, these constraints are called 
first-class constraints by definition.
On the other hand, when the Poisson brackets of these constraints among themselves 
do not vanish even on the constrained subspace, we call them second-class constraints.
In the present case, $\pi_{\phi}(\bm{r})$ and 
$\bm{\nabla}_{\bm{r}}\cdot\bm{\pi}(\bm{r})$ commute with each other,
and the commutation relation between $\bm{\nabla}_{\bm{r}}\cdot\bm{\pi}(\bm{r})$ 
and the Hamiltonian generates no additional constraint.
So the present system has two of first-class constraints,
\begin{subequations}
\begin{eqnarray}
\chi_{1}(\bm{r}) & \equiv & \pi_{\phi}(\bm{r})\approx0,\\
\chi_{2}(\bm{r}) & \equiv & \bm{\nabla}_{\bm{r}}\cdot\bm{\pi}(\bm{r})\approx0.
\end{eqnarray}
\end{subequations}
In order to make a canonical formalism for such a constrained system,
all the first-class constraints are transformed to be second class 
by introducing gauge fixing conditions.
Here we take the following gauge conditions,
\begin{subequations}
\begin{eqnarray}
\chi_{3}(\bm{r}) & \equiv & \phi(\bm{r})\approx0,\\
\chi_{4}(\bm{r}) & \equiv & \bm{\nabla}_{\bm{r}}\tensor{\epsilon}(\bm{r})\bm{A}(\bm{r})\approx0.
\end{eqnarray}
\end{subequations}
Then the commutation relations between the original constraints and 
the gauge conditions are represented by
\begin{subequations}
\begin{eqnarray}
\tensor{\mathcal{C}}(\bm{r},\bm{r}')
& = & \{\chi_{\alpha}(\bm{r}),\chi_{\beta}(\bm{r}')\}_{\mathrm{P}}
\nonumber\\
& = &\left(\begin{array}{cc}
0 & -\tensor{C}(\bm{r},\bm{r}')\\
\tensor{C}(\bm{r},\bm{r}') & 0\end{array}\right),
\\
\tensor{C}(\bm{r},\bm{r}') &=&  
\left(\begin{array}{cc}
\delta(\bm{r}-\bm{r}') & 0\\
0 & -\bm{\nabla}_{\bm{r}}
\tensor{\epsilon}(\bm{r})\bm{\nabla}_{\bm{r}}\delta(\bm{r}-\bm{r}')
\end{array}\right).
\end{eqnarray}
\end{subequations}

Introducing Lagrange multipliers for the constraints including the gauge conditions, 
we redefine the Hamiltonian as
\begin{eqnarray}
H & = & 
H_{0}+\int d\bm{r}\bm{\lambda}(\bm{r})\cdot\bm{\chi}(\bm{r}),
\end{eqnarray}
where  
$\bm{\lambda}(\bm{r}) 
= [\lambda_{1}(\bm{r}),\lambda_{2}(\bm{r}),\lambda_{3}(\bm{r}),\lambda_{4}(\bm{r})]$ and
$\bm{\chi}(\bm{r}) = [\chi_{1}(\bm{r}),\chi_{2}(\bm{r}),\chi_{3}(\bm{r}),\chi_{4}(\bm{r})]$.
The Lagrange multipliers are determined by the conditions 
$\{\bm{\chi}(\bm{r}),H\}_{\mathrm{P}}\approx0$
and given by
\begin{eqnarray}
\bm{\lambda}(\bm{r}) 
& = & 
-\int d\bm{r}'
\tensor{\mathcal{C}}^{-1}(\bm{r},\bm{r}')\{\bm{\chi}(\bm{r}'),H_{0}\}_{\mathrm{P}},
\end{eqnarray}
where 
\begin{subequations}
\begin{eqnarray}
\tensor{\mathcal{C}}^{-1}(\bm{r},\bm{r}') & = & \left(\begin{array}{cc}
0 & \tensor{C}^{-1}(\bm{r},\bm{r}')\\
-\tensor{C}^{-1}(\bm{r},\bm{r}')& 0
\end{array}\right),
\\
\tensor{C}^{-1}(\bm{r},\bm{r}')  
&=& 
\left(\begin{array}{cc}
\delta(\bm{r}-\bm{r}') & 0\\
0 & g(\bm{r},\bm{r}')\end{array}\right),
\end{eqnarray}
\end{subequations}
and $g(\bm{r},\bm{r}')$ satisfies
\begin{eqnarray}
&&\bm{\nabla}_{\bm{r}}
\tensor{\epsilon}(\bm{r})
\bm{\nabla}_{\bm{r}} g(\bm{r},\bm{r}')
=
\bm{\nabla}_{\bm{r}'}
\tensor{\epsilon}^{T}(\bm{r}')
\bm{\nabla}_{\bm{r}'} g(\bm{r},\bm{r}')
\nonumber\\
&&=-\delta(\bm{r}-\bm{r}').
\end{eqnarray}

As a preparation for the quantum theory,
we introduce the Dirac bracket defined by
\begin{eqnarray}
&&\{ F,G\}_{\mathrm{D}} 
\nonumber\\
&&=  \{ F,G\}_{\mathrm{P}}
-\int d\bm{r}d\bm{r}'
\{ F,\bm{\chi}(\bm{r})\}_{\mathrm{P}}
\tensor{\mathcal{C}}^{-1}(\bm{r},\bm{r}')\{\bm{\chi}(\bm{r}'),G\}_{\mathrm{P}}
.\nonumber\\
\end{eqnarray}
Especially for $\bm{A}(\bm{r})$ and $\bm{\pi}(\bm{r})$,
we obtain the following relation,
\begin{eqnarray}
&&\{ A^{i}(\bm{r}),\pi_{j}(\bm{r}')\}_{\mathrm{D}} 
\nonumber\\
&& =  \delta^{i}_{j}\delta(\bm{r}-\bm{r}')
-\sum_{k}\nabla^{i}_{\bm{r}}\tensor{\epsilon}^{T}_{jk}(\bm{r}')\nabla^{k}_{\bm{r}'}g(\bm{r},\bm{r}'),
\end{eqnarray}
and this leads to the relation between the physical observables, 
\begin{eqnarray}
\{B_{i}(\bm{r}),D_{j}(\bm{r}')\}_{\mathrm{D}}
&=&\sum_{k}\epsilon_{ijk}\nabla^{k}_{\bm{r}}\delta(\bm{r}-\bm{r}'),
\label{eq:comm-BD-classical}
\end{eqnarray}
where $\epsilon_{ijk}$ is 
the completely antisymmetric tensor
defined using $\epsilon_{xyz}=1$.
It is noted that this relation is the same as
that in the vacuum,
while that between $\bm{E}(\bm{r})$
and $\bm{H}(\bm{r})$ is not the case.
The equations of motion are derived as
\begin{subequations}
\begin{eqnarray}
\partial_{t}\bm{A}(\bm{r},t)
&=& \{\bm{A}(\bm{r},t),H\}_{\mathrm{D}}
\approx\tensor{\epsilon}^{-1}(\bm{r})\bm{\pi}(\bm{r},t),
\\
\partial_{t}\bm{\pi}(\bm{r},t) 
& = & \{\bm{\pi}(\bm{r},t),H\}_{\mathrm{D}}
\nonumber\\
&=&-\bm{\nabla}_{\bm{r}}\times
\left[\tensor{\mu}^{-1}(\bm{r})[\bm{\nabla}_{\bm{r}}\times\bm{A}(\bm{r},t)]\right].
\end{eqnarray}
\end{subequations}
The above equations are equivalent
to the Maxwell equations,
\begin{subequations}
\begin{eqnarray}
&&\partial_{t}\bm{D}(\bm{r},t) 
 = 
\bm{\nabla}_{\bm{r}}\times\bm{H}(\bm{r},t) 
,\\
&&\partial_{t}\bm{B}(\bm{r},t)
 =  
-\bm{\nabla}_{\bm{r}}\times\bm{E}(\bm{r},t) 
,\\
&&
\bm{\nabla}_{\bm{r}}\cdot\bm{D}(\bm{r})  =
\bm{\nabla}_{\bm{r}}\cdot\bm{B}(\bm{r}) =  0.
\end{eqnarray}
\end{subequations}

The present system is straightforwardly quantized by
the identification as $i\{F,G\}_{\mathrm{D}}\to [F,G]$.
Especially, the basic commutation relation is quantized as follows,
\begin{eqnarray}
[B_{i}(\bm{r}),D_{j}(\bm{r}')]
& = & i\sum_{k}\epsilon_{ijk}\nabla^{k}_{\bm{r}}\delta(\bm{r}-\bm{r}').
\label{eq:comm-BD}
\end{eqnarray}
Here we introduce some quantum operators
which are useful to check the property of a wavepacket. 
They are the Hamiltonian $H$,
the center of the position $\bm{\mathcal{R}}$ weighted by energy density,
the energy current (the Poynting vector) $\bm{\mathcal{P}}$
and the rotation of energy current $\bm{\mathcal{J}}$,
which are respectively defined by
\begin{subequations}
\begin{eqnarray}
H &=&
\frac{1}{2}\int d\bm{r}
\left[\bm{E}(\bm{r})\cdot\bm{D}(\bm{r})+\bm{H}(\bm{r})\cdot\bm{B}(\bm{r})\right]
,\\
\bm{\mathcal{R}} &=&
\frac{1}{2}\int d\bm{r}
\:\bm{r}\left[\bm{E}(\bm{r})\cdot\bm{D}(\bm{r})+\bm{H}(\bm{r})\cdot\bm{B}(\bm{r})\right]
,\\
\bm{\mathcal{P}}&=&
\frac{1}{2}\int d\bm{r}
\left[\bm{E}(\bm{r})\times\bm{H}(\bm{r})-\bm{H}(\bm{r})\times\bm{E}(\bm{r})\right]
,\\ 
\bm{\mathcal{J}}&=&
\frac{1}{2}\int d\bm{r}
\:\bm{r}\times\left[\bm{E}(\bm{r})\times\bm{H}(\bm{r})-\bm{H}(\bm{r})\times\bm{E}(\bm{r})\right].
\nonumber\\
\end{eqnarray}
\end{subequations}
It should be noted that the last two operators are different from
the momentum and angular momentum operators defined by
\begin{subequations}
\begin{eqnarray}
\bm{P}&=&
\frac{1}{2}\int d\bm{r}
\left[\bm{D}(\bm{r})\times\bm{B}(\bm{r})-\bm{B}(\bm{r})\times\bm{D}(\bm{r})\right]
,\\ 
\bm{J}&=&
\frac{1}{2}\int d\bm{r}
\:\bm{r}\times\left[\bm{D}(\bm{r})\times\bm{B}(\bm{r})-\bm{B}(\bm{r})\times\bm{D}(\bm{r})\right],
\nonumber\\
\end{eqnarray}
\end{subequations}
while $\bm{\mathcal{P}}$ and $\bm{\mathcal{J}}$
are conceptually close to $\bm{P}$ and $\bm{J}$, respectively.
This is because $\bm{\mathcal{P}}$ and $\bm{\mathcal{J}}$
are not necessarily proportional to $\bm{P}$ and $\bm{J}$.
In other words, $\bm{\mathcal{P}}$ and $\bm{\mathcal{J}}$ 
do not necessarily satisfy the algebra of the momentum and the angular momentum.
Therefore, in a system with
translational and rotational symmetries,
what should be conserved are $\bm{P}$ and $\bm{J}$,
rather than $\bm{\mathcal{P}}$ and $\bm{\mathcal{J}}$.
Actually it has been experimentally confirmed that
$\bm{J}$ is conserved in a dielectric medium with rotational symmetry \cite{Kristensen}.
In spite of these shortcomings of $\bm{\mathcal{P}}$ and $\bm{\mathcal{J}}$, 
when a system has no continuous translational nor rotational symmetry,
we focus on $\bm{\mathcal{P}}$ and $\bm{\mathcal{J}}$.
This is because $\bm{\mathcal{P}}$ and $\bm{\mathcal{J}}$ have 
relatively simple expressions even in a periodic system
as shown in Appendix~\ref{sec:periodic-system}.
Especially, a part of $\bm{\mathcal{J}}$ 
suggests a close relation between the internal rotation and 
the Berry curvature in a photonic system,
as well as in the quantum Hall system 
where the internal rotation of an spinless electron
is originated by the cyclotron motion \cite{Chang-Niu}.
However, when a system has continuous translational and rotational symmetries,
we focus on $\bm{P}$ and $\bm{J}$.
This is the case in
the reflection/refraction problem at a flat interface in Sec.~\ref{sec:Imbert}.
The list of physical observables including the above operators
both in electronic and photonic systems are given in Table \ref{tbl:Operators}
for comparison.
\begin{table*}[hbt]
\caption{\label{tbl:Operators}Operators relevant to wavepacket dynamics.}
\begin{ruledtabular}
\begin{tabular}{lll}
&
\textbf{Electronic system}
&
\textbf{Photonic system}
\\
\hline
$
H
$
&
$\int d\bm{r}\:\psi^{\dagger}(\bm{r})\hat{H}(\bm{r})\psi(\bm{r})$
&
$\frac{1}{2}\int d\bm{r}\:\left[\bm{E}(\bm{r})\cdot\bm{D}(\bm{r})
+\bm{H}(\bm{r})\cdot\bm{B}(\bm{r})\right]$\\
$\bm{R}$ &
$\int d\bm{r}\:\bm{r}\psi^{\dagger}(\bm{r})\psi(\bm{r})$
& undefined\\
$\bm{P}$ &
$\int d\bm{r}\:\psi^{\dagger}(\bm{r})
\left[-i\bm{\nabla}_{\bm{r}}-e\bm{A}(\bm{r})+e\bm{B}\times\bm{r}\right]\psi(\bm{r})$
& 
$\frac{1}{2}\int d\bm{r}\:\left[\bm{D}(\bm{r})\times\bm{B}(\bm{r})-\bm{B}(\bm{r})\times\bm{D}(\bm{r})\right]$\\
$\bm{J}$ &
$\int d\bm{r}\:\psi^{\dagger}(\bm{r})\left[\bm{r}\times(-i\bm{\nabla}_{\bm{r}})+\hat{\bm{s}}\right]\psi(\bm{r})$
& 
$\frac{1}{2}\int d\bm{r}\:\bm{r}\times\left[\bm{D}(\bm{r})\times\bm{B}(\bm{r})
-\bm{B}(\bm{r})\times\bm{D}(\bm{r})\right]$\\
$\bm{I}$ &
$\int d\bm{r}\:\psi^{\dagger}(\bm{r})e\hat{v}(\bm{r})\psi(\bm{r})$
& 
undefined\\
$\bm{M}$ &
$\int d\bm{r}\:\psi^{\dagger}(\bm{r})\left[\frac{e}{2}\:\bm{r}\times\hat{v}(\bm{r})+g\mu_{B}\hat{\bm{s}}\right]\psi(\bm{r})$
& 
undefined\\
$\bm{\mathcal{R}}$ &
$\frac{1}{2}\int d\bm{r}\:\psi^{\dagger}(\bm{r})\left\{\hat{H}(\bm{r}),\bm{r}\right\}\psi(\bm{r})$
&
$\frac{1}{2}\int d\bm{r}\:\bm{r}\left[\bm{E}(\bm{r})\cdot\bm{D}(\bm{r})
+\bm{H}(\bm{r})\cdot\bm{B}(\bm{r})\right]
$\\
$\bm{\mathcal{P}}$ 
&
$\frac{i}{2}\int d\bm{r}\left[H,\psi^{\dagger}(\bm{r})
\left\{\hat{H}(\bm{r}),\bm{r}\right\}
\psi(\bm{r})\right]$
&
$\frac{1}{2}\int d\bm{r}\:\left[\bm{E}(\bm{r})\times\bm{H}(\bm{r})
-\bm{H}(\bm{r})\times\bm{E}(\bm{r})\right]$\\
$\bm{\mathcal{J}}$ 
&
$\frac{i}{2}\int d\bm{r}\:\bm{r}\times
\left[H,\psi^{\dagger}(\bm{r})
\left\{\hat{H}(\bm{r}),\bm{r}\right\}
\psi(\bm{r})\right]$
& 
$\frac{1}{2}\int d\bm{r}\:\bm{r}\times\left[\bm{E}(\bm{r})\times\bm{H}(\bm{r})
-\bm{H}(\bm{r})\times\bm{E}(\bm{r})\right]$\\
\hline
\multicolumn{3}{l}{
$H$: Hamiltonian, \quad
$\hat{H}(\bm{r})$: first-quantized Hamiltonian
}\\
\multicolumn{3}{l}{
$\bm{R}$: position (dipole moment),\quad
$\bm{P}$: (pseudo) momentum,\quad
$\bm{J}$: angular momentum,\quad
$\hat{s}$: spin matrix
}\\
\multicolumn{3}{l}{
$\bm{I}$: charge current,\quad
$\bm{M}$: magnetic moment,\quad
$\hat{v}(\bm{r})$: first-quantized velocity
}\\
\multicolumn{3}{l}{
$\bm{\mathcal{R}}$: position weighted by energy density,\quad
$\bm{\mathcal{P}}$: energy current,\quad
$\bm{\mathcal{J}}$: rotation of energy current
}\\
\end{tabular}
\end{ruledtabular}
\end{table*}

Finally, it should be noted that the optical Hall effect comes 
from the particle-wave duality of an optical wavepacket 
and the geometrical/topological property of a wavefunction.
Therefore, this effect can be observed in a macroscopic wavepacket of light
described by classical electrodynamics,
when a wavepacket under consideration is approximately coherent.
In this sense, the second quantization is not always necessary.
The second quantization is adopted, for convenience,
to calculate a motion of a wavepacket on an equal footing with 
that of an electronic system, as shown in Sec.~\ref{sec:EOM}.
As long as we consider an approximately coherent wavepacket 
in a single particle approximation of quantum theory of photon
or in a linear approximation of classical electrodynamics,
results obtained by both formalisms coincide with each other
as shown in Sec.~\ref{sec:Imbert}.
Detailed remarks on the relation between quantum and classical
pictures of the optical Hall effect
is given in Appendix~\ref{sec:quantum-classical}.

\subsection{\label{sec:eigenfunctions}Eigenfunctions in a periodic system}
Here we introduce eigenfunctions in a periodic system, 
\begin{eqnarray}
\bm{\Phi}^{F}_{n\lambda\bm{k}}(\bm{r},t) 
&=& e^{-iE_{n\bm{k}}t}\bm{\Phi}^{F}_{n\lambda\bm{k}}(\bm{r})
=
\frac{e^{i\bm{k}\cdot\bm{r}-iE_{n\bm{k}}t}}{\sqrt{2E_{n\bm{k}}}}
\bm{U}^{F}_{n\lambda\bm{k}}(\bm{r}),
\nonumber\\
\end{eqnarray}
where $F=E$ or $H$. The symbols $n$, $\lambda$ and $\bm{k}$ represent
the band index, the index for degenerate modes in the $n$-th band
and the lattice momentum, respectively,
and $E_{n\bm{k}}$ is the energy eigenvalue of the $n$-th band,
which may be degenerate.
It should be noted that the band index $n$ is not needed
in locally isotropic systems without periodic structure,
but we must keep the index $\lambda$
to distinguish different polarization states.
$\bm{U}^{E}_{n\lambda\bm{k}}(\bm{r})$ 
and $\bm{U}^{H}_{n\lambda\bm{k}}(\bm{r})$ are Bloch functions
for electric field and magnetic field, respectively.
It should be noted that the lattice momentum $\bm{k}$ will be restricted to
the first Brillouin zone in the rest of this paper.
The eigenfunctions satisfy the Maxwell equations,
\begin{subequations}
\begin{eqnarray}
\tensor{\epsilon}(\bm{r})\partial_{t}\bm{\Phi}^{E}_{n\lambda\bm{k}}(\bm{r},t)
&=& \bm{\nabla}_{\bm{r}}\times\bm{\Phi}^{H}_{n\lambda\bm{k}}(\bm{r},t) ,
\label{eq:Maxwell-PhiEH}
\\
\tensor{\mu}(\bm{r})\partial_{t}\bm{\Phi}^{H}_{n\lambda\bm{k}}(\bm{r},t)
&=& 
-\bm{\nabla}_{\bm{r}}\times\bm{\Phi}^{E}_{n\lambda\bm{k}}(\bm{r},t) ,
\label{eq:Maxwell-PhiHE}
\\
\bm{\nabla}_{\bm{r}}\tensor{\epsilon}(\bm{r})\bm{\Phi}^{E}_{n\lambda\bm{k}}(\bm{r},t)  &=& 
\bm{\nabla}_{\bm{r}}\tensor{\mu}(\bm{r})\bm{\Phi}^{H}_{n\lambda\bm{k}}(\bm{r},t)  =  0,
\end{eqnarray}
\end{subequations}
and they lead to the following eigen equations,
\begin{subequations}
\begin{eqnarray}
\bm{\nabla}_{\bm{r}}\times
\left[\tensor{\mu}^{-1}\bm{\nabla}\times\bm{\Phi}^{E}_{n\lambda\bm{k}}(\bm{r})\right]
&=&
\tensor{\epsilon}(\bm{r})E^{2}_{n\bm{k}}\bm{\Phi}^{E}_{n\lambda\bm{k}}(\bm{r}),
\label{eq:eigen-PhiE}
\\
\bm{\nabla}_{\bm{r}}\times
\left[\tensor{\epsilon}^{-1}\bm{\nabla}\times\bm{\Phi}^{H}_{n\lambda\bm{k}}(\bm{r})\right]
&=& 
\tensor{\mu}(\bm{r})E^{2}_{n\bm{k}}\bm{\Phi}^{H}_{n\lambda\bm{k}}(\bm{r}).
\label{eq:eigen-PhiH}
\end{eqnarray}
\end{subequations}
In the case of $\tensor{\epsilon}^{T}(\bm{r})=\tensor{\epsilon}(\bm{r})$ 
and $\tensor{\mu}^{T}(\bm{r})=\tensor{\mu}(\bm{r})$,
we can orthonormalize the Bloch functions with the same lattice momentum $\bm{k}$ as,
\begin{subequations}
\begin{eqnarray}
\int_{\mathrm{WS}}\frac{d\bm{r}}{v_{\mathrm{WS}}}\:
\bm{U}^{E*}_{n\lambda\bm{k}}(\bm{r})
\tensor{\epsilon}(\bm{r})\bm{U}^{E}_{n'\lambda'\bm{k}}(\bm{r})
&=&\delta_{nn'}\delta_{\lambda\lambda'},
\label{eq:orthonormal-UE}
\\
\int_{\mathrm{WS}}\frac{d\bm{r}}{v_{\mathrm{WS}}}\:
\bm{U}^{H*}_{n\lambda\bm{k}}(\bm{r})
\tensor{\mu}(\bm{r})\bm{U}^{H}_{n'\lambda'\bm{k}}(\bm{r})
&=&\delta_{nn'}\delta_{\lambda\lambda'},
\label{eq:orthonormal-UH}
\end{eqnarray}
\end{subequations}
where the domain of integration is the unit cell
with the volume $v_{\mathrm{WS}}$.
The orthonormality for the eigen functions
will be discussed later.

We introduce the 
Fourier transformation,
\begin{subequations}
\begin{eqnarray}
\bm{U}^{F}_{n\lambda\bm{k}}(\bm{G})
&=& \int_{\mathrm{WS}}\frac{d\bm{r}}{v_{\mathrm{WS}}}\:
e^{-i\bm{G}\cdot\bm{r}}\bm{U}^{F}_{n\lambda\bm{k}}(\bm{r})
,\\
\tensor{\epsilon}(\bm{G},\bm{G}')
&=& \int_{\mathrm{WS}}\frac{d\bm{r}}{v_{\mathrm{WS}}}\:
e^{-i(\bm{G}-\bm{G}')\cdot\bm{r}}\tensor{\epsilon}(\bm{r})
,\\
\tensor{\mu}(\bm{G},\bm{G}')
&=& \int_{\mathrm{WS}}\frac{d\bm{r}}{v_{\mathrm{WS}}}\:
e^{-i(\bm{G}-\bm{G}')\cdot\bm{r}}\tensor{\mu}(\bm{r}),
\end{eqnarray}
\end{subequations}
where $F=E$ or $H$.
$\bm{G}$ represents a reciprocal lattice vector.
In terms of the above representations in the Fourier space,
we introduce the following compact representation
for the latter convenience,
\begin{eqnarray}
|U\rangle
&=& [\bm{U}(\bm{G}_{0}), \bm{U}(\bm{G}_{1}), \bm{U}(\bm{G}_{2}), \cdots], 
\end{eqnarray}
and the tensors,
\begin{subequations}
\begin{eqnarray}
&&\bm{P}_{\bm{k}}(\bm{G},\bm{G}')
= (\bm{k}+\bm{G})\delta(\bm{G},\bm{G}') = \bm{K}\delta(\bm{G},\bm{G}')
,\nonumber\\
\\
&&\left[S_{i}\right]_{jk}(\bm{G},\bm{G}') 
= -i\epsilon_{ijk}\delta(\bm{G},\bm{G}')
,\\
&&\Xi^{E}_{\bm{k}}
=\bm{P}_{\bm{k}}\cdot\bm{S}\tensor{\mu}^{-1}\bm{P}_{\bm{k}}\cdot\bm{S}
\label{eq:XiE}
,\\
&&\Xi^{H}_{\bm{k}}
=\bm{P}_{\bm{k}}\cdot\bm{S}\tensor{\epsilon}^{-1}\bm{P}_{\bm{k}}\cdot\bm{S}
\label{eq:XiH}
,
\end{eqnarray}
\end{subequations}
where we have introduced the abbreviation $\bm{K}=\bm{k}+\bm{G}$.
The inner product of the Bloch functions and
the algebra of the above tensors are represented as,
\begin{subequations}
\begin{eqnarray}
\langle U|V\rangle 
&=& 
\int_{\mathrm{WS}}\frac{d\bm{r}}{v_{\mathrm{WS}}}
\bm{U}^{*}(\bm{r})\cdot\bm{V}(\bm{r})
,\\
\langle U|i\bm{S}|V\rangle 
&=& 
\int_{\mathrm{WS}}\frac{d\bm{r}}{v_{\mathrm{WS}}}
\bm{U}^{*}(\bm{r})\times\bm{V}(\bm{r})
,\\
-i\bm{P}_{\bm{k}}\cdot\bm{S}|U\rangle 
&=& 
[\bm{K}\times\bm{U}(\bm{G}_{0}), \bm{K}\times\bm{U}(\bm{G}_{1}), \cdots]
.
\end{eqnarray}
\end{subequations}
Thus the orthonormality is rewritten as,
\begin{subequations}
\begin{eqnarray}
\langle U^{E}_{n\lambda\bm{k}}|\tensor{\epsilon}|U^{E}_{n'\lambda'\bm{k}}\rangle
&=& \delta_{nn'}\delta_{\lambda\lambda'},
\\
\langle U^{H}_{n\lambda\bm{k}}|\tensor{\mu}|U^{H}_{n'\lambda'\bm{k}}\rangle
&=& \delta_{nn'}\delta_{\lambda\lambda'}.
\end{eqnarray}
\end{subequations}

By this notation convention, the Maxwell equations
for the Bloch functions
are represented 
in the following compact forms,
\begin{subequations}
\begin{eqnarray}
\tensor{\epsilon} E_{n\bm{k}}|U^{E}_{n\lambda\bm{k}}\rangle
&=& i\bm{P}_{\bm{k}}\cdot\bm{S}|U^{H}_{n\lambda\bm{k}}\rangle,
\label{eq:Maxwell-UEH}\\
\tensor{\mu} E_{n\bm{k}}|U^{H}_{n\lambda\bm{k}}\rangle
&=& -i\bm{P}_{\bm{k}}\cdot\bm{S}|U^{E}_{n\lambda\bm{k}}\rangle,
\label{eq:Maxwell-UHE}\\
\langle K|\tensor{\epsilon}|U^{E}_{n\lambda\bm{k}}\rangle
&=& 
\langle K|\tensor{\mu}|U^{H}_{n\lambda\bm{k}}\rangle 
= 0,\label{eq:Maxwell-KUE-KUH}
\end{eqnarray}
\end{subequations}
where
\begin{eqnarray}
|K\rangle
&=& [0,\cdots, 0, \bm{k}+\bm{G}, 0, \cdots].
\end{eqnarray}
From Eqs.~(\ref{eq:Maxwell-UEH}) and (\ref{eq:Maxwell-UHE}),
we can easily derive the following equations,
\begin{subequations}
\begin{eqnarray}
\Xi^{E}_{\bm{k}}|U^{E}_{n\lambda\bm{k}}\rangle
&=&
\tensor{\epsilon} E^{2}_{n\bm{k}}|U^{E}_{n\lambda\bm{k}}\rangle,
\label{eq:eigen-UE}\\
\Xi^{H}_{\bm{k}}|U^{H}_{n\lambda\bm{k}}\rangle
&=&
\tensor{\mu} E^{2}_{n\bm{k}}|U^{H}_{n\lambda\bm{k}}\rangle.
\label{eq:eigen-UH}
\end{eqnarray}
\end{subequations}

In relativistic systems, 
the orthonormality for the eigenfunctions
are conventionally represented in terms of
the inner product defined by
\begin{eqnarray}
(f|g) 
& = & i\int d\bm{r}\left[
\bm{f}^{*}(\bm{r},t)\cdot[\partial_{t}\bm{g}(\bm{r},t)]
-[\partial_{t}\bm{f}^{*}(\bm{r},t)]\cdot\bm{g}(\bm{r},t)
\right],
\nonumber\\
\label{eq:inner-Phi}
\end{eqnarray}
and we obtain the following orthonormality
relation as shown in Appendix~\ref{sec:orthonormality},
\begin{subequations}
\begin{eqnarray}
&&(\Phi^{E}_{n\lambda\bm{k}}|\tensor{\epsilon}
|\Phi^{E}_{n'\lambda'\bm{k}'})
 =  \delta_{nn'}\delta_{\lambda\lambda'}\tilde{\delta}(\bm{k}-\bm{k}'),
\label{eq:orthonormal-PhiE}
\\
&&(\Phi^{H}_{n\lambda\bm{k}}|\tensor{\mu}
|\Phi^{H}_{n'\lambda'\bm{k}'})
 =  \delta_{nn'}\delta_{\lambda\lambda'}\tilde{\delta}(\bm{k}-\bm{k}'),
\label{eq:orthonormal-PhiH}
\\
&&(\Phi^{E*}_{n\lambda\bm{k}}|\tensor{\epsilon}|\Phi^{E}_{n'\lambda'\bm{k}'})
 =  (\Phi^{H*}_{n\lambda\bm{k}}|\tensor{\mu}|\Phi^{H}_{n'\lambda'\bm{k}'}) = 0.
\label{eq:orthogonality-Phi}
\end{eqnarray}
\end{subequations}
where 
$\tilde{\delta}(\bm{k}-\bm{k}')
=(2\pi)^{3}\delta(\bm{k}-\bm{k}')$.
It should be noted that
we have used $\tensor{\epsilon}^{T}(\bm{r})=\tensor{\epsilon}(\bm{r})$ 
and $\tensor{\mu}^{T}(\bm{r})=\tensor{\mu}(\bm{r})$
in the derivation of the above relations.
This orthonormality is required
to expand the electric and magnetic fields in terms 
of the eigenfunctions as follows,
\begin{subequations}
\begin{eqnarray}
\bm{E}(\bm{r},t) 
& = & \sum_{n,\lambda}\int_{\mathrm{BZ}}d\bm{k}E_{n\bm{k}}
\nonumber\\
&&\times
\left[
\bm{\Phi}^{E}_{n\lambda\bm{k}}(\bm{r},t)a_{n\lambda\bm{k}}
+\bm{\Phi}^{E*}_{n\lambda\bm{k}}(\bm{r},t)a_{n\lambda\bm{k}}^{\dagger}
\right]
,\\
\bm{H}(\bm{r},t) 
& = & \sum_{n,\lambda}\int_{\mathrm{BZ}}d\bm{k}E_{n\bm{k}}
\nonumber\\
&&\times
\left[
\bm{\Phi}^{H}_{n\lambda\bm{k}}(\bm{r},t)a_{n\lambda\bm{k}}
+\bm{\Phi}^{H*}_{n\lambda\bm{k}}(\bm{r},t)a_{n\lambda\bm{k}}^{\dagger}
\right]
,
\end{eqnarray}
\end{subequations}
where the $\bm{k}$-integration is 
over the first Brillouin zone, i.e.,
\begin{eqnarray}
\int_{\mathrm{BZ}}d\bm{k} &=&
\int_{\bm{k}\in\text{1st BZ}}\frac{d\bm{k}}{(2\pi)^{3}}.
\end{eqnarray}
The operators $a_{n\lambda\bm{k}}$ and $a^{\dagger}_{n\lambda\bm{k}}$ are defined by
\begin{subequations}
\begin{eqnarray}
a_{n\lambda\bm{k}} &=& 
\frac{1}{E_{n\bm{k}}}(\Phi^{E}_{n\lambda\bm{k}}|\tensor{\epsilon}|E)
= \frac{1}{E_{n\bm{k}}}(\Phi^{H}_{n\lambda\bm{k}}|\tensor{\mu}|H)
\nonumber\\
&=& \int d\bm{r}\left[
\bm{\Phi}^{E*}_{n\lambda\bm{k}}(\bm{r})\cdot\bm{D}(\bm{r})
+\bm{\Phi}^{H*}_{n\lambda\bm{k}}(\bm{r})\cdot\bm{B}(\bm{r})
\right],
\nonumber\\
\\
a^{\dagger}_{n\lambda\bm{k}} &=& 
\frac{1}{E_{n\bm{k}}}(E|\tensor{\epsilon}|\Phi^{E}_{n\lambda\bm{k}})
= \frac{1}{E_{n\bm{k}}}(H|\tensor{\mu}|\Phi^{H}_{n\lambda\bm{k}})
\nonumber\\
&=& \int d\bm{r}\left[
\bm{D}(\bm{r})\cdot\bm{\Phi}^{E}_{n\lambda\bm{k}}(\bm{r})
+\bm{B}(\bm{r})\cdot\bm{\Phi}^{H}_{n\lambda\bm{k}}(\bm{r})
\right],
\nonumber\\
\end{eqnarray}
\end{subequations}
By using Eq.~(\ref{eq:comm-BD}),
the following commutation relation is obtained,
\begin{eqnarray}
\left[a_{n\lambda\bm{k}},a_{n'\lambda'\bm{k}'}^{\dagger}\right]
&=&
\delta_{nn'}\delta_{\lambda\lambda'}\tilde{\delta}(\bm{k}-\bm{k}')
.
\end{eqnarray}

\begin{table*}[hbt]
\caption{\label{tbl:Berry}Berry connection and curvature.}
\begin{ruledtabular}
\begin{tabular}{lll}
&
\textbf{Electronic system}
&
\textbf{Photonic system}
\\
\hline
Bloch function &
$|U_{n\lambda\bm{k}}\rangle$
&
$|U^{E}_{n\lambda\bm{k}}\rangle$,
\quad $|U^{H}_{n\lambda\bm{k}}\rangle$\\
Normalization 
&
$\langle U_{n\lambda\bm{k}}|U_{n'\lambda'\bm{k}}\rangle
=\delta_{nn'}\delta_{\lambda\lambda'}$
&
$
\langle U^{E}_{n\lambda\bm{k}}|\tensor{\epsilon}|U^{E}_{n'\lambda'\bm{k}}\rangle
=\langle U^{H}_{n\lambda\bm{k}}|\tensor{\mu}|U^{H}_{n'\lambda\bm{k}}\rangle
=\delta_{nn'}\delta_{\lambda\lambda'}
$\\
Berry connection
&
$
\left[\bm{\Lambda}_{n\bm{k}}\right]_{\lambda\lambda'}=
-i\langle U_{n\lambda\bm{k}}|\bm{\nabla}_{\bm{k}}U_{n\lambda'\bm{k}}\rangle
$
&
$
\bm{\Lambda}_{n\bm{k}}
=\frac{1}{2}\left[\bm{\Lambda}^{E}_{n\bm{k}}+\bm{\Lambda}^{H}_{n\bm{k}}\right]
$
\\
&&
$
\left[\bm{\Lambda}^{E}_{n\bm{k}}\right]_{\lambda\lambda'}=
-i\langle U^{E}_{n\lambda\bm{k}}|\tensor{\epsilon}|\bm{\nabla}_{\bm{k}}U^{E}_{n\lambda'\bm{k}}\rangle
$\\
&&
$
\left[\bm{\Lambda}^{H}_{n\bm{k}}\right]_{\lambda\lambda'}=
-i\langle U^{H}_{n\lambda\bm{k}}|\tensor{\mu}|\bm{\nabla}_{\bm{k}}U^{H}_{n\lambda'\bm{k}}\rangle
$
\\
Berry curvature
&
$\bm{\Omega}_{n\bm{k}} 
= 
\bm{\nabla}_{\bm{k}}\times\bm{\Lambda}_{n\bm{k}}+i\bm{\Lambda}_{n\bm{k}}\times\bm{\Lambda}_{n\bm{k}}$
&
$
\bm{\Omega}_{n\bm{k}} 
= 
\bm{\nabla}_{\bm{k}}\times\bm{\Lambda}_{n\bm{k}}+i\bm{\Lambda}_{n\bm{k}}\times\bm{\Lambda}_{n\bm{k}}
$
\end{tabular}
\end{ruledtabular}
\end{table*}

\subsection{\label{sec:EOM}Equations of motion}
In order to see the effect of geometrical phase on
the trajectory of a wavepacket,
a driving force is needed \cite{note-force}.
This is because the geometrical effect is given by
the vector product between the driving force and the Berry curvature
as we shall see later.
In an electronic system, a driving force is most conventionally 
produced by the gradient of electric potential.
The counterpart in a photonic system is given by the gradient of 
$\tensor{\epsilon}(\bm{r})$ or $\tensor{\mu}(\bm{r})$.
Of course, a periodic structure itself gives a gradient.
However, this effect is exactly taken into account 
by considering optical Bloch states
as shown in Appendix~\ref{sec:periodic-system}
where readers can find details about the basic features 
of an optical wavepacket in a periodic system.
Thus we regard the deviation from a periodic structure as a driving force
for the Bloch states and treat it perturbatively.
Here, the perturbation is introduced 
as a modulation superimposed onto
the periodic structure of $\tensor{\epsilon}(\bm{r})$ and $\tensor{\mu}(\bm{r})$ as
\begin{eqnarray}
&&\tensor{\epsilon}^{-1}(\bm{r}) \to \gamma^{2}_{\epsilon}(\bm{r})\tensor{\epsilon}^{-1}(\bm{r}),
\quad
\tensor{\mu}^{-1}(\bm{r}) \to \gamma^{2}_{\mu}(\bm{r})\tensor{\mu}^{-1}(\bm{r}),
\label{eq:modulation}
\nonumber
\\
\end{eqnarray}
where $\gamma_{\epsilon}(\bm{r})$ and $\gamma_{\mu}(\bm{r})$ are scalar functions,
and we call them ``the modulation functions'' hereafter.
This kind of modulation does not change the local symmetries of 
$\tensor{\epsilon}(\bm{r})$ and $\tensor{\mu}(\bm{r})$,
and does not violate the energy conservation.
We summarize the definitions of Berry connection and curvature in Table~\ref{tbl:Berry}
and the main results obtained here, i.e.,
the equations of motion for an optical wavepacket,
in Table~{\ref{tbl:EOM}.
Appendix~\ref{sec:modulated-system} supplements
details about the commutation relations and expectation values of various operators
which are needed to derive the equations of motion in a modulated system.

Now we derive the equations of motion for the dynamics of an optical wavepacket.
An exact wavefunction $|\Psi\rangle$ satisfies
the Schr{\" o}dinger equation,
\begin{eqnarray}
i\frac{d}{dt}|\Psi\rangle
&=&H|\Psi\rangle.
\end{eqnarray}
This equation is derived by applying the variational principle to the quantity
\cite{Jackiw},
\begin{eqnarray}
L&=&\langle \Psi |i\frac{d}{dt}-H|\Psi\rangle
\end{eqnarray}
It is natural to consider that
the trajectory of the wavepacket is determined in terms of 
the effective Lagrangian which is given by replacing
$|\Psi\rangle$ with a variational wavepacket, $|W\rangle$,
characterized only by the centers $\bm{r}_{c}$, $\bm{k}_{c}$
of the position and wavevector, respectively,
and the polarization state \cite{Pattanayak}.
Although $|W\rangle$ can be brought closer to $|\Psi\rangle$
by enlarging the number of variational parameters,
those concerning the details of the wavepacket are neglected here.
This approximation is justified when a modulation is weak and slowly varying,
and has been successfully applied to the quantum Hall system
and gives the semiclassical understanding of the motion of magnetic Bloch states 
\cite{Chang-Niu,Sundaram-Niu,note-second-quantization}.

In general, the modulation may mix the creation and annihilation operators.
Here we consider the situation in which the modulation is 
sufficiently weak and time-independent, and this mixing is negligible.
Therefore, the approximated wavepacket can be constructed as
\begin{subequations}
\begin{eqnarray}
&&|W\rangle =
\int_{\mathrm{BZ}}d\bm{k}\:
w(\bm{k},\bm{k}_{c},\bm{r}_{c},z_{c},t)
\sum_{\lambda}z_{c\lambda}a^{\dagger}_{n\lambda\bm{k};\bm{r}_{c}}|0\rangle
,\nonumber\\
&&w(\bm{k},\bm{k}_{c},\bm{r}_{c},z_{c},t)
=
w_{r}(\bm{k}-\bm{k}_{c})e^{-i\vartheta(\bm{k},\bm{r}_{c},z_{c},t)},
\end{eqnarray}
\end{subequations}
where $w_{r}(\bm{k}-\bm{k}_{c})$ is a real function, and 
$w_{r}(\bm{k}-\bm{k}_{c})$ and $z_{c\lambda}$ satisfy
the normalization conditions,
$\int_{\mathrm{BZ}}d\bm{k}\:
w^{2}_{r}(\bm{k}-\bm{k}_{c})=1$
and $\sum_{\lambda}|z_{c\lambda}|^{2}=1$, respectively.
We assume $w_{r}(\bm{k}-\bm{k}_{c})$ has a sharp
peak around 
$\bm{k}_{c} =\int_{\mathrm{BZ}}d\bm{k}\:
w^{2}_{r}(\bm{k}-\bm{k}_{c})\bm{k}$.
Here we require that the center of wavepacket, $\bm{r}_{c}$, 
is self-consistently determined by
\begin{eqnarray}
\bm{r}_{c}
&=&
\int_{\mathrm{BZ}}d\bm{k}\:
w^{2}_{r}(\bm{k}-\bm{k}_{c})
\left[\bm{\nabla}_{\bm{k}}\vartheta(\bm{k},\bm{r}_{c},z_{c},t)
-(z_{c}|\bm{\Lambda}_{n\bm{k}}|z_{c})\right],
\nonumber\\
\label{eq:rc-perturbed}
\end{eqnarray}
where $\bm{\Lambda}_{n\bm{k}}$ is the Berry connection defined by
\begin{subequations}
\begin{eqnarray}
\bm{\Lambda}_{n\bm{k}}
&=& 
\frac{1}{2}
\left[\bm{\Lambda}^{E}_{n\bm{k}}+\bm{\Lambda}^{H}_{n\bm{k}}\right]
,\label{eq:Lambda}\\
\left[\bm{\Lambda}^{E}_{n\bm{k}}\right]_{\lambda\lambda'}
&=&
-i\langle U^{E}_{n\lambda\bm{k}}|\tensor{\epsilon}
|\bm{\nabla}_{\bm{k}}U^{E}_{n\lambda'\bm{k}}\rangle
, \label{eq:Lambda-E}\\
\left[\bm{\Lambda}^{H}_{n\bm{k}}\right]_{\lambda\lambda'}
&=&
-i\langle U^{H}_{n\lambda\bm{k}}|\tensor{\mu}
|\bm{\nabla}_{\bm{k}}U^{H}_{n\lambda'\bm{k}}\rangle
, \label{eq:Lambda-H}
\end{eqnarray}
\end{subequations}
and we introduced the abbreviation, 
\begin{eqnarray}
(z_{c}|M|z_{c}) &=& 
\sum_{\lambda,\lambda'} z^{*}_{c\lambda}M_{\lambda\lambda'}z_{c\lambda'}.
\end{eqnarray}
The annihilation and creation operators of approximate eigen modes are defined by
\begin{widetext}
\begin{subequations}
\begin{eqnarray}
a_{n\lambda\bm{k};\bm{r}_{c}}
&=& \int d\bm{r}\biggl[
\sqrt{\frac{\gamma_{\epsilon}(\bm{r}_{c})}{\gamma_{\mu}(\bm{r}_{c})}}
\bm{\Phi}^{E*}_{n\lambda\bm{k}}(\bm{r})\cdot\bm{D}(\bm{r})
+\sqrt{\frac{\gamma_{\mu}(\bm{r}_{c})}{\gamma_{\epsilon}(\bm{r}_{c})}}
\bm{\Phi}^{H*}_{n\lambda\bm{k}}(\bm{r})\cdot\bm{B}(\bm{r})
\biggr]
,\\
a^{\dagger}_{n\lambda\bm{k};\bm{r}_{c}}
&=& \int d\bm{r}\biggl[
\sqrt{\frac{\gamma_{\epsilon}(\bm{r}_{c})}{\gamma_{\mu}(\bm{r}_{c})}}
\bm{D}(\bm{r})\cdot\bm{\Phi}^{E}_{n\lambda\bm{k}}(\bm{r})
+\sqrt{\frac{\gamma_{\mu}(\bm{r}_{c})}{\gamma_{\epsilon}(\bm{r}_{c})}}
\bm{B}(\bm{r})\cdot\bm{\Phi}^{H}_{n\lambda\bm{k}}(\bm{r})
\biggr].
\end{eqnarray}
\end{subequations}
\end{widetext}
These operators
satisfy the same commutation relation as that in a periodic system,
\begin{eqnarray}
\left[a_{n\lambda\bm{k};\bm{r}_{c}},
a^{\dagger}_{n'\lambda'\bm{k}';\bm{r}_{c}}\right]
&=&
\delta_{nn'}\delta_{\lambda\lambda'}\tilde{\delta}(\bm{k}-\bm{k}').
\end{eqnarray}
The approximate eigen modes depend on the variable $\bm{r}_{c}$.
Thus we must consider also the operator
$\bm{\nabla}_{\bm{r}_{c}}a^{\dagger}_{n\lambda\bm{k};\bm{r}_{c}}$
when estimating the effective Lagrangian.
However, we can show that the contribution from 
$\bm{\nabla}_{\bm{r}_{c}}a^{\dagger}_{n\lambda\bm{k};\bm{r}_{c}}$
vanishes,
by using 
$
\left[a_{n\lambda\bm{k};\bm{r}_{c}},
\bm{\nabla}_{\bm{r}_{c}}a^{\dagger}_{n'\lambda'\bm{k}';\bm{r}_{c}}\right]=0$
and $a_{n\lambda\bm{k};\bm{r}_{c}}|0\rangle=0$.

As was mentioned above, we consider the situation in which
the mixing of the approximate annihilation and creation operators
is negligible and assume that the relation
$a_{n\lambda\bm{k};\bm{r}_{c}}|0\rangle = 0$, holds.
The expectation values of $H$ and $\bm{\mathcal{R}}$ are estimated 
by the derivative expansion with respect to
$\gamma_{\epsilon}(\bm{r})$ and $\gamma_{\mu}(\bm{r})$
as
\begin{subequations}
\begin{eqnarray}
\langle W|H|W\rangle
&\cong&\mathcal{E}_{n\bm{k}_{c};\bm{r}_{c};z_{c}}
\\
\langle W|\bm{\mathcal{R}}|W\rangle
&\cong&
E_{n\bm{k}_{c};\bm{r}_{c}}
\left[\bm{\nabla}_{\bm{k}_{c}}\vartheta(\bm{k}_{c},\bm{r}_{c},z_{c},t)
-(z_{c}|\bm{\Lambda}_{n\bm{k}_{c}}|z_{c})\right],
\nonumber\\
\end{eqnarray}
\end{subequations}
where $E_{n\bm{k};\bm{r}_{c}}
=\gamma_{\epsilon}(\bm{r}_{c})\gamma_{\mu}(\bm{r}_{c})E_{n\bm{k}}$,
\begin{eqnarray}
\frac{\mathcal{E}_{n\bm{k}_{c};\bm{r}_{c};z_{c}}}
{E_{n\bm{k}_{c};\bm{r}_{c}}}
&=& 1
-\left[\nabla_{\bm{r}_{c}}
\ln\frac{\gamma_{\epsilon}(\bm{r}_{c})}{\gamma_{\mu}(\bm{r}_{c})}\right]
\cdot
(z_{c}|\bm{\Delta}_{n\bm{k}_{c}}|z_{c})
,
\label{eq:modulated-E}
\end{eqnarray}
and 
\begin{equation}
\bm{\Delta}_{n\bm{k}}
=
\frac{1}{2}\left[
\bm{\Lambda}^{E}_{n\bm{k}}-\bm{\Lambda}^{H}_{n\bm{k}}\right].
\end{equation}
This function $\bm{\Delta}_{n\bm{k}}$ is a difference between 
the Berry connections of the electric and magnetic parts. 
Although $(z_{c}|\bm{\Lambda}^{E,H}_{n\bm{k}}|z_{c})$ depends 
on the representations of eigen modes,
$(z_{c}|\bm{\Delta}_{n\bm{k}}|z_{c})$ does not.
This issue is related to the gauge transformation in $\bm{k}$-space
given in Appendix~\ref{sec:gauge-transformation}.
The property of $(z_{c}|\bm{\Delta}_{n\bm{k}}|z_{c})$
is also discussed in Appendix~\ref{sec:internal-rotation}.

In the above evaluations,
we assumed that the shape of $w^{2}_{r}(\bm{k}-\bm{k}_{c})$
is sufficiently sharp
compared to the slow variations of $E_{n\bm{k};\bm{r}_{c}}$ and 
$\bm{\Lambda}_{n\bm{k}}(\bm{k})$ around $\bm{k}_{c}$,
and neglected terms which depend on the shape of 
$w^{2}_{r}(\bm{k}-\bm{k}_{c})$.
Therefore, even with the perturbative modulation,
we can regard $\bm{r}_{c}$ defined by Eq.~(\ref{eq:rc-perturbed})
as the center of gravity 
$\langle W|\bm{\mathcal{R}}|W\rangle/\langle W|H|W\rangle$.
In the present approximation, we assume that the modulation
is so weak and smooth that we can neglect the second order derivatives
of the modulation functions in the effective Lagrangian and
the equations of motion, which we shall derive later.
In the equations of motion,
the difference between $\bm{r}_{c}$ and the center of gravity 
appears as higher derivatives than the original derivatives.
Therefore we may neglect 
the difference between $\bm{r}_{c}$ and the center of gravity
due to the derivatives of the modulation functions,
while we cannot neglect the correction due to the first derivatives
in $\langle W|H| W\rangle$.

Here we introduce the effective Lagrangian 
in order to derive the equations
of motion of a wavepacket.
\begin{eqnarray}
L_{\mathrm{eff}}& = & \langle W|i\frac{d}{dt}-H|W\rangle.
\label{eq:eff-Lagrangian}
\end{eqnarray}
The first term on the right-hand side of Eq.~(\ref{eq:eff-Lagrangian})
is calculated as follows,
\begin{eqnarray}
&&\langle W|i\frac{d}{dt}|W\rangle
\nonumber\\
&&\cong
\bm{k}_{c}\cdot\dot{\bm{r}}_{c}
-\dot{\bm{k}}_{c}\cdot(z_{c}|\bm{\Lambda}_{n\bm{k}_{c}}|z_{c})
+i(z_{c}|\dot{z}_{c})
\nonumber\\
&&\quad
+\frac{d}{dt}\left[
\int_{\mathrm{BZ}}d\bm{k}\:
w^{2}_{r}(\bm{k}-\bm{k}_{c})
\vartheta(\bm{k},\bm{r}_{c},z_{c},t)-\bm{k}_{c}\cdot\bm{r}_{c}\right].
\nonumber\\
\end{eqnarray}
Neglecting the total time-derivative,
we obtain the final form of the effective Lagrangian,
\begin{eqnarray}
L_{\mathrm{eff}} & \cong & \bm{k}_{c}\cdot\dot{\bm{r}}_{c}
-\dot{\bm{k}}_{c}\cdot(z_{c}|\bm{\Lambda}_{n\bm{k}_{c}}|z_{c})
+i(z_{c}|\dot{z}_{c})
-\mathcal{E}_{n\bm{k}_{c};\bm{r}_{c};z_{c}}.
\nonumber\\
\end{eqnarray}
From this Lagrangian, the equations of motion are derived as follows,
\begin{subequations}
\begin{eqnarray}
\dot{\bm{r}}_{c}
&=& \bm{\nabla}_{\bm{k}_{c}}\mathcal{E}_{n\bm{k}_{c};\bm{r}_{c};z_{c}}
+\dot{\bm{k}}_{c}\times(z_{c}|\bm{\Omega}_{n\bm{k}_{c}}|z_{c})
\nonumber\\
&&\qquad
-i(z_{c}|\left[\bm{f}^{\Delta}_{c}\cdot\bm{\Delta}_{n\bm{k}_{c}},\bm{\Lambda}_{n\bm{k}_{c}}\right]|z_{c})
\label{eq:EOM-r},\\
\dot{\bm{k}}_{c}
&=& -\bm{\nabla}_{\bm{r}_{c}}\mathcal{E}_{n\bm{k}_{c};\bm{r}_{c};z_{c}}
,\label{eq:EOM-k}\\
|\dot{z}_{c})
&=&-i\left[\dot{\bm{k}}_{c}\cdot\bm{\Lambda}_{n\bm{k}_{c}}
+\bm{f}^{\Delta}_{c}\cdot\bm{\Delta}_{n\bm{k}_{c}}\right]
|z_{c}),\label{eq:EOM-z}
\end{eqnarray}
\end{subequations}
where
\begin{equation}
\bm{f}^{\Delta}_{c}
=-\left[\nabla_{\bm{r}_{c}}\ln\frac{\gamma_{\epsilon}(\bm{r}_{c})}{\gamma_{\mu}(\bm{r}_{c})}\right]
E_{n\bm{k}_{c};\bm{r}_{c}},
\end{equation}
and $\bm{\Omega}_{n\bm{k}}$ is the Berry curvature defined by
\begin{eqnarray}
\bm{\Omega}_{n\bm{k}} &=& \bm{\nabla}_{\bm{k}}\times\bm{\Lambda}_{n\bm{k}}
+i\bm{\Lambda}_{n\bm{k}}\times\bm{\Lambda}_{n\bm{k}}.
\label{eq:Omega}
\end{eqnarray}

It should be noted that the above equations of motion satisfy
the energy conservation, i.e., 
\begin{eqnarray}
\frac{d}{dt}\mathcal{E}_{n\bm{k}_{c};\bm{r}_{c};z_{c}}&=&0.
\end{eqnarray}
Rigorously speaking, a Lagrange multiplier 
is needed for the constraint $(z_{c}|z_{c})=\sum_{\lambda}|z_{c\lambda}|^{2}=1$
in the derivation of the equations of motion,
while the constraint is implicitly imposed here.
Therefore, in the above equations of motion,
we should consider that the above constraint is always imposed.

In generic cases with periodic structures,
we cannot analytically evaluate 
$\bm{\Lambda}_{n\bm{k}}$, $\bm{\Omega}_{n\bm{k}}$
nor $\bm{\Delta}_{n\bm{k}}$.
However, in principle, we can numerically calculate them
and thus solve the equations of motion of a wavepacket
subject to a modulation superimposed onto a periodic structure.
Appendix~\ref{sec:internal-rotation} presents
formulae for $\bm{\Omega}_{n\bm{k}}$ and $\bm{\Delta}_{n\bm{k}}$
as well as the internal rotation of a wavepacket,
which are useful for numerical calculations.
Although the relation, 
$(z_{c}|\bm{\Lambda}^{E}_{n\bm{k}}|z_{c})=(z_{c}|\bm{\Lambda}^{H}_{n\bm{k}}|z_{c})$,
is not generally proved,
this can be confirmed at least for some systems 
with locally isotropic $\epsilon(\bm{r})$ and $\mu(\bm{r})$,
e.g., for the elliptically polarized light in systems without periodic structure
and for the TM modes in two-dimensional photonic crystals where $\mu$ is constant.
In this case with $\bm{\Delta}_{n\bm{k}}=0$,
we can replace the perturbed energy as 
$\mathcal{E}_{n\bm{k}_{c};\bm{r}_{c};z_{c}} \to
E_{n\bm{k}_{c};\bm{r}_{c}}=\gamma_{\epsilon}(\bm{r}_{c})\gamma_{\mu}(\bm{r}_{c})E_{n\bm{k}_{c}}$.

Finally, it is noted that 
the optical Hall effect is originated by the second term
on the right-hand side of Eq.~(\ref{eq:EOM-r}),
which is sometimes called ``anomalous velocity''.
The anomalous velocity is the vector product of
the Berry curvature $(z_{c}|\bm{\Omega}_{n\bm{k}_{c}}|z_{c})$
and the driving force $\dot{\bm{k}}_{c}$, i.e., 
the gradient of a superimposed modulation.
Therefore, both of the driving force and the Berry curvature
are needed for this phenomena.
In an electronic system under a strong magnetic field, 
it is pointed out that the Berry curvature
is closely related to the internal rotation of an electronic wavepacket \cite{Chang-Niu}.
Thus, a similar relation is also expected in a photonic system.
Actually, in Appendices~\ref{sec:periodic-system} and \ref{sec:internal-rotation},
we can see the close relation between the Berry curvature
and internal rotation of an optical wavepacket.

\begin{table*}
\caption{\label{tbl:EOM}Equations of motion of optical wavepacket.}
\begin{ruledtabular}
\begin{tabular}{ll}
Modulation &
$\epsilon^{-1}(\bm{r})\to\gamma^{2}_{\epsilon}(\bm{r})\epsilon^{-1}(\bm{r})$,\quad
$\mu^{-1}(\bm{r})\to\gamma^{2}_{\mu}(\bm{r})\mu^{-1}(\bm{r})$\\
Effective Lagrangian
&
$L_{\mathrm{eff}}  \cong  \bm{k}_{c}\cdot\dot{\bm{r}}_{c}
-\dot{\bm{k}}_{c}\cdot(z_{c}|\bm{\Lambda}_{n\bm{k}_{c}}|z_{c})
+i(z_{c}|\dot{z}_{c})
-\mathcal{E}_{n\bm{k}_{c};\bm{r}_{c};z_{c}}$
\\
Equations of motion
&
$
\dot{\bm{r}}_{c}
= \bm{\nabla}_{\bm{k}_{c}}\mathcal{E}_{n\bm{k}_{c};\bm{r}_{c};z_{c}}
+\dot{\bm{k}}_{c}\times(z_{c}|\bm{\Omega}_{n\bm{k}_{c}}|z_{c})
-i(z_{c}|\left[\bm{f}^{\Delta}_{c}
\cdot\bm{\Delta}_{n\bm{k}_{c}},\bm{\Lambda}_{n\bm{k}_{c}}\right]|z_{c})
$
\\
&
$
\dot{\bm{k}}_{c}
= -\bm{\nabla}_{\bm{r}_{c}}\mathcal{E}_{n\bm{k}_{c};\bm{r}_{c};z_{c}}
$
\\
&
$
|\dot{z}_{c})
=-i\left[
\dot{\bm{k}}_{c}\cdot\bm{\Lambda}_{n\bm{k}_{c}}
+\bm{f}^{\Delta}_{c}\cdot\bm{\Delta}_{n\bm{k}_{c}}
|z_{c})
\right]
$
\\
\hline
Perturbed energy
&
$
\mathcal{E}_{n\bm{k}_{c};\bm{r}_{c};z_{c}}
= \left[1
-\frac{1}{2}\left[\nabla_{\bm{r}_{c}}\ln\frac{\gamma_{\epsilon}(\bm{r}_{c})}{\gamma_{\mu}(\bm{r}_{c})}\right]
\cdot
(z_{c}|
\left[\bm{\Lambda}^{E}_{n\bm{k}_{c}}-\bm{\Lambda}^{H}_{n\bm{k}_{c}}\right]
|z_{c})
\right]E_{n\bm{k}_{c};\bm{r}_{c}}
= E_{n\bm{k}_{c};\bm{r}_{c}}+\bm{f}^{\Delta}_{c}\cdot(z_{c}|\bm{\Delta}_{n\bm{k}_{c}}|z_{c})
$
\\
&
$E_{n\bm{k}_{c};\bm{r}_{c}}
= \gamma_{\epsilon}(\bm{r}_{c})\gamma_{\mu}(\bm{r}_{c})E_{n\bm{k}_{c}}$,
\quad
$\bm{f}^{\Delta}_{c}
=-\left[\nabla_{\bm{r}_{c}}\ln\frac{\gamma_{\epsilon}(\bm{r}_{c})}{\gamma_{\mu}(\bm{r}_{c})}\right]
E_{n\bm{k}_{c};\bm{r}_{c}}$,
\quad
$\bm{\Delta}_{n\bm{k}_{c}}
=
\frac{1}{2}\left[\bm{\Lambda}^{E}_{n\bm{k}_{c}}-\bm{\Lambda}^{H}_{n\bm{k}_{c}}\right]
$\\
Energy conservation
&
$\frac{d}{dt}\mathcal{E}_{n\bm{k}_{c};\bm{r}_{c};z_{c}} =0$
\end{tabular}
\end{ruledtabular}
\end{table*}

\section{\label{sec:applications}Applications}
\subsection{\label{sec:Imbert}Transverse shift in reflection and refraction}
As an application of the theoretical framework developed in Sec.~\ref{sec:formalism}, 
we consider the case with locally isotropic $\epsilon(\bm{r})$ and $\mu(\bm{r})$.
In this case, we can write down the equations of motion (\ref{eq:EOM-r})-(\ref{eq:EOM-z}) 
in simple forms. For this purpose, we first calculate the gauge field 
$\bm{\Lambda}_{\bm{k}}$ and the Berry curvature $\bm{\Omega}_{\bm{k}}$. 
In a helicity basis, the eigenvectors for the right($+$)- and left($-$)-circular
polarizations can be written as
\begin{equation}
\bm{U}_{\pm,\bm{k}}^{E}=\frac{1}{\sqrt{2\epsilon}}(\bm{e}_{\theta}\pm 
i\bm{e}_{\phi}),\ \ 
\bm{U}_{\pm,\bm{k}}^{H}=\frac{1}{\sqrt{2\mu}}(\bm{e}_{\phi}\mp
i\bm{e}_{\theta}),
\end{equation}
where $\bm{e}_{\theta,\phi}$ with $\bm{e}_{k}=\bm{k}/k$
are the orthogonal unit vectors in the spherical coordinate 
of the $\bm{k}$-space.
After some calculations we obtain 
\begin{equation}
\bm{\Lambda}_{\bm{k}}^{E}=\bm{\Lambda}_{\bm{k}}^{H}=
-\frac{\cos\theta}{k\sin\theta}
\sigma_{3}{\bm{e}_{\phi}},
\end{equation}
which then yields \cite{note-Omega}
\begin{equation}
\bm{\Lambda}_{\bm{k}}=-\frac{\cos\theta}{k\sin\theta}
\sigma_{3}{\bm{e}_{\phi}},
\ \ 
\bm{\Omega}_{\bm{k}}=
\frac{\bm{k}}{k^{3}}\sigma_{3}, \ \ 
\bm{\Delta}_{\bm{k}}= 0.
\label{eq:Omega-isotropic}
\end{equation}
Thus the equations of motion are simplified as follows,
\begin{subequations} 
\begin{eqnarray}
\dot{\bm{r}}_{c}
&=& v(\bm{r}_{c})
\frac{\bm{k}_{c}}{k_{c}}
+\dot{\bm{k}}_{c}\times(z_{c}|\bm{\Omega}_{\bm{k}_{c}}|z_{c})
,\label{eq:EOM-r-isotropic}
\\
\dot{\bm{k}}_{c}
&=& -\left[\bm{\nabla}_{\bm{r}_{c}}v(\bm{r}_{c})\right]
k_{c}
, \label{eq:EOM-k-isotropic}\\
|\dot{z}_{c})
&=&-i\dot{\bm{k}}_{c}\cdot\bm{\Lambda}_{\bm{k}_{c}}|z_{c}),
\label{eq:EOM-z-isotropic}
\end{eqnarray}
\end{subequations}
where $v(\bm{r})=1/\sqrt{\epsilon(\bm{r})\mu(\bm{r})}$, and
$|z)=[z_{+},z_{-}]$ is represented in a helicity basis.
As is analogous to the Hall effect in electronic systems, 
the second term on the right-hand side of Eq.~(\ref{eq:EOM-r-isotropic}) describes 
the optical Hall effect induced by a modulation of refractive index \cite{MSN}. 
The equation for $|z_{c})$,
Eq.~(\ref{eq:EOM-z-isotropic}), describes a phase shift 
by the directional change of 
propagation discussed in Refs.~\cite{Chiao-Wu,Tomita-Chiao, Berry-II}.
This equation gives the solution
$|z_{c}^{\mathrm{out}})
=[e^{-i\Theta}z_{+}^{\mathrm{in}},e^{i\Theta}z_{-}^{\mathrm{in}}]$,
where $|z_{c}^{\mathrm{in}})
=[z_{+}^{\mathrm{in}}, z_{-}^{\mathrm{in}}]$
is the initial state of polarization.
$\Theta$ is a solid angle made by the trajectory of momentum: 
$\Theta=\oint d\bm{k}\cdot[\bm{\Lambda}_{\bm{k}}]_{++}
=\int_{S}d\bm{S}_{\bm{k}}\cdot[\bm{\Omega}_{\bm{k}}]_{++}$ 
where $d\bm{S}_{\bm{k}}$ is the surface 
element in $\bm{k}$-space and $S$ 
is a surface surrounded by the trajectory.
Our approach can be easily generalized to treat systems with periodic structures
on the same footing, and offers a powerful tool for applications
compared with the eikonal approximation \cite{Liberman-Zeldovich}.

The simplest example of the optical Hall effect is realized 
as the transverse shift at the interface refraction and reflection.
There have been a number of studies on the shifts within and
out of the incident plane at the total reflection.
The former is well known as
the Goos-H{\" a}nchen effect \cite{Goos-Hanchen,Jackson}
and has been explained in terms of the evanescent wave penetrating
into the forbidden region.
The latter one, which is referred to as the Imbert-Fedorov shift,
was interpreted by F.~I.~Fedorov \cite{Fedorov}
as an analog of the Goos-H{\" a}nchen effect
and was observed experimentally by C.~Imbert
using multiple total reflections \cite{Imbert},
followed by a number of theoretical approaches 
\cite{Boulware,Ashby-Miller,Schilling}.
Furthermore, it was pointed out that the shift out of the incident plane
could also occur in partial reflection and refraction \cite{Schilling,Fedoseev-I,Fedoseev-II}.
However, some of the theoretical predictions for the amount of shift
contradict each other. One reason is an experimental difficulty for a measurement of
the tiny shift, as the shift is only a fraction of a wavelength.
It was only recent that the Imbert-Fedorov shift is measured for a single total reflection
\cite{Pillon}.
Thus the physical mechanism for the transverse shift is still controversial.

\begin{figure}[hbt]
\includegraphics[scale=0.25]{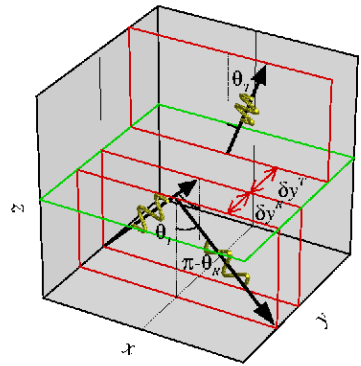}
\caption{
Transverse shift of light beams in the refraction
and reflection at an interface.
}
\label{fig:configuration}
\end{figure}
In our previous paper \cite{MSN}, we calculated this transverse shift by using
the conservation of the $z$-component of TAM 
{\it for individual photons}, which follows from the equations of motion 
(\ref{eq:EOM-r})-(\ref{eq:EOM-z}) applied to this interface problem.
Bliokh {\it et al.} \cite{Bliokh-PRL} then questioned the result, 
claiming that it does not match their result of the shift for elliptically 
polarized Gaussian beams.  
Here we show that, in all cases with generic polarizations,
an identical result for the transverse shift of each beam
is given by the following different approaches,
(i) analytic evaluation of wavepacket dynamics,
(ii) TAM conservation for individual photons in Ref.~\cite{MSN},
and (iii) numerically exact simulation of wavepacket dynamics.
It agrees with a result by classical electrodynamics,
as presented in Appendix~\ref{sec:Maxwell-shift}.
In Sec.~\ref{sec:remarks}, 
we shall resolve the inconsistency between the identical result 
by the approaches (i)-(iii) and that given in Ref.~\cite{Bliokh-PRL}.
Thereby the validity of our theory presented here is completely guaranteed.

Our equations of motion are not directly applicable
to the refraction and reflection problem at a sharp interface
since they require the slowly varying conditions
$|\bm{\nabla} \ln\epsilon|,|\bm{\nabla}\ln\mu|\ll k$.
Indeed, our equations of motion do not correctly describe
a splitting of an incident wavepacket 
into reflected and transmitted wavepackets
nor changes of their polarization states at the interface.
However, in a case with a flat interface,
a simple extension of our theory works well, 
as will be explained in the following.
Here, as shown in Fig.~\ref{fig:configuration},
we consider the case where 
the incident beam comes from the region with $x < 0$ and $z < 0$ 
along the plane of $y=\text{const.}$,
and the interface is the $z=0$ plane.

\subsubsection{analytic evaluation}
Far from the interface, $\bm{r}_{c}$ in Eq.~(\ref{eq:rc-perturbed})
is easily estimated as
\begin{eqnarray}
\bm{r}_{c}|_{t\to\pm\infty} &\cong& 
\bm{\nabla}_{\bm{k}^{A}}\vartheta^{A}(\bm{k}^{A},t\to\pm\infty)
-(z^{A}|\bm{\Lambda}_{\bm{k}^{A}}|z^{A}),
\label{eq:rc-asymptotic}
\end{eqnarray}
where $A=I$ for $t\to-\infty$, $A=T$ or $R$ for $t\to\infty$.
The momenta $\bm{k}^{I,T,R}$ and the polarization states $|z^{I,T,R})$ are 
those of the incident ($I$), transmitted ($T$)
and reflected ($R$) beams, respectively.
Due to the wavepacket splitting at the interface,
our semiclassical equations of motion do not tell us
the values of $|z^{A})$ and $\vartheta^{A}$.
Hence, for these variables, we borrow the results
of reflection/refraction of a plane wave.
With a natural choice of a wavepacket presented in Appendix~\ref{sec:Maxwell-shift},
the $y$-component of the first term in Eq.~(\ref{eq:rc-asymptotic})
is unchanged at the interface.
Thus, the transverse shift comes only from the second term as
\begin{eqnarray}
\delta y^{A} 
&=& 
-(z^{A}|\bm{\Lambda}_{\bm{k}^{A}}|z^{A})
+(z^{I}|\bm{\Lambda}_{\bm{k}^{I}}|z^{I}).
\label{eq:delta_y-Lambda}
\end{eqnarray}
where $A=T$ or $R$.
Substituting  Eq.~(\ref{eq:Omega-isotropic})
in each Berry connection $(z^{A}|\bm{\Lambda}_{\bm{k}^{A}}|z^{A})$
($A=I,T,R$), 
we obtain the following equation for the transverse shift 
\begin{eqnarray}
\delta y^{A} 
&=& \frac{1}{k^{I}\sin\theta_{I}}
\left[
(z^{A}|\sigma_{3}|z^{A})\cos\theta_{A}-(z^{I}|\sigma_{3}|z^{I})\cos\theta_{I}
\right],
\nonumber\\
\label{eq:delta_y}
\end{eqnarray}
where $A=T$ or $R$, $\theta_{I,T,R}$ are the angles 
between the positive $z$-axis and the propagating directions
of the incident, transmitted and reflected beams, respectively.

\subsubsection{total angular momentum conservation}
Then, what is the physical meaning of the above result?
Firstly, it should be noted that $(z|\sigma_{3}|z)=|z_{+}|^{2}-|z_{-}|^{2}$ 
represents the magnitude of spin polarization in the direction of $\bm{k}_{c}$,
i.e., $(z|\sigma_{3}|z)=\pm1$ for 
right/left-circular polarizations,
$(z|\sigma_{3}|z)=0$ for linear polarizations,
and $|(z|\sigma_{3}|z)| < 1$ for elliptic polarizations.
Therefore, it is intuitively expected that this phenomena is closely related to 
the angular momentum of wavepacket.
For a system with rotational symmetry around the $z$-axis, 
the equations of motion lead to the conservation of 
the $z$-component of the following TAM,
\begin{eqnarray}
\bm{j}_{c} &=& \bm{r}_{c}\times \bm{k}_{c}
+(z_{c}|\sigma_{3}|z_{c})\frac{\bm{k}_{c}}{k_{c}}.
\label{eq:TAM-isotropic}
\end{eqnarray}
This conservation 
is expected to hold even in the case of a sharp interface,
because it is based on the rotational symmetry around the $z$-axis. 
Actually, from Eq.~(\ref{eq:delta_y}) for the transverse shift
and Eq.~(\ref{eq:TAM-isotropic}) for the TAM,
we can reach the conservation of 
the $z$-component of the TAM {\it for each of individual photons}
\begin{equation}
j_{z}^{I}=j_{z}^{T},\ j_{z}^{I}=j_{z}^{R},
\label{eq:TAM-photon}
\end{equation}
where $\bm{j}^{I,T,R}$ are the TAM of incident, 
transmitted and reflected beams, respectively.
It just makes sense that the incident beam is regarded 
as a collection of photons; each photon is reflected 
or transmitted stochastically at the interface. 
As we shall see in Sec.~\ref{sec:remarks}
(and Appendix \ref{sec:Maxwell-shift} in detail),
Eq.~(\ref{eq:delta_y}) for the transverse shift
is consistent with the result derived 
in classical electrodynamics.
In this sense, this photon picture
is implicitly incorporated already 
in classical electrodynamics.

Inversely, assuming the conservation of TAM for individual photons,
we can derive the transverse shift as Eq.~(\ref{eq:delta_y}).
This is what we have done in our previous paper \cite{MSN}.
This derivation of the transverse shift
is akin to the derivation of Snell's law 
based on the particle picture of light
in which the refracted and reflected angles is obtained
from the conservation of energy and momentum 
(parallel to the interface) for individual photons.

\subsubsection{\label{sec:simulation}numerical simulation}
\begin{figure*}[hbt]
\includegraphics[scale=0.30]{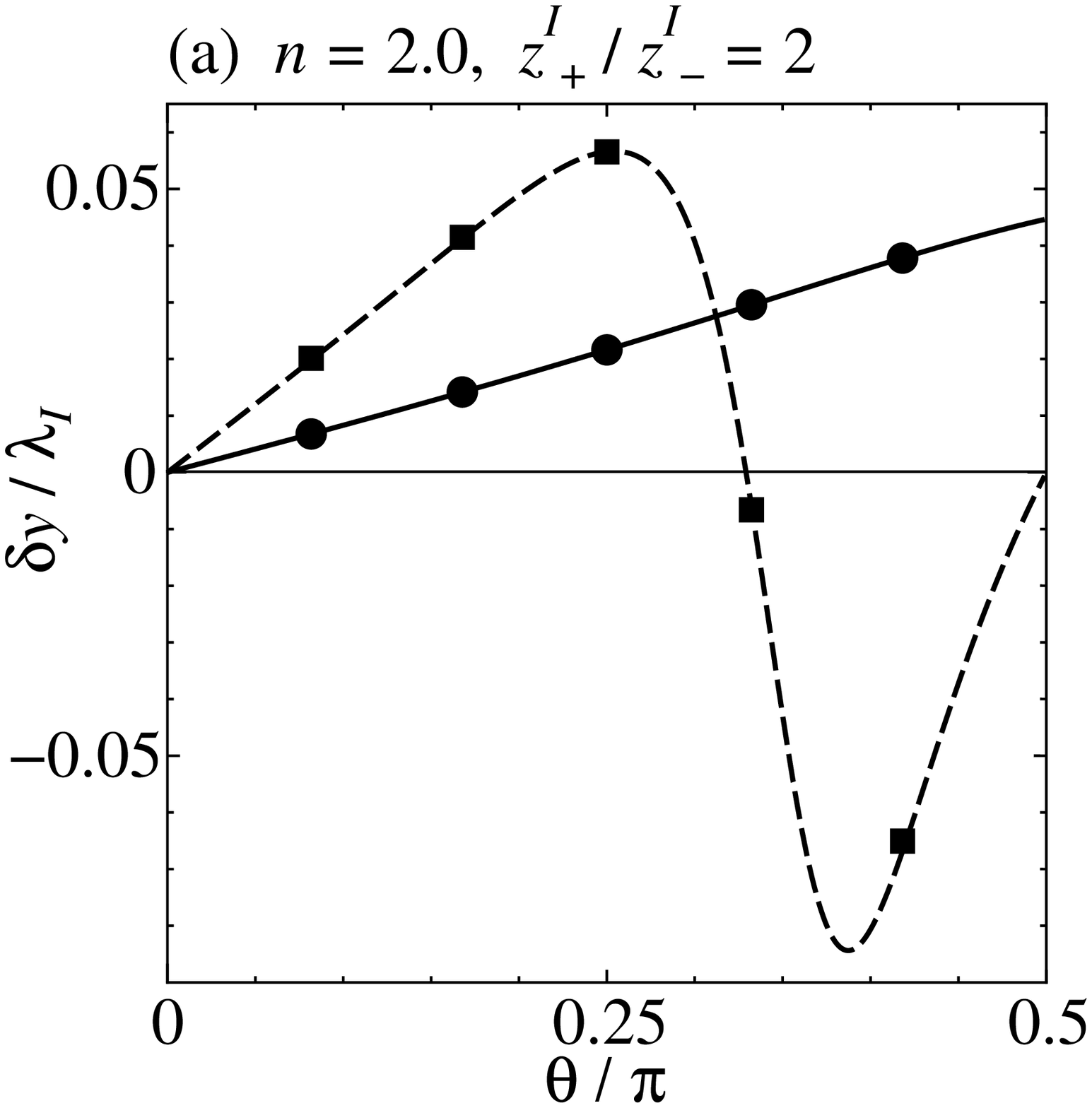}
\includegraphics[scale=0.30]{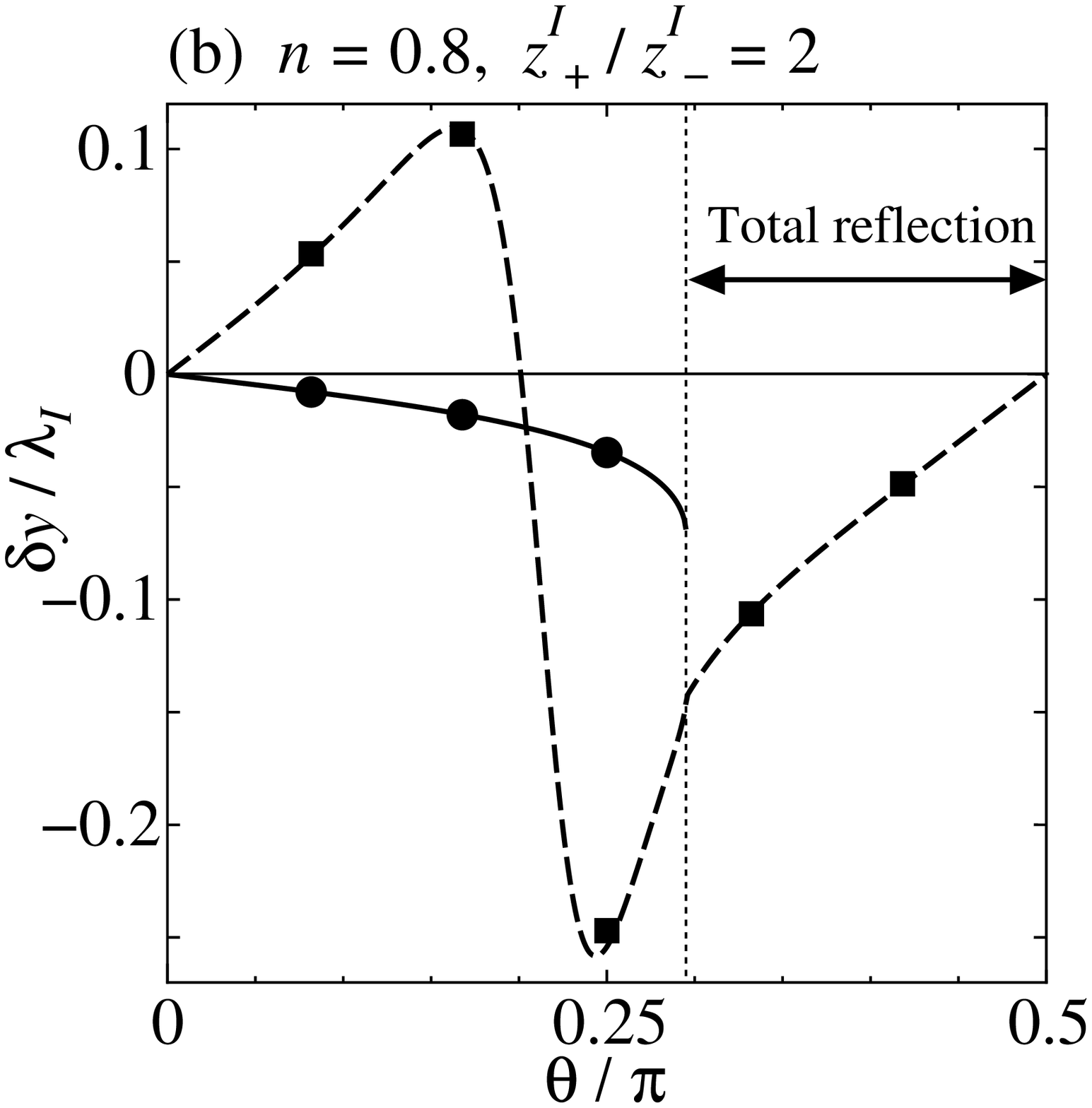}
\includegraphics[scale=0.30]{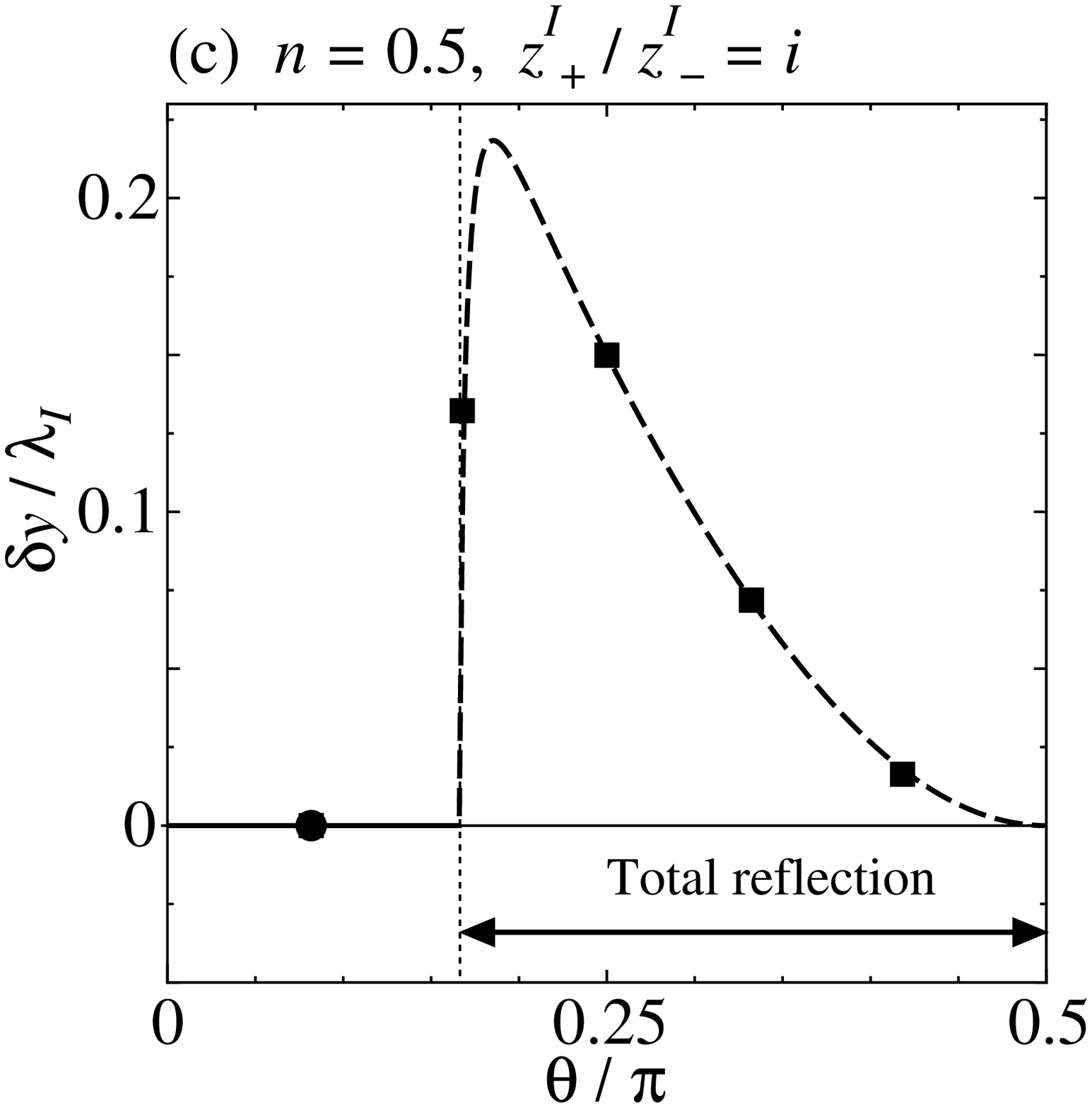}
\caption{
Shifts of reflected and transmitted beams.
$n$ is the relative refractive index of the upper medium
with respect to the lower medium.
$\lambda_{I}$ is the wave length of incident light
in the lower medium.
The solid and dashed lines represent the analytic results
Eq.~(\ref{eq:delta_y}) for transmitted and reflected beams
respectively.
The filled circles and squares are the results of simulations
for transmitted and reflected beams.
}
\label{fig:shift}
\end{figure*}
In order to verify our theory quantitatively,
we check the property of the transverse shift
in more detail.
To obtain $(z^{A}|\sigma_{3}|z^{A})$, 
we decompose the incident wave as
\begin{equation}
|z^{I})=\frac{z^{I}_{+}+z^{I}_{-}}{\sqrt{2}}|p)
+\frac{i(z^{I}_{+}-z^{I}_{-})}{\sqrt{2}}|s),
\end{equation}
where $|p)=\frac{1}{\sqrt{2}}[1,1]$ and $|s)=\frac{1}{\sqrt{2}}[-i,i]$ 
represent the $p$- and $s$-polarized states. 
Straightforward calculation yields 
\begin{eqnarray}
&&(z^{A}|\sigma_{3}|z^{A})
\nonumber\\
&&=
\frac{2\left[
(z^{I}|\sigma_{3}|z^{I})\Re(A^{*}_{p}A_{s})+(z^{I}|\sigma_{2}|z^{I})\Im(A^{*}_{p}A_{s})
\right]}
{\left[1+(z^{I}|\sigma_{1}|z^{I})\right]|A_{p}|^2
+\left[1-(z^{I}|\sigma_{1}|z^{I})\right]|A_{s}|^2},
\nonumber\\
\end{eqnarray}
with $A=T$ or $R$, and 
$T_{p}$ and $T_{s}$ ($R_{p}$ and $R_{s}$)
are the amplitude transmission (reflection) coefficients 
for $p$- and $s$-polarization, respectively.

When we focus on the partial reflection and refraction,
$A_p$ and $A_s$ are real, and 
Eq.~(\ref{eq:delta_y}) is rewritten as follows,
\begin{eqnarray}
&&\delta y^{A} 
= \frac{(z^{I}|\sigma_{3}|z^{I})}{k^{I}\tan\theta_{I}}\nonumber\\
&&\ \ \cdot
\left[
\frac{2A_{p}A_{s}\cos\theta_{A}/\cos\theta_I}
{\left[1+(z^{I}|\sigma_{1}|z^{I})\right]A_{p}^2
+\left[1-(z^{I}|\sigma_{1}|z^{I})\right]A_{s}^2}
-1
\right],
\label{eq:Imbert-partial}
\end{eqnarray}
where $A = R$ or $T$.
This means that the incident beams with 
$|z) =[e^{\mp i\frac{\phi}{2}}\cos\frac{\theta}{2}, e^{\pm i\frac{\phi}{2}}\sin\frac{\theta}{2}]$,
where $\theta$ and $\phi$ represent the spherical coordinate of the Poincar{\'e} sphere,
cause the shift of the same magnitude and the same direction with each other, 
i.e., the shift independent of the sign of $\phi$.
In addition, the incident beams with 
$|z) = [e^{\mp i\frac{\phi}{2}}\sin\frac{\theta}{2}, e^{\pm i\frac{\phi}{2}}\cos\frac{\theta}{2}]$
cause the shift of the same magnitude as the above beams,
but of the opposite direction to them.
In the partial reflection and refraction, no shift is observed for
the incident beam with linear polarization. 

On the other hand, in the total reflection, we have $|R_{p}|=|R_{s}|=1$, 
and the shift of the reflected beam
is represented by
\begin{eqnarray}
\delta y^{R} 
&=& -\frac{1}{k^{I}\tan\theta_{I}}
\bigl[
(z^{I}|\sigma_{3}|z^{I})[\Re(R^{*}_{p}R_{s})+1]
\nonumber\\
&&\hspace{2cm}
+(z^{I}|\sigma_{2}|z^{I})\Im(R^{*}_{p}R_{s})
\bigr].
\label{eq:Imbert-total}
\end{eqnarray}
In particular,
for the incident beam with linear polarization 
$|z) =[\frac{e^{-i\frac{\phi}{2}}}{\sqrt{2}}, \frac{e^{i\frac{\phi}{2}}}{\sqrt{2}}]$,
the shift is the same magnitude and the same direction for $\phi=\alpha$ and 
$\phi=\pi-\alpha$.
The direction is reversed by the replacement $\phi\to-\phi$ without change of the magnitude.

We have confirmed all the above features quantitatively
by numerically solving Maxwell equations for wavepackets.
In Ref.~\cite{MSN}, we have presented 
the results only for the incident beam with
right-circular polarization. 
Here, to complete the argument, 
we present the results of the numerical simulations for more generic cases.
Figure~\ref{fig:shift} shows the shifts
for the incident beams with 
the elliptical polarization $z^{I}_{+}/z^{I}_{-}=2$ 
at the interfaces with relative refractive indices (a) $n = 2.0$, (b) $n=0.8$,
and (c) for the incident beam with linear polarization $z^{I}_{+}/z^{I}_{-}=i$ 
with $n = 0.5$. 
(We take the value of magnetic permeability common in both media
upper and lower the interface,
i.e., $\mu_{1}=\mu_{2}$, in these simulations.)
The solid and dashed lines represent the analytic results
Eq.~(\ref{eq:delta_y}) for transmitted and reflected beams,
respectively.
The filled circles and squares are the results of simulations
for transmitted and reflected beams.
We note that, in Fig.~\ref{fig:shift}(c), 
the shift for the linearly polarized beam is nonzero only 
for a region of total reflection, in accordance with our analytic result.
In all cases, the numerical results excellently agree with Eq.~(\ref{eq:delta_y}), 
thus verifying our theory.
(We have confirmed this consistency also in cases 
where both of permittivity and permeability
are different in two media upper and lower the interface,
i.e., $\epsilon_{1}\neq\epsilon_{2}$ and $\mu_{1}\neq\mu_{2}$.)

Finally we should comment on our constitution method of a set of 
incident, transmitted and reflected wavepackets,
which is an exact solution to Maxwell equations.
In each numerical simulation, we have constructed 
an elliptically-polarized incident wavepacket
as a superposition of plane waves with a common polarization state,
i.e., $|z^{I})$ which is independent of $\bm{k}$.
This is a natural definition of incident wavepacket.
Otherwise, the concept ``an elliptically-polarized incident wavepacket''
gets fuzzy, and the linear composition from and decomposition to
different orthonormal bases of incident wavepackets are violated.
This is because incident wavepackets with different functions of 
$\{|z^{I}(\bm{k}))\}$'s for constituent plane waves 
can have the same mean polarization state $|z^{I}(\bm{k}^{I}))$.
Imposing the exact boundary conditions,
transmitted and reflected wavepackets are automatically generated.
For partial reflection, a single incident wavepacket split into 
reflected and transmitted wavepackets after reflection/refraction 
at the interface. The position of each wavepacket is estimated
when each wavepacket is far from the interface.
It should be noted that the numerical simulations exactly
take into account the changes of shapes of wavepackets,
while the analytic evaluation assumes the sharpness of
a weight function for the superposition.

\subsection{\label{sec:remarks}Remarks on other theories}
Recently, Bliokh {\it et al.} calculated the shift for an elliptic
Gaussian incident beam in classical electrodynamics, and 
their result disagrees with that obtained from our theory \cite{Bliokh-PRL}. 
They attributed the difference 
to a ``fallacy" in our TAM conservation for individual photons. 
We explain below in detail that our theory is totally  
free from the criticism. 
To prove this, it is enough to show that Eq.~(\ref{eq:delta_y})
is equivalent
to the transverse shift evaluated in classical electrodynamics,
i.e., the result by Fedoseev \cite{Fedoseev-I,Fedoseev-II}.
His procedure of calculation is as follows. 
First construct a wavepacket by a linear superposition of plane waves. 
By taking into account the exact boundary conditions of electromagnetic fields
(Eqs.~(\ref{eq:BC-E}) and (\ref{eq:BC-H})) at a flat interface, 
as in the textbooks of optics or classical electrodynamics 
\cite{Born, Jackson},
one can construct transmitted and reflected wavepackets
as an exact solution to Maxwell equations.
The center of each wavepacket is defined as 
an average position weighted by each energy density.
The result of this calculation by classical electrodynamics
\cite{Fedoseev-I,Fedoseev-II}
is identical with that by our theory
(Eqs.~(\ref{eq:Imbert-partial})
and (\ref{eq:Imbert-total})
derived from Eq.~(\ref{eq:delta_y}))
in the second quantized formalism.
The details are presented in Appendix \ref{sec:Maxwell-shift}.
Hence we checked that the following three approaches give 
the same transverse shift for each wavepacket; 
(i) analytic evaluation of wavepacket dynamics
both in classical and quantum-mechanical formalisms,
(ii) TAM conservation for individual photons 
(Eq.~(\ref{eq:TAM-photon})), 
(iii) numerically exact simulation of wavepacket dynamics.

There remains an inconsistency between the identical result 
obtained by (i)-(iii) and one by Bliokh {\it et al.} 
in Ref.~\cite{Bliokh-PRL}.
One reason is an inappropriate boundary condition
for a set of wavepackets in paraxial approximation.
This boundary condition is different from the correct one:
\begin{subequations}
\begin{eqnarray}
&&\bm{t}\cdot
\left[\bm{E}^{I}(\bm{r},t)+\bm{E}^{R}(\bm{r},t)\right]
= \bm{t}\cdot\bm{E}^{T}(\bm{r},t),
\label{eq:BC-E}
\\
&&\bm{t}\cdot
\left[\bm{H}^{I}(\bm{r},t)+\bm{H}^{R}(\bm{r},t)\right]
= \bm{t}\cdot\bm{H}^{T}(\bm{r},t),
\label{eq:BC-H}
\end{eqnarray}
\end{subequations}
where $\bm{t}$ is an arbitrary unit vector parallel to the  interface,
$\bm{E}^{A}$ and $\bm{H}^{A}$ ($A=I,T,R$ ) 
are electric and magnetic fields
of incident ($I$), transmitted ($T$) and reflected ($R$) beams, respectively.
Another reason for this contradiction
comes from the definition of a center of wavepacket in Ref.~\cite{Bliokh-PRL}.
The methods (i)-(iii) commonly use the position averaged 
with a weight of energy density. In Ref.~\cite{Bliokh-PRL}, on the other
hand, the center is defined as a center of the wavepacket 
{\it projected onto} its mean polarization state.
The center of the wavepacket in the former definition, 
i.e. the position averaged by 
the energy density, can be easily measured by 
photon counting, as employed in two measurements on the Imbert-Fedorov shift 
\cite{Imbert,Pillon},
while the latter definition requires counting of photons projected 
onto a specified polarization. 
The agreement between totally different approaches,
i.e., (i)-(iii) and classical electrodynamics, 
suggests that our definition is a natural one. 

Finally we should comment on
the relation between the conservation laws of TAM
in the wave and particle pictures of light.
In Ref.~\cite{Bliokh-PRL}, it is claimed that,
for an incident beam with an elliptic polarization,
the conservation of TAM for individual photons (Eq.~(\ref{eq:TAM-photon}))
is inconsistent with the conservation of TAM for whole beams,
\begin{equation}
j_{z}^{I}=R^{2}j_{z}^{R}+T^{2}\frac{n_2 \mu_1\cos\theta^{T}}{
n_1 \mu_2 \cos\theta^{I}}j_{z}^{T}.
\label{eq:TAM-wave}
\end{equation}
However, as we shall show below,
Eq.~(\ref{eq:TAM-photon})
is a sufficient condition for Eq.~(\ref{eq:TAM-wave}).
From Fresnel formulae, we have
\begin{subequations}
\begin{eqnarray}
1&=&R_{p}^{2}+T_{p}^{2}\frac{n_2 \mu_1\cos\theta_{T}}{
n_1 \mu_2 \cos\theta_{I}},
\\
1&=&R_{s}^{2}+T_{s}^{2}\frac{n_2 \mu_1\cos\theta_{T}}{
n_1 \mu_2 \cos\theta_{I}},
\end{eqnarray}
\end{subequations}
for the $p$- and $s$-polarized beams, and also
\begin{equation}
1=R^{2}+T^{2}\frac{n_2 \mu_1\cos\theta_{T}}{
n_1 \mu_2 \cos\theta_{I}},
\end{equation}
for a beam with an arbitrary polarization, 
where 
\begin{equation}
R^{2}=\frac{R_p^2+|m|^2R_s^2}{1+|m|^{2}},\ \ T^{2}=\frac{T_p^2+|m|^2T_s^2}{1+|m|^{2}},
\end{equation}
and $m=z_s /z_p$, $|z)=z_{p}|p)+z_{s}|s)$.
The above formula represents 
the conservation of energy flow or equivalently
the conservation of the number of photons.
One can easily see that the above formula and Eq.~(\ref{eq:TAM-photon}) 
yields Eq.~(\ref{eq:TAM-wave}).
To summarize, the TAM conservation for individual photons
(Eq.~(\ref{eq:TAM-photon}))
has neither contradiction 
nor inconsistency with other theories.

\subsection{\label{sec:photonic-crystal}Two-dimensional photonic crystal}
We consider two-dimensional photonic crystals, 
where $\tensor{\epsilon}(\bm{r})$ and $\tensor{\mu}(\bm{r})$ are 
scalar variables $\epsilon(\bm{r})$ and $\mu(\bm{r})$
which are periodically modulated in the $xy$-plane. 
Along the $z$-direction $\epsilon(\bm{r})$ and $\mu(\bm{r})$ 
are assumed to be uniform.
However, it is noted that in general there may appear ordinary degeneracies
at symmetric points and
accidental degeneracies at some specific points
in the Brillouin zone.
Around these points, the semiclassical argument
based on the adiabaticity
would not be a good approximation, 
and it is needed to seriously incorporate the dynamics of the wavepacket.
We restrict ourselves to bands without degeneracy
here for simplicity.
In this case, the inversion symmetry of the periodic structure must be broken
in order for a band to have nonzero Berry curvature.
This is because the Fourier transformation of 
$\epsilon(\bm{r})$ and $\mu(\bm{r})$ are real-valued, 
when a system has the inversion symmetry.

For a wavepacket constructed from a non-degenerate band,
the equations of motion in Eqs.~(\ref{eq:EOM-r})-(\ref{eq:EOM-z})
are reduced to the following ones,
\begin{subequations}
\begin{eqnarray}
\dot{\bm{r}}_{c}
&=& \bm{\nabla}_{\bm{k}_{c}}\mathcal{E}_{n\bm{k}_{c};\bm{r}_{c}}
+\dot{\bm{k}}_{c}\times \bm{\Omega}_{n\bm{k}_{c}},
\label{eq:EOM-r_single-band}
\\
\dot{\bm{k}}_{c}
&=& -\bm{\nabla}_{\bm{r}_{c}}\mathcal{E}_{n\bm{k}_{c};\bm{r}_{c}},
\\
\dot{z}_{c}
&=&-i\left[\dot{\bm{k}}_{c}\cdot\bm{\Lambda}_{n\bm{k}_{c}}
+\bm{f}^{\Delta}_{c}\cdot\bm{\Delta}_{n\bm{k}_{c}}\right]
z_{c},
\end{eqnarray}
\end{subequations}
In the above equations of motion, the most important and controllable
quantity is the second term on the right-hand side of 
Eq.~(\ref{eq:EOM-r_single-band}),
i.e., the Berry curvature $\bm{\Omega}_{n\bm{k}_{c}}$.
The anomalous velocity 
$\dot{\bm{k}}_{c}\times \bm{\Omega}_{n\bm{k}_{c}}$
of the optical wavepacket leads to the optical Hall effect.
Compared with this term, the other correction due to 
$\bm{\Delta}_{n\bm{k}_{c}}$ are small as shown below
(see Appendix~\ref{sec:Delta} also).
The parameter $z_{c}$ becomes a simple complex number
and just represents a phase shift.
Thus an optimal design for the enhancement of the optical Hall effect
is equivalent to the enhancement of the magnitude of the Berry curvature.
In  the present case, the Berry curvature comes from 
an interband effect due to a periodic structure without inversion symmetry,
and roughly scales as the inverse square of a band splitting
(see Eqs.~(\ref{eq:OmegaE}) and (\ref{eq:OmegaH})).
Therefore, we can expect the enhancement of the optical Hall effect
for wavepackets constructed from Bloch waves around nearly degenerate
points in the Brillouin zone.
In two-dimensional photonic crystals, 
Bloch waves propagating along the $xy$-plane ($k_{z}=0$)
are classified into
the transverse magnetic (TM) and 
the transverse electric (TE) modes.
In other words, the Maxwell equations for the Bloch functions 
(\ref{eq:Maxwell-UEH})-(\ref{eq:Maxwell-KUE-KUH}) decouple into 
two sets of equations, one for the TM and the other for the TE modes,
and the problem of wavepacket dynamics can be more simplified.
Appendix~\ref{sec:Omega-2D} gives useful formulae for the Berry curvature 
$\bm{\Omega}_{n\bm{k}}$ and $\bm{\Delta}_{n\bm{k}}$ in such modes.
As for other modes and more generic case with degenerate bands,
we must use formulae given in Appendix~\ref{sec:internal-rotation}.

\begin{figure}[hbt]
$\begin{array}{cc}
\includegraphics[scale=0.2]{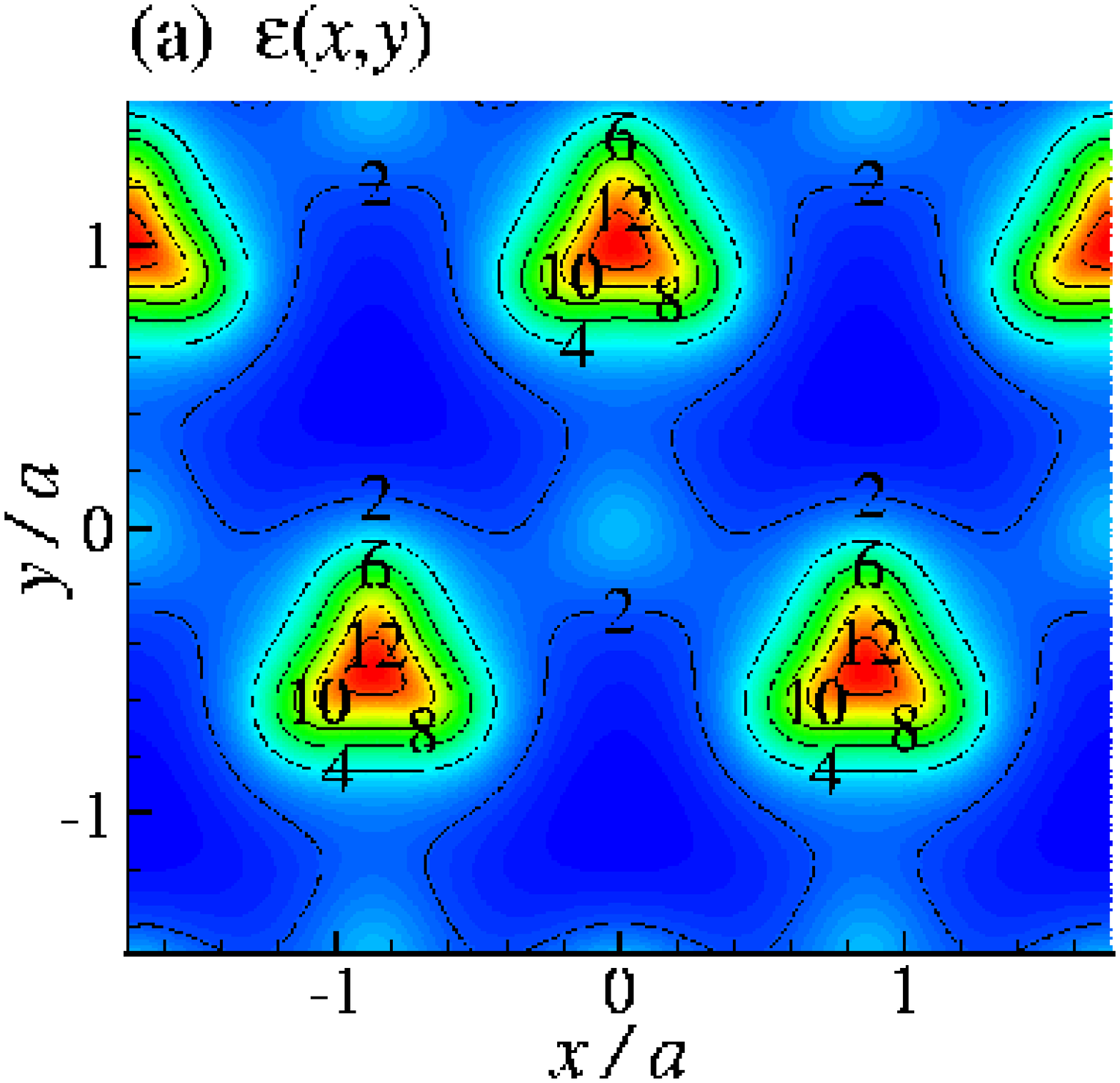}
& \includegraphics[scale=0.25]{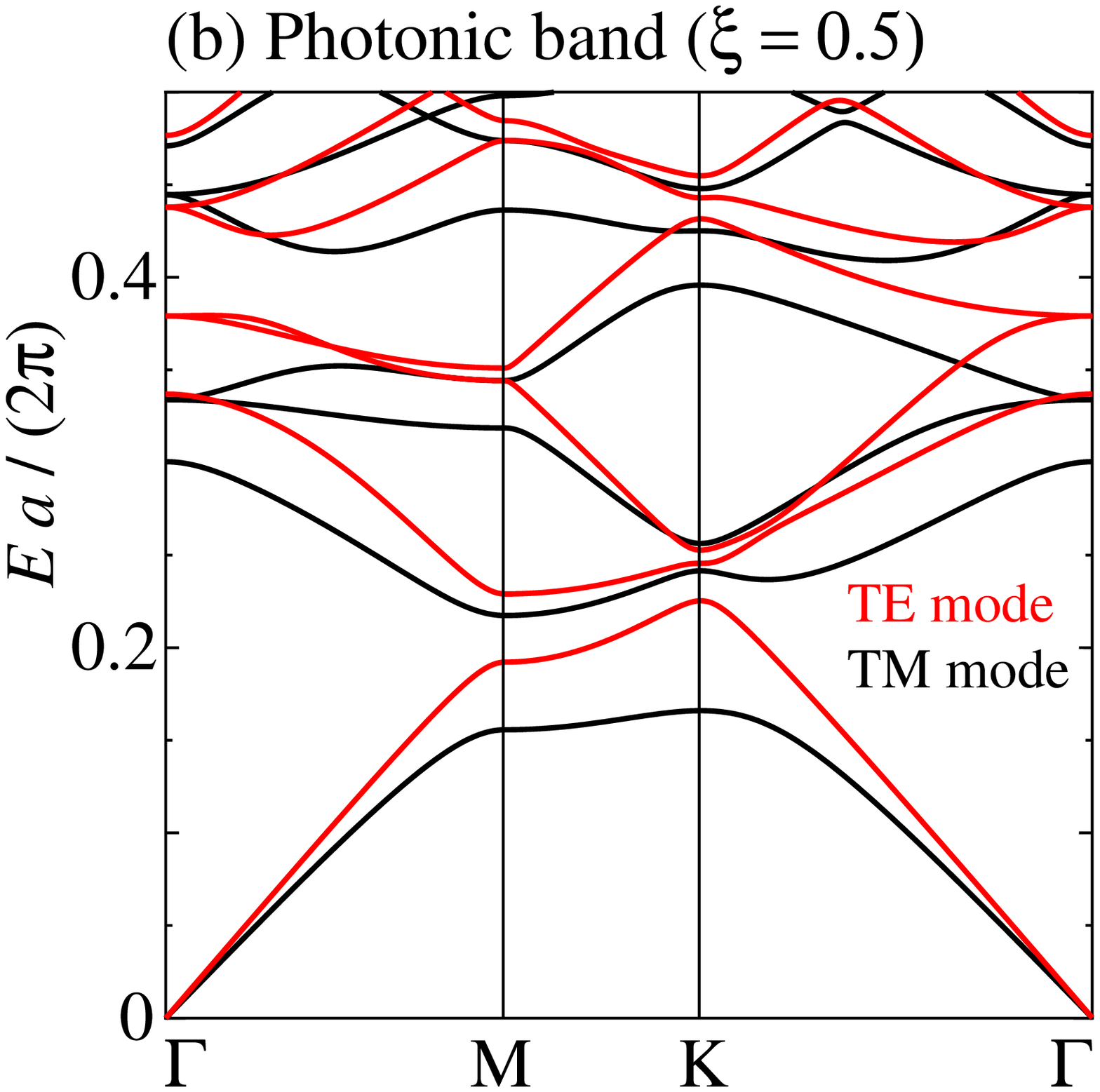}
\end{array}$
\caption{
(a) Dielectric function and (b) band structure of 
a two-dimensional photonic crystal.
The Brillouin zone is shown in Figs. \ref{fig:flux-spin} and
\ref{fig:seom}(b)
}
\label{fig:band}
\end{figure}
We present examples of
the Berry curvatures and the internal rotations
of non-degenerate bands
in the two-dimensional photonic crystal
with $\mu=\mu_{0}$ and
\begin{eqnarray}
\epsilon^{-1}(\bm{r})
&=&
\frac{4}{3(5+12|\xi|+8\xi^{2})}
\nonumber\\
&&\times
\sum_{i=1}^{3}
\Biggl[
\left[
\xi-\cos(\bm{b}_{i}\cdot\bm{r}+\frac{2\pi}{3})
\right]^{2}
\nonumber\\
&&\quad
+\left[
\xi+\cos(\bm{b}_{i}\cdot\bm{r}-\frac{2\pi}{3})
\right]^{2}
\Biggr],
\end{eqnarray}
where 
$\bm{b}_{1}
=(\frac{2\pi\sqrt{3}}{3a},
-\frac{2\pi}{3a})$,
$\bm{b}_{2}=(0,\frac{4\pi}{3a})$,
$\bm{b}_{3}=-\bm{b}_{1}-\bm{b}_{2}$,
and $a$ is the lattice constant.
It is noted that, for $0<|\xi|<1$, 
$\xi$ represents the degree of inversion-symmetry breaking.
The spatial distribution of $\epsilon(\bm{r})$ 
and the band structure of TM and TE modes
are shown in Fig.~\ref{fig:band}(a) and (b), respectively.

Figure \ref{fig:flux-spin} shows
the Berry curvatures and the internal rotations
of the first and second bands of TM and TE modes.
The internal rotation of an optical wavepacket is defined
by $(z_{c}|\bm{\mathcal{S}}_{n\bm{k}_{c}}|z_{c}) =
\langle W|\bm{\mathcal{J}}|W\rangle
-\bm{r}_{c}\times E_{n\bm{k}_{c}}\bm{\nabla}_{\bm{k}_{c}}E_{n\bm{k}_{c}}$,
where $\langle W|\bm{\mathcal{J}}|W\rangle$
is the total rotation of energy current 
and the second term represents the rotation of the center of gravity
(see Appendices~\ref{sec:periodic-system}, \ref{sec:internal-rotation}
and \ref{sec:Omega-2D}).
We can clearly see the correlation between them
in each band except for their relative sign.
The relative sign are roughly determined by a factor 
$\delta E = (E_{\mathrm{TM(TE)}\:n\bm{k}}-E_{\mathrm{TM(TE)}\:m\bm{k}})$
at nearly degenerate points $\bm{k}$,
where $n$ and $m$ represent band indices 
of nearly degenerate bands.
This is because the Berry curvature and
the internal rotation are proportional to 
$1/\delta E^{2}$ and $1/\delta E$, respectively
(see Appendix~\ref{sec:Omega-2D}).
It is expected that this internal rotation
is closely related to a physical angular momentum.
Therefore, these results suggest that
we can generate photonic modes with angular momentum by using
photonic crystals without inversion symmetry.
\begin{figure}[hbt]
$\begin{array}{cc}
\includegraphics[scale=0.18]{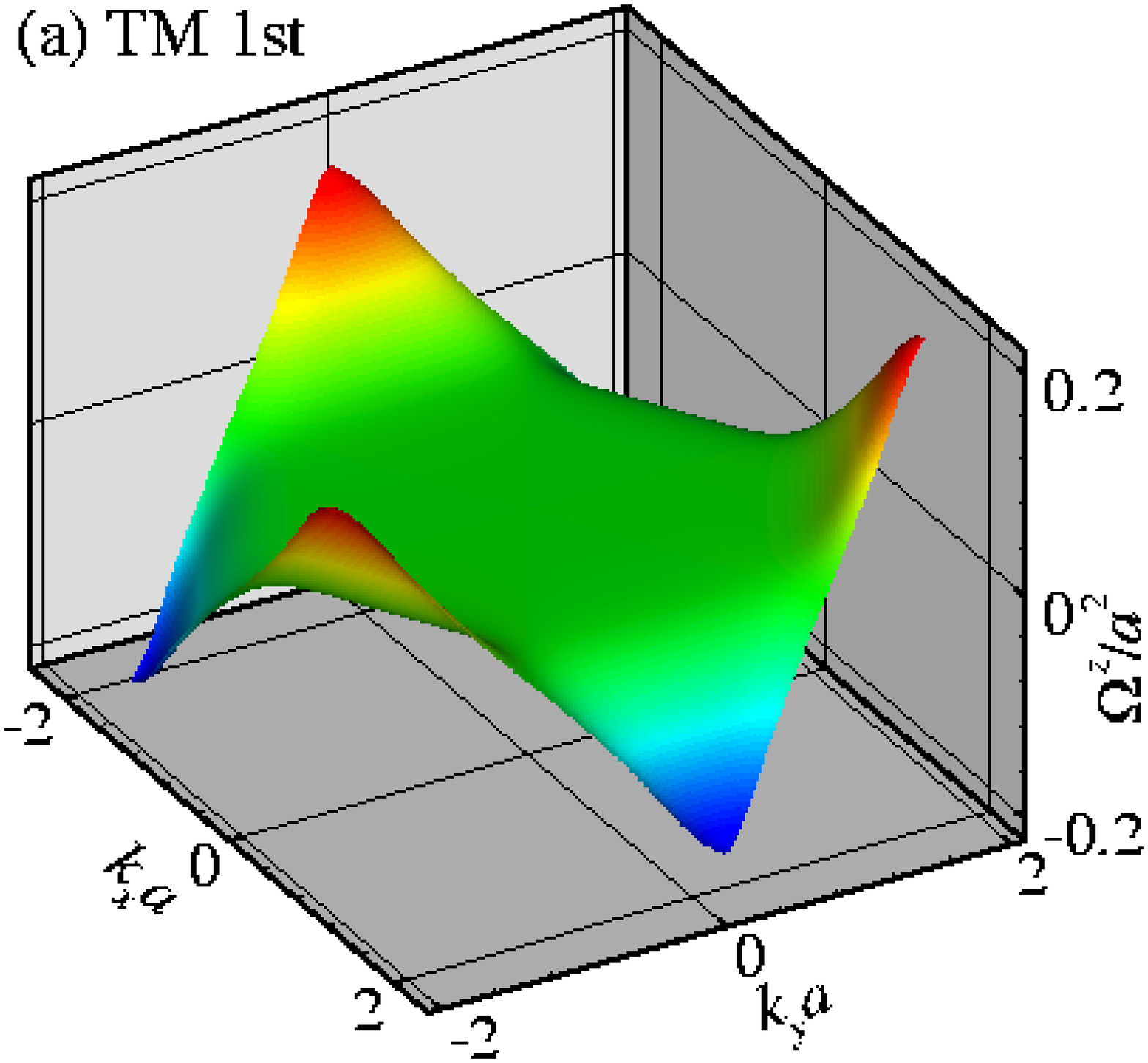}
& \includegraphics[scale=0.18]{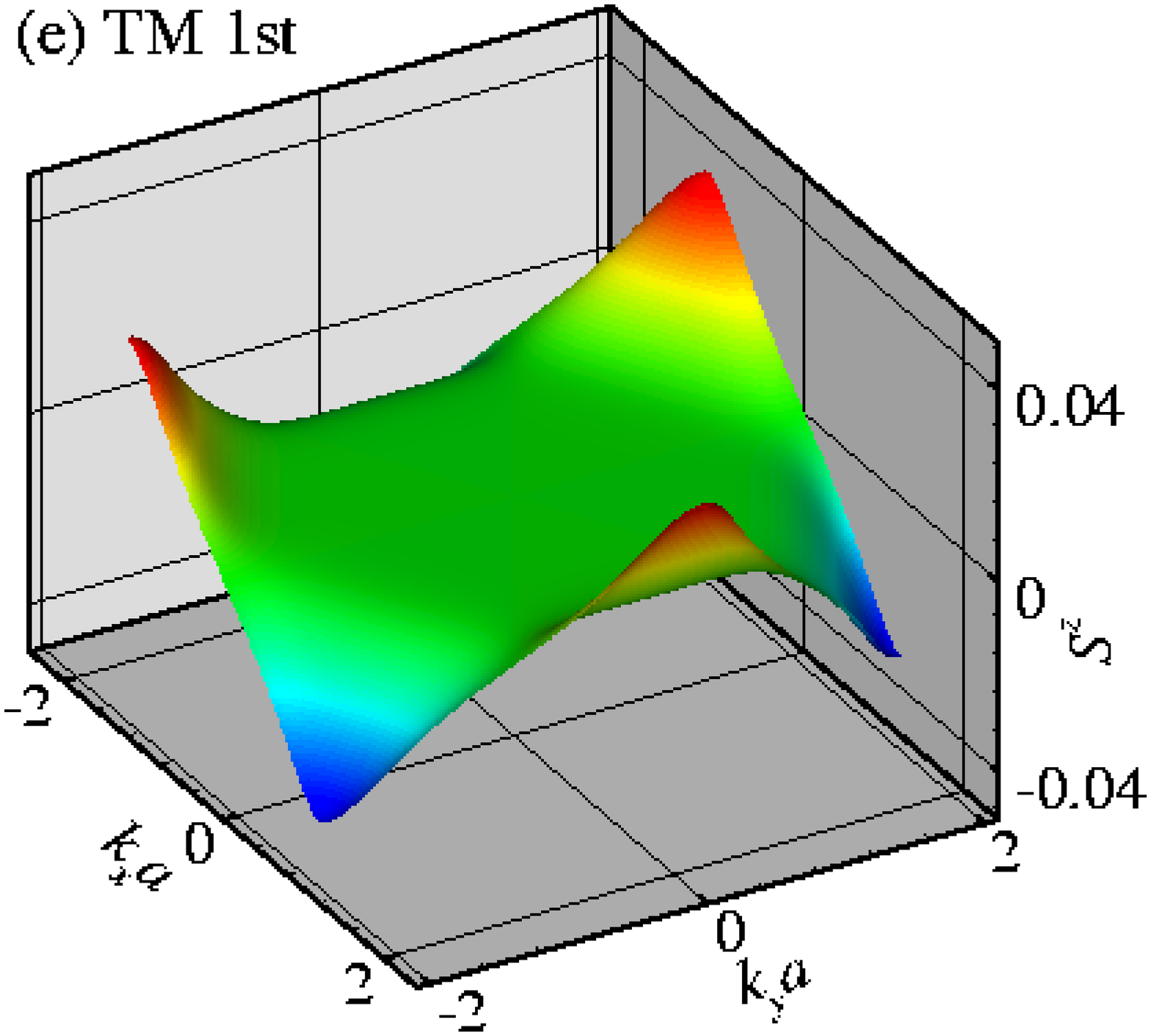}
\\
\includegraphics[scale=0.18]{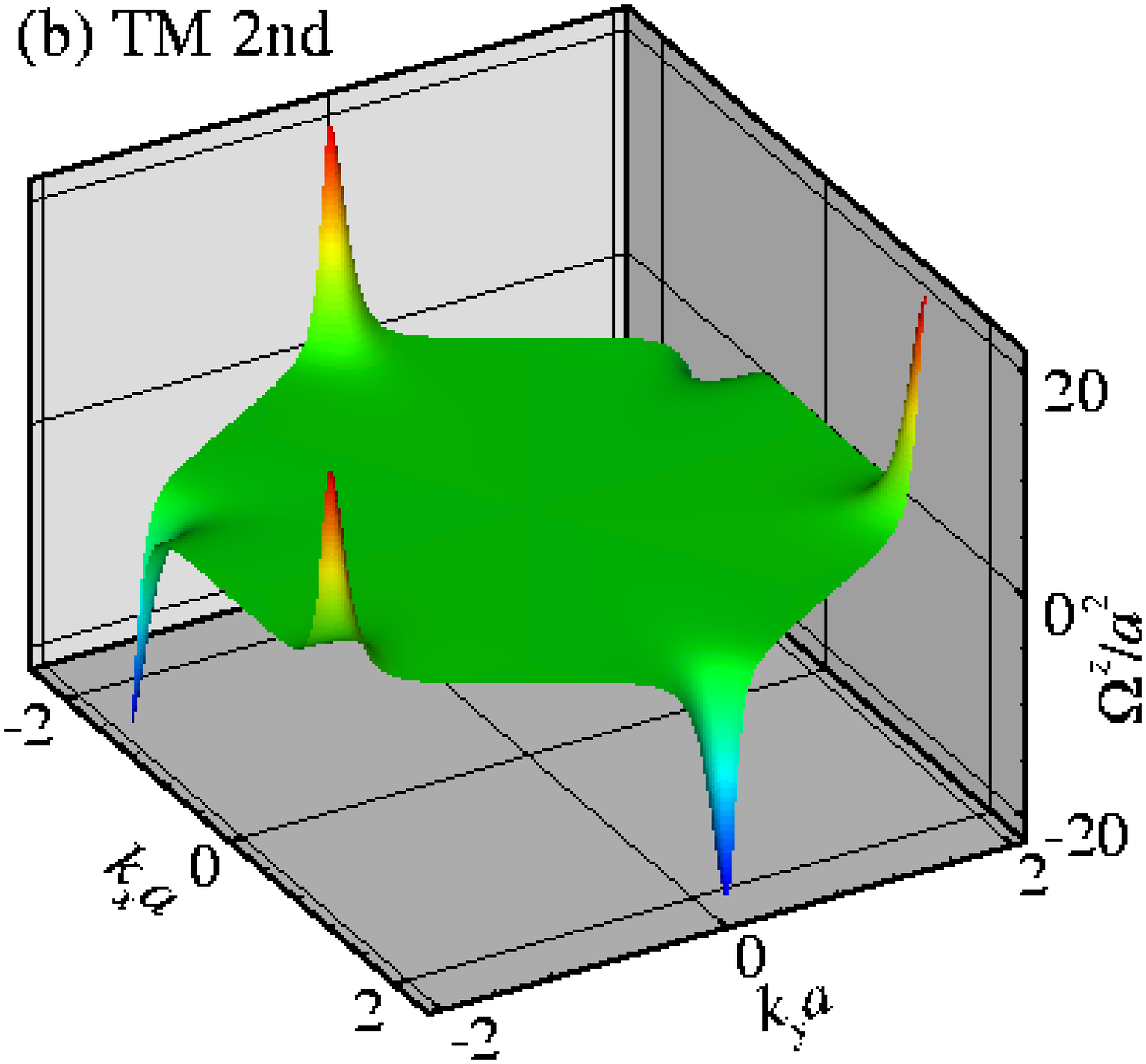}
& \includegraphics[scale=0.18]{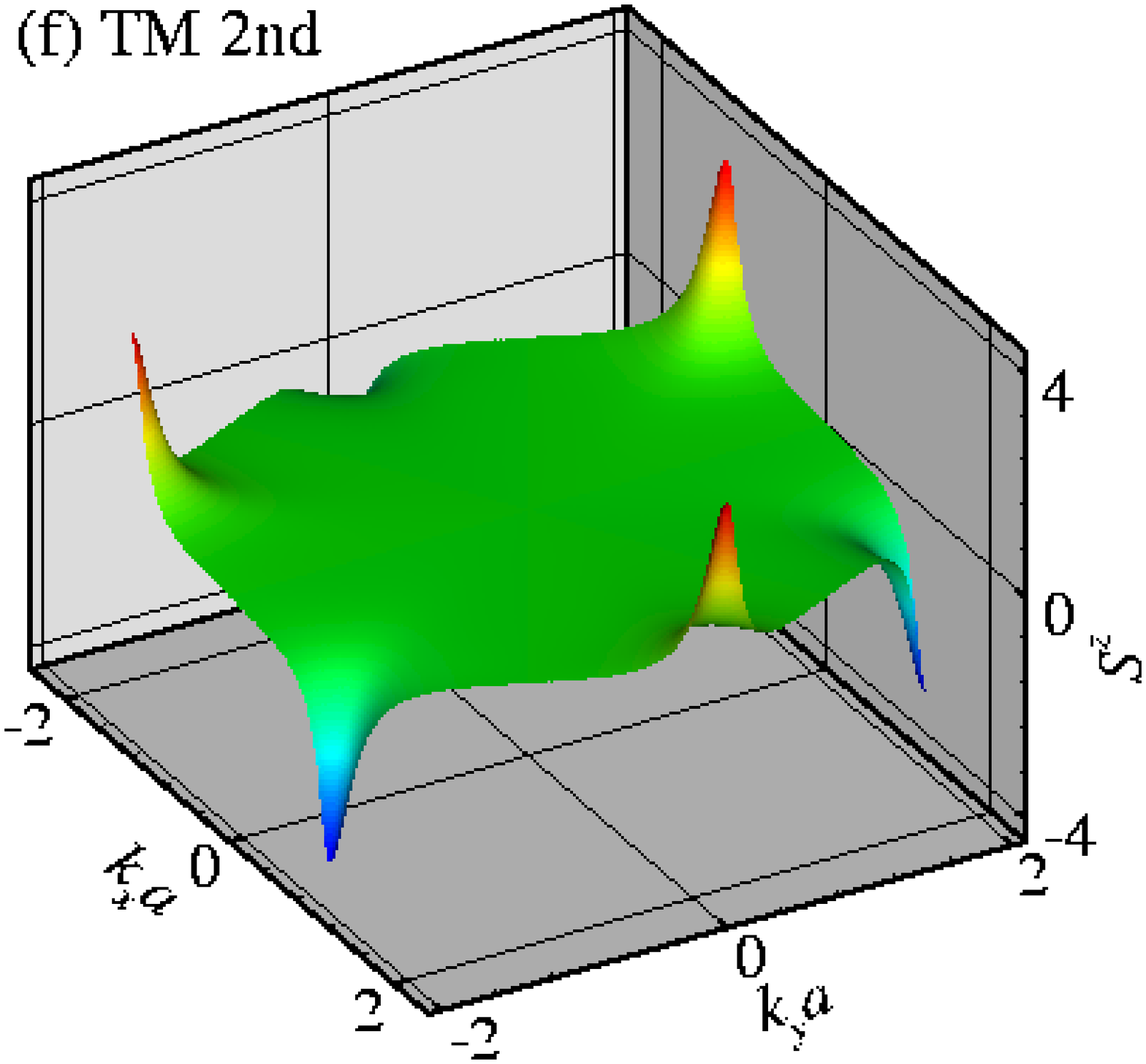}
\\
\includegraphics[scale=0.18]{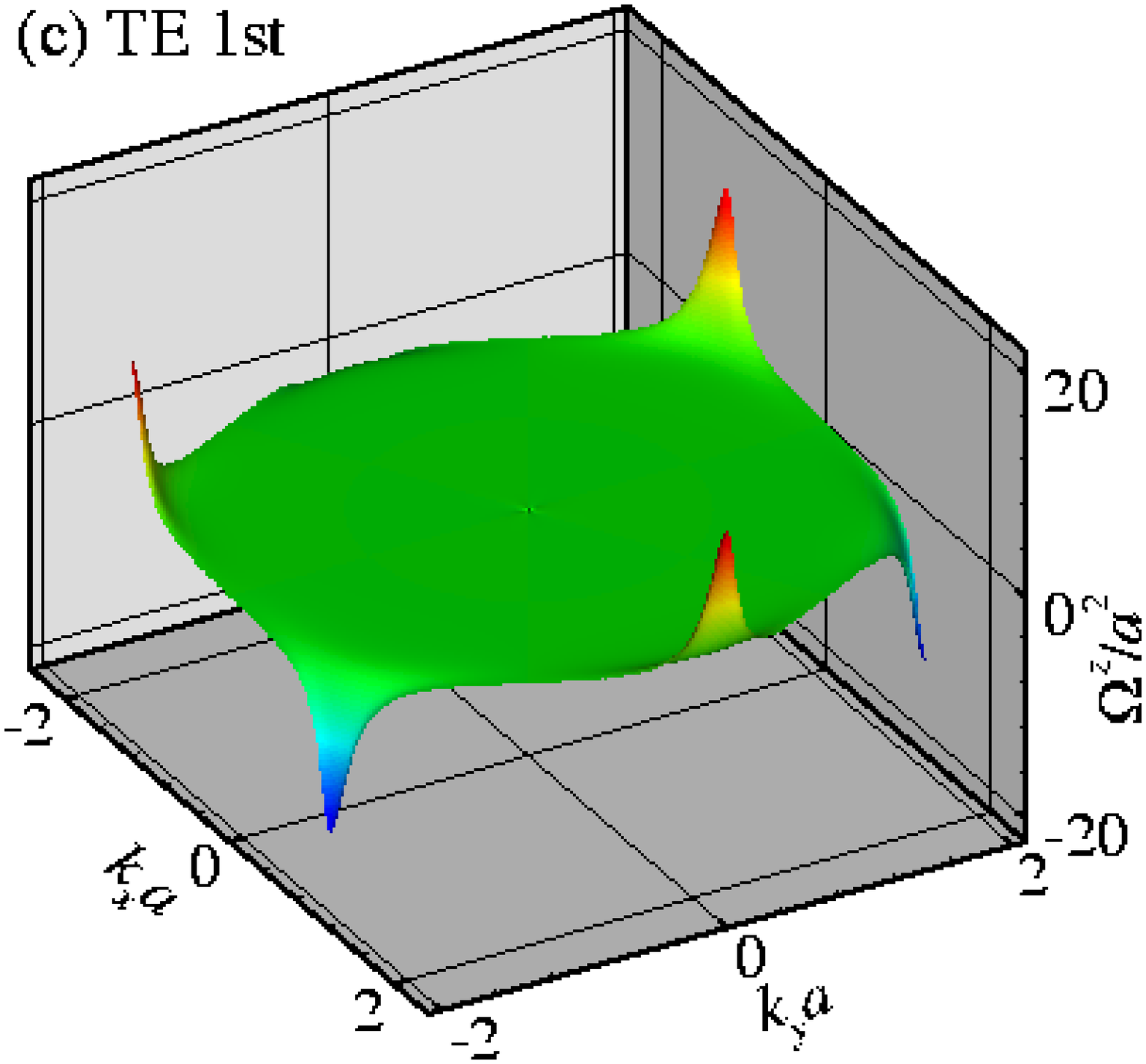}
& \includegraphics[scale=0.18]{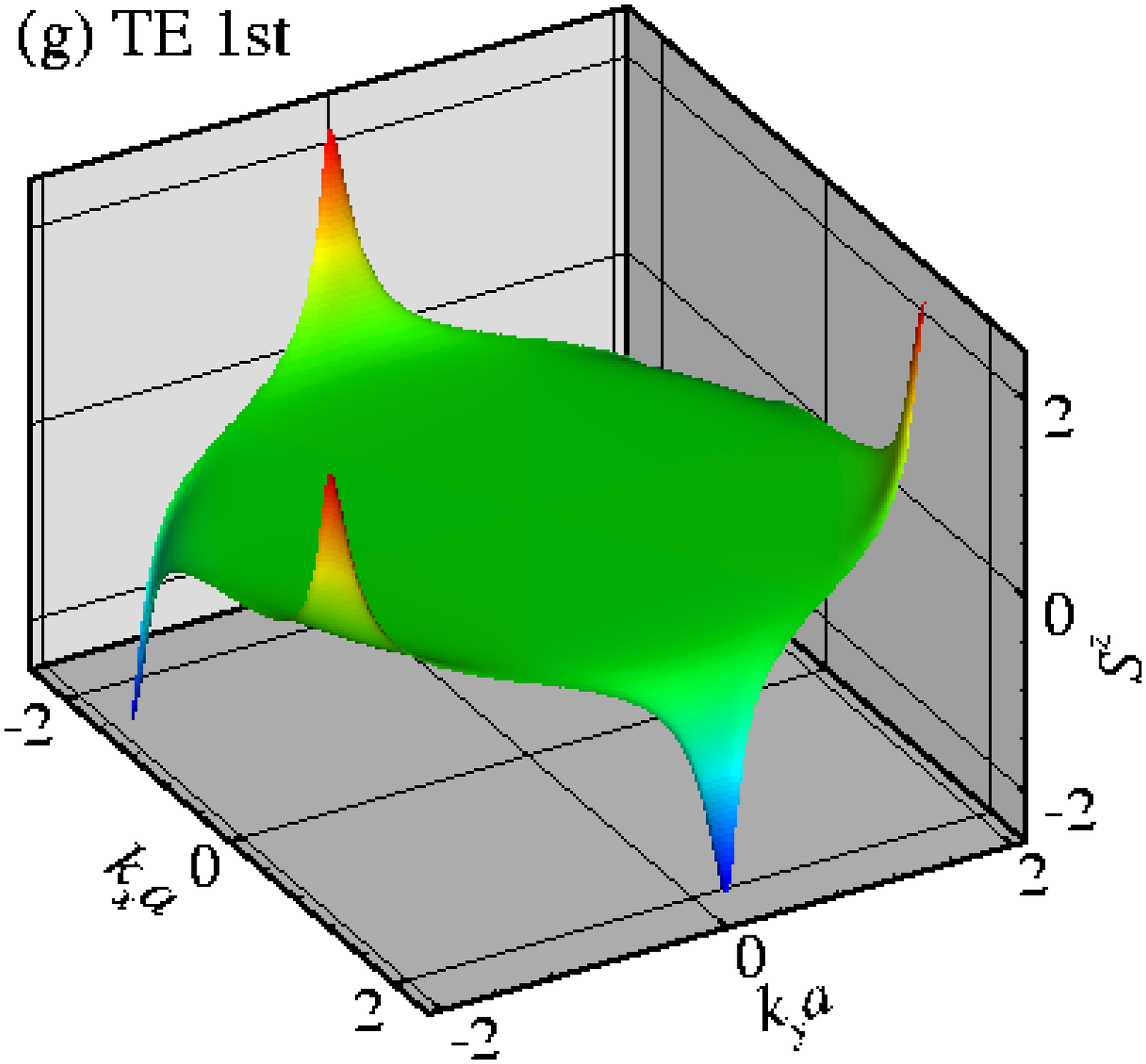}
\\
\includegraphics[scale=0.18]{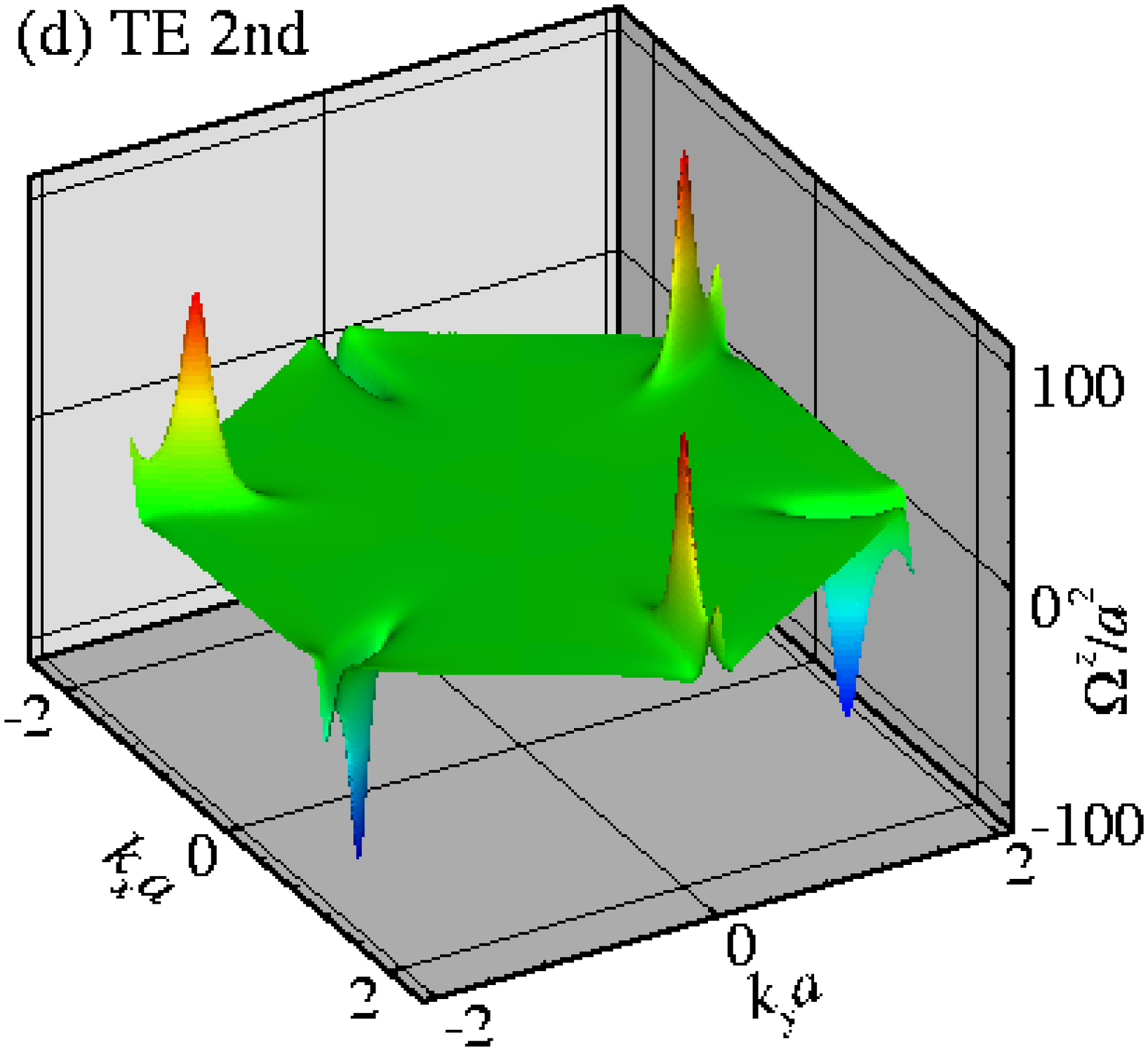}
& \includegraphics[scale=0.18]{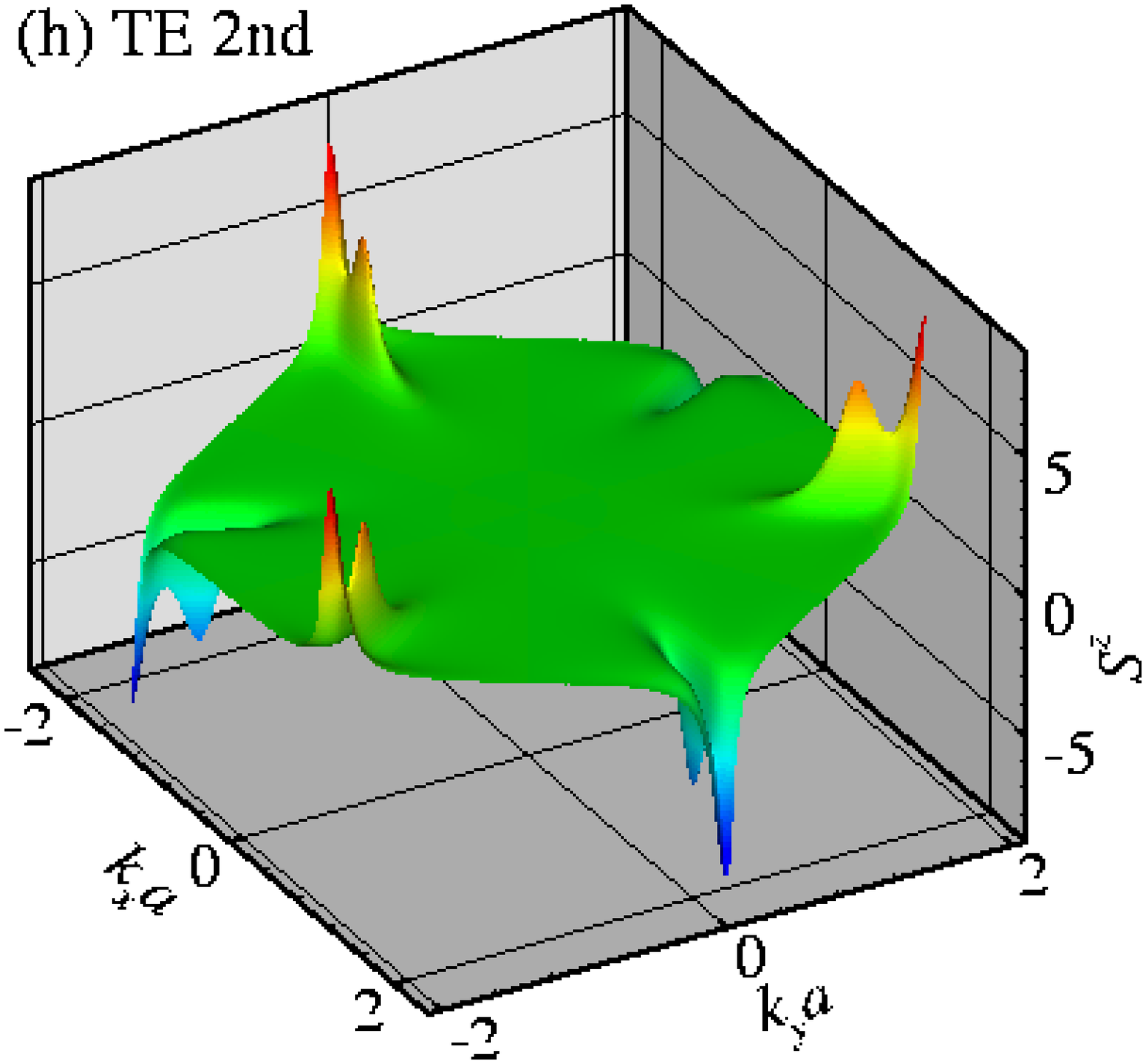}
\end{array}$
\caption{
Berry curvatures (a,b,c,d) and 
the internal rotations (e,f,g,h)
of the first and second bands of TM and TE modes
in the two-dimensional photonic crystal ($\xi=0.5$).
}
\label{fig:flux-spin}
\end{figure}

Before considering the motion of wavepackets in this photonic crystal,
we should comment on $\bm{\Delta}_{\mathrm{TM}\:n\bm{k}}$ and 
$\bm{\Delta}_{\mathrm{TE}\:n\bm{k}}$,
which give corrections to energy dispersions and group velocities
of TM and TE modes.
Because $\mu(\bm{r})=\mu_{0}$ in the present case, 
it follows from Eq.~(\ref{eq:DeltaTM})
in Appendix~\ref{sec:Omega-2D} 
that $\bm{\Delta}_{\mathrm{TM}\:n\bm{k}}=0$.
Thus, when a modulation is applied only to the dielectric permittivity
as $1/\epsilon(\bm{r})\to \gamma^{2}_{\epsilon}(\bm{r})/\epsilon(\bm{r})$,
the energy of the TM mode is just rescaled by the factor 
$\gamma_{\epsilon}(\bm{r}_{c})$, i.e., 
$E_{\mathrm{TM}\:n\bm{k}_{c}}\to
E_{\mathrm{TM}\:n\bm{k}_{c};\bm{r}_{c}}=
\gamma_{\epsilon}(\bm{r}_{c})E_{\mathrm{TM}\:n\bm{k}_{c}}$.
On the other hand, $\bm{\Delta}_{\mathrm{TE}\:n\bm{k}}$ is nonzero. 
From Eq.~(\ref{eq:modulated-E}), additional corrections 
appear in the energy dispersions of TE modes.
However, as shown in Appendix~\ref{sec:Delta}, 
we can see $\bm{\Delta}_{\mathrm{TE}\:n\bm{k}}\lesssim 0.1 a$.
Thus these corrections
are estimated to be at most a few percent as long as 
the modulation is sufficiently weak, i.e., 
$|a\bm{\nabla}_{\bm{r}_{c}}\ln\gamma_{\epsilon}(\bm{r}_{c})|\ll 1$.
In the similar argument, we can also neglect
corrections to the group velocities of TE modes
compared to their anomalous velocities,
at least in the present photonic crystal.
All the details of this issue is given in Appendix~\ref{sec:Delta}.

\begin{figure}[hbt]
$\begin{array}{cc}
\includegraphics[scale=0.3]{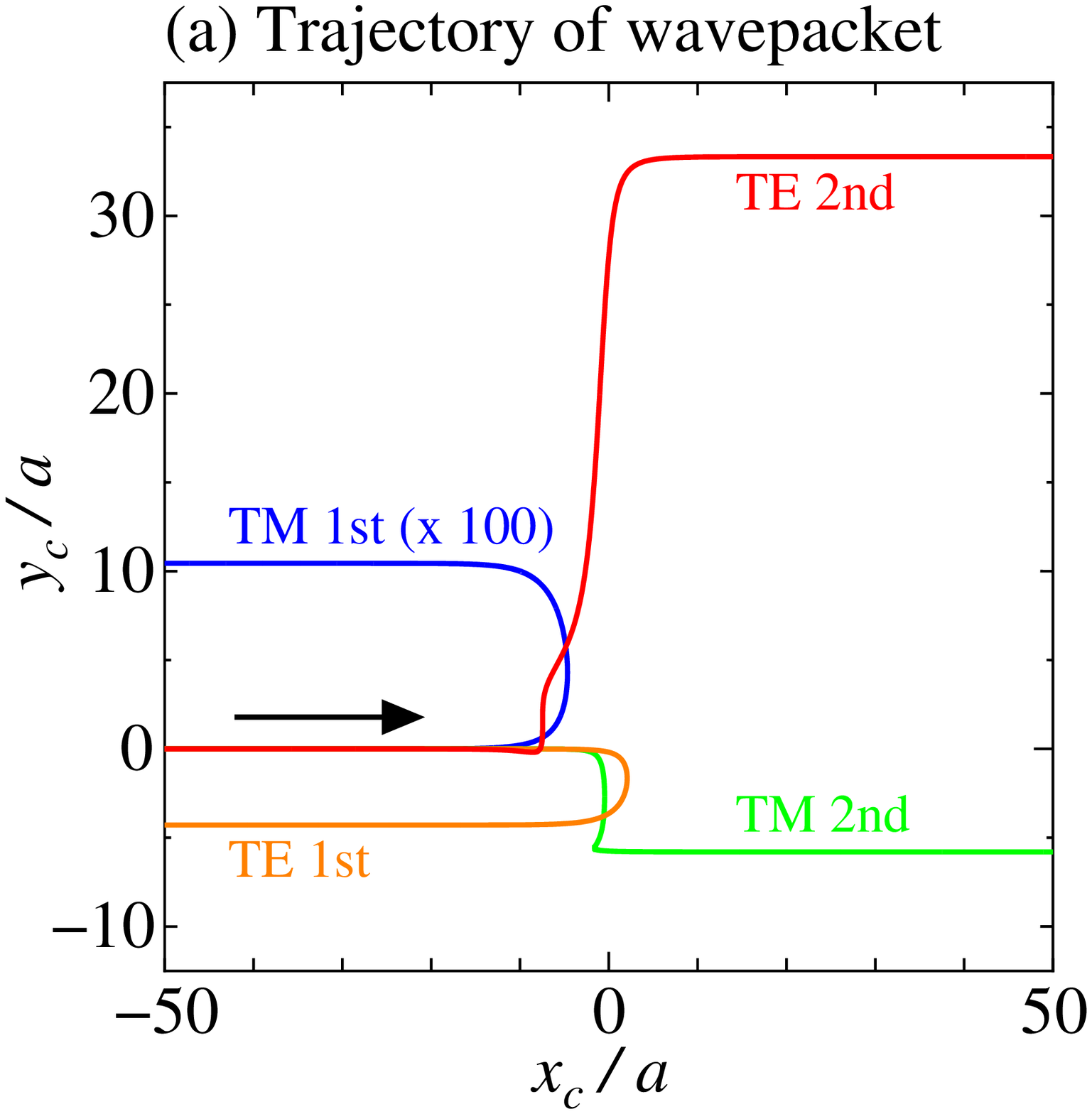}
& \includegraphics[scale=0.3]{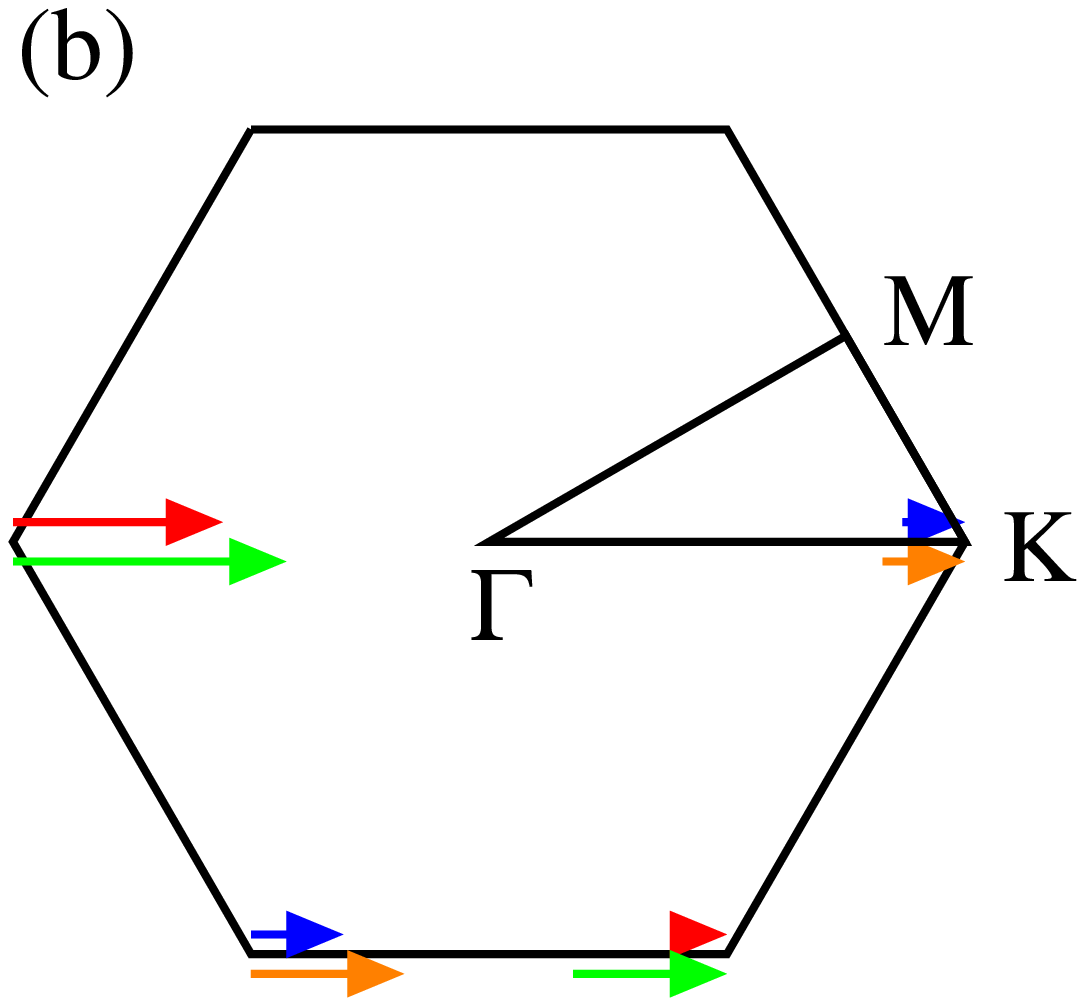}
\end{array}$
\caption{
Trajectories of wavepackets in
(a) real and (b) momentum spaces. 
The color of each arrow in (b)
corresponds to that of each line in (a).
The momentum-space trajectories in the figure are 
drawn with appropriate shifts
from their actual ones which are 
on the line of $k_{y}=0$
or the horizontal Brillouin-zone boundary.
}
\label{fig:seom}
\end{figure}
Now we consider the motions of wavepackets 
constructed from TM and TE modes.
It is noted that these wavepackets 
are extended in the $z$-direction, 
because the $z$-components of their momentum
are fixed as $k_{z}=0$.
From Fig.~\ref{fig:flux-spin}, 
we can see that $\bm{\Omega}_{\bm{k}}$ is strongly enhanced
near the corners of the Brillouin zone.
This enhancement is interpreted as a two-dimensional cut of 
the monopole structure in an extended space 
including parameters \cite{MN}, e.g., $\xi$ in the present case.
Therefore, we set the initial $\bm{k}_{c}$ near the corners of the Brillouin zone
in order to make the effect of anomalous velocity prominent.
We superimpose the following modulation onto the periodic structure
as $1/\epsilon(\bm{r})\to \gamma^{2}_{\epsilon}(\bm{r})/\epsilon(\bm{r})$,
\begin{eqnarray}
\frac{1}{\gamma_{\epsilon}(\bm{r})}
&=&
\frac{1}{2}\left[(\tilde{n}+1)+(\tilde{n}-1)\tanh\frac{x}{w}\right],
\end{eqnarray}
where $\tilde{n}>0$ represents a relative refractive index 
multiplied to the periodic structure in the region $x\to\infty$,
and $w$ is the mean width of the modulation.
Here we take $\tilde{n}=1.2$ and $w=5a$
which satisfy the condition of weak and slowly-varying modulation.
The obtained trajectories are shown in Fig.~\ref{fig:seom}.
It is found that the shift of $\bm{r}_{c}$
reaches to dozens of times the lattice constant especially 
for the wavepacket constructed from the TE second band.

Finally we note that, 
also in more generic systems than discussed above,
this effect can be enhanced considerably 
by designing crystal structures.
The Berry curvature around a nearly degenerate point
is determined mostly by the splitting, 
$2|m_{g}|$, between neighboring bands.
Note that the sign of $m_{g}$ depends on details 
of wavefunctions around the nearly degenerate point, 
while its magnitude is determined only by the splitting.
Suppose that at $\bm{k}=\bm{k}_{0}$ another band 
comes very close in energy to the one considered.
The Berry curvature around $\bm{k}_{0}$ is evaluated as
\begin{equation}
\Omega_{z}\sim\frac{v^{2} m_{g}}{
(v^{2}|\bm{k}-\bm{k}_{0}|^{2}+m_{g}^{2})^{3/2}},
\end{equation}
where $v$ is a nominal velocity around $\bm{k}_{0}$.
Thus when the light traverses near $\bm{k}_{0}$,
the shift is estimated as
\begin{equation}
\delta y_{c}\sim 
-\frac{v}
{\sqrt{v^{2}\kappa^{2}+m_{g}^{2}}}
\mathrm{sgn}[m_{g}\nabla_{x_{c}}\gamma(x_c)]
,
\end{equation}
where $\kappa$ is the minimum value of $|\bm{k}-\bm{k}_{0}|$ when
$\bm{k}$ traverses near $\bm{k}_{0}$. 
Therefore the shift is larger for smaller $|m_{g}|$ and $\kappa$.
This argument gives an intelligent explanation of
a relation among relative magnitudes and signs 
of the sifts in Fig.~\ref{fig:seom}.

\section{\label{sec:Discussion}Discussion}
We have presented in detail the derivation of 
the equations of motion for an optical wavepacket
within the unitary theory. In our formalism, the equations are derived in
the same fashion as those of electronic systems, thereby 
the similarities between them are evident.
This suggests a broad concept, i.e., 
topological Hall effect driven by the geometrical mechanism, 
which ranges over a wide area of 
physics such as electronic, acoustic, hydrodynamical, relativistic 
and photonic phenomena. 
For example, this optical Hall effect is referred to as 
``the optical Magnus effect" \cite{Liberman-Zeldovich,Bliokh},
in analogy with the Magnus effect, which is 
a transverse aerodynamic effect on rotating objects.
On the other hand, the relation of this effect to
the geometrical effect on a spinning particle in general relativity was
recently pointed out \cite{Duval}.
They derived the similar equations of motion to ours
by considering the motion of a spinning particle
in the space with a metric $g_{ij}(\bm{r})=n(\bm{r})\delta_{ij}$. 
Their argument reminds us that, in the early stage of the study on general relativity,
Einstein had tried to formulate the theory by generalizing 
the speed of light in vacuum.
This issue might be related to the deep question of 
the dual nature between the force and the velocity in the dynamics.

Here we should mention the effects called
the photonic Hall effect \cite{photonic-Hall-th, photonic-Hall-ex}
and the magnetically induced deflection due to the Pitaefskii magnetization
\cite{Landau, Rikken-Tiggelen, comment-I, reply-I, comment-II, reply-II},
both of which are observed
in Faraday-active media subject to external magnetic fields.
The former effect takes place in a random medium,
and is theoretically interpreted by
the magnetically induced off-diagonal components
of a diffusion tensor \cite{photonic-Hall-th} and 
experimentally proved to be due to 
the magnetically induced changes in the optical 
properties of scatterers \cite{photonic-Hall-ex}.
The latter effect is observed in a homogeneous medium \cite{Rikken-Tiggelen},
and is interpreted by the magnetically induced change 
in the dispersion relation of each mode
due to the Pitaevskii magnetization \cite{Landau}.
(Additional remarks on these effects 
are given in Appendix~\ref{sec:magneto-deflection}.)
On the other hand, the optical Hall effect 
is caused by the anomalous velocity due to 
the geometrical propriety of a wavepacket,
which appears without external magnetic field nor scatterers.

As focused in Appendix~\ref{sec:internal-rotation}, there is a close relation
between the Berry curvature and the internal rotation.
It is physically expected that an internal rotation can be related to
an internal angular momentum.
From this viewpoint, the various modes of Laguerre-Gauss beam, 
which have internal orbital angular momenta \cite{OAM}, 
are of a particular interest.
The Imbert-Fedorov effect is expected for these modes
as well as for circularly polarized states, and 
theoretical and experimental investigations on this problem
has been done recently \cite{Fedoseev-LG,Dasgupta,Sasada}.
In Sec.~\ref{sec:photonic-crystal},
we have shown that there appear
photonic modes with internal rotations
in a two-dimensional photonic crystal
without inversion symmetry.
Angular momentum corresponding to this kind of internal rotation
would be detected by measuring a torque 
taken by a photonic crystal
when we inject a linearly polarized light into the crystal
(through a buffer layer if needed).
In addition, when a photonic crystal is composed of
Faraday-active media and subject to an external magnetic field,
there would take place the magnetically induced deflection 
due to the Pitaevskii magnetization caused by 
this kind of generic internal rotation,
rather than by the spin of circular polarization.

Lastly, we make a remark on the relevance of the quantum nature 
in the geometrical/topological properties discussed in this paper. 
Although we have formulated the theory of an optical wavepacket
in the quantum-mechanical formalism in order to clarify 
its connection to that of an electronic wavepacket,
the phenomena itself is based on duality between 
real-space coordinates and momenta which is common in wave dynamics.
Therefore, the topological Hall effect is generic in both quantum mechanics
and classical wave dynamics. 
(See the argument in Appendix~\ref{sec:quantum-classical}.) 
Actually, we can extend the argument presented here 
to other kinds of wavepacket dynamics, e.g., dynamics of sonic wavepacket
which is mentioned in Ref.~\cite{Liberman-Zeldovich}.
The sonic Hall effect in phononic crystals 
\cite{phononic-crystal-I, phononic-crystal-II}
could be enhanced in the same manner as
the optical Hall effect in photonic crystals.
It should be noted that the spin of the constituent particles 
is not always necessary; 
even a scalar wave can also have an internal rotation
and a Berry curvature due to a periodic structure
breaking inversion and/or time-reversal symmetries.

\appendix
\section{\label{sec:quantum-classical}Quantum or classical?}
In this paper, we employ a quantum-mechanical formalism
in order to formulate our theory of a photonic system
on an equal footing with that of an electronic system.
The photonic system is described by an effective model where
a dielectric medium is regarded as a classical object.
This quantization procedure of photons is corresponding to
that of electrons described by an effective model,
e.g., a model with an effective mass-matrix
and/or an effective (periodic) potential.
In an electronic system, an electron is treated as a quantum object,
and the equations of motion for a semiclassical wavepacket
are derived from an effective Lagrangian.
In a photonic system, the counterpart of this effective Lagrangian
is most naturally represented in a second-quantized formalism
with keeping its close connection to that in an electronic system.
This is because, in the Maxwell theory, we cannot define a positive-definite 
probability density, while we can define a positive-definite energy density.
(The second quantization is adopted to define 
the quantum wavefunction of a photon in Sec.~\ref{sec:EOM},
which cannot be directly represented 
by the field strength of an electromagnetic field.) 
However, as long as we consider an approximately coherent wavepacket,
the center of wavepacket coincide with the center of gravity
as shown in Sec.~\ref{sec:EOM}.
This fact enables us to link our quantum-mechanical theory
and classical electrodynamics.

One may wonder whether the optical Hall effect is quantum or classical.
We cannon answer this question in a single sentence,
but give some remarks on it as follows.
This effect comes from the particle-wave duality of an optical wavepacket
and the geometrical/topological property of a wavefunction.
Therefore, similar geometrical/topological effects 
are expected in various kinds of
quantum/classical and microscopic/macroscopic wave dynamics,
when wavepackets under consideration are approximately coherent.
Indeed, we can formulate a theory for this class of phenomena
based on a classical wave dynamics of a macroscopic system,
while its direct connection to electronic systems is not necessarily clear.
The confusion represented by the above question
is mainly due to the situation in which we sometimes refer to 
wave equations for photonic systems as classical Maxwell equations
and those for electronic systems as quantum Schr{\" o}dinger equations,
while both photons and electrons are quantum objects.
The origin of this situation comes from
two of advantages of photonic systems
which lead to the success of classical Maxwell theory.
In contrast to electronic systems,
the statistics of photons is bosonic,
and effective self-interactions between photons 
are usually very weak.
However, as long as we treat a photon and an electron
in a single particle approximation,
we can formulate both theories on an equal footing.

It is beyond the scope of this paper
to fix the terminologies ``quantum'' and ``classical''
common in photonic and electronic systems.
Although we treat a photon as a quantum object through to 
the end of this paper, 
we refer to results obtained purely by wave dynamics of light 
as those obtained by classical electrodynamics.
As long as we consider an approximately coherent wavepacket 
in a single particle approximation of quantum theory of photon
or in a linear approximation of classical electrodynamics,
results obtained by both formalisms coincide with each other
as shown in Sec.~\ref{sec:Imbert}.

\section{\label{sec:orthonormality}Orthonormality of eigenfunctions}
The orthonormality, Eqs.~(\ref{eq:orthonormal-PhiE}) and (\ref{eq:orthonormal-PhiH})
is approved with the orthonomality of Bloch functions,
Eqs.~(\ref{eq:orthonormal-UE}) and (\ref{eq:orthonormal-UH}).
For example, Eq.~(\ref{eq:orthonormal-PhiE}) can be shown by using
the following relation,
\begin{eqnarray}
&&(\Phi^{E}_{n\lambda\bm{k}}|\tensor{\epsilon}
|\Phi^{E}_{n'\lambda'\bm{k}'})
\nonumber\\
&&=(E_{n\bm{k}}+E_{n'\bm{k}'})
\int d\bm{r} \bm{\Phi}^{E*}_{n\lambda\bm{k}}(\bm{r},t)
\tensor{\epsilon}(\bm{r})\bm{\Phi}^{E}_{n'\lambda'\bm{k}'}(\bm{r},t)
\nonumber\\
&&=\frac{E_{n\bm{k}}+E_{n'\bm{k}'}}{2\sqrt{E_{n\bm{k}}E_{n'\bm{k}'}}}
e^{i(E_{n\bm{k}}-E_{n'\bm{k}'})t}
\nonumber\\
&&\quad\times
\int d\bm{r}
\:e^{-i(\bm{k}-\bm{k}')\cdot\bm{r}}
\bm{U}^{E*}_{n\lambda\bm{k}}(\bm{r})
\tensor{\epsilon}(\bm{r})\bm{U}^{E}_{n'\lambda'\bm{k}'}(\bm{r})
\nonumber\\
&&=\frac{E_{n\bm{k}}+E_{n'\bm{k}'}}{2\sqrt{E_{n\bm{k}}E_{n'\bm{k}'}}}
e^{i(E_{n\bm{k}}-E_{n'\bm{k}'})t}
\nonumber\\
&&\quad\times
\sum_{\bm{a}}\int_{\mathrm{WS}} d\bm{r}
\:e^{-i(\bm{k}-\bm{k}')\cdot(\bm{a}+\bm{r})}
\bm{U}^{E}_{n\lambda\bm{k}}(\bm{r})
\tensor{\epsilon}(\bm{r})\bm{U}^{E}_{n'\lambda'\bm{k}'}(\bm{r})
\nonumber\\
&&=\frac{E_{n\bm{k}}+E_{n'\bm{k}'}}{2\sqrt{E_{n\bm{k}}E_{n'\bm{k}'}}}
e^{i(E_{n\bm{k}}-E_{n'\bm{k}'})t}
\sum_{\bm{G}}\tilde{\delta}(\bm{k}-\bm{k}'+\bm{G})
\nonumber\\
&&\quad\times
\int_{\mathrm{WS}} \frac{d\bm{r}}{v_{\mathrm{WS}}}
\:e^{-i(\bm{k}-\bm{k}')\cdot\bm{r}}
\bm{U}^{E}_{n\lambda\bm{k}}(\bm{r})
\tensor{\epsilon}(\bm{r})\bm{U}^{E}_{n'\lambda'\bm{k}'}(\bm{r}),
\end{eqnarray}
where $\bm{a}$ represents an arbitrary lattice vector.
Since the lattice momentum $\bm{k}$ and $\bm{k}'$ are in the first Brillouin zone,
we can reach the result
\begin{eqnarray}
&&(\Phi^{E}_{n\lambda\bm{k}}|\tensor{\epsilon}
|\Phi^{E}_{n'\lambda'\bm{k}'})
\nonumber\\
&&=\frac{E_{n\bm{k}}+E_{n'\bm{k}}}{2\sqrt{E_{n\bm{k}}E_{n'\bm{k}}}}
e^{i(E_{n\bm{k}}-E_{n'\bm{k}})t}
\tilde{\delta}(\bm{k}-\bm{k}')
\langle U^{E}_{n\lambda\bm{k}}|
\tensor{\epsilon}|U^{E}_{n'\lambda'\bm{k}}\rangle
\nonumber\\
&&=\delta_{nn'}\delta_{\lambda\lambda'}\tilde{\delta}(\bm{k}-\bm{k}'),
\end{eqnarray}
where we have used Eq.~(\ref{eq:orthonormal-UE}).
In the same manner, Eq.~(\ref{eq:orthonormal-PhiH}),
can be also proved by using Eq.~(\ref{eq:orthonormal-UH}).

Next we prove the orthogonality, Eq.~(\ref{eq:orthogonality-Phi}).
From the definition of the inner product, Eq.~(\ref{eq:inner-Phi}),
we can show
\begin{subequations} 
\begin{eqnarray}
&&(\Phi^{E*}_{n\lambda\bm{k}}|\tensor{\epsilon}|\Phi^{E}_{n'\lambda'\bm{k}'})
\nonumber\\
&&=(E_{n'\bm{k}'}-E_{n\bm{k}})
\int d\bm{r} \bm{\Phi}^{E}_{n\lambda\bm{k}}(\bm{r},t)
\tensor{\epsilon}(\bm{r})\bm{\Phi}^{E}_{n'\lambda'\bm{k}'}(\bm{r},t),
\nonumber\\
\\
&&(\Phi^{H*}_{n\lambda\bm{k}}|\tensor{\mu}|\Phi^{H}_{n'\lambda'\bm{k}'})
\nonumber\\
&&=(E_{n'\bm{k}'}-E_{n\bm{k}})
\int d\bm{r} \bm{\Phi}^{H}_{n\lambda\bm{k}}(\bm{r},t)
\tensor{\mu}(\bm{r})\bm{\Phi}^{H}_{n'\lambda'\bm{k}'}(\bm{r},t).
\nonumber\\
\end{eqnarray}
\end{subequations}
In the case of $E_{n\bm{k}}=E_{n'\bm{k}'}$,
it is clear that Eq.~(\ref{eq:orthogonality-Phi}) is approved.
Thus, in what follows, we consider the case of $E_{n\bm{k}}\neq E_{n'\bm{k}'}$.
In this case, we can easily show 
$\int d\bm{r} \bm{\Phi}^{E}_{n\lambda\bm{k}}(\bm{r},t)
\tensor{\epsilon}(\bm{r})\bm{\Phi}^{E}_{n'\lambda'\bm{k}'}(\bm{r},t)= 0$
and $\int d\bm{r} \bm{\Phi}^{H}_{n\lambda\bm{k}}(\bm{r},t)
\tensor{\mu}(\bm{r})\bm{\Phi}^{H}_{n'\lambda'\bm{k}'}(\bm{r},t)
= 0$ from the relations,
\begin{subequations}
\begin{eqnarray}
(E^{2}_{n'\bm{k}'}-E^{2}_{n\bm{k}})
\int d\bm{r} \bm{\Phi}^{E}_{n\lambda\bm{k}}(\bm{r},t)
\tensor{\epsilon}(\bm{r})\bm{\Phi}^{E}_{n'\lambda'\bm{k}'}(\bm{r},t)
&=& 0,
\nonumber\\
\\
(E^{2}_{n'\bm{k}'}-E^{2}_{n\bm{k}})
\int d\bm{r} \bm{\Phi}^{H}_{n\lambda\bm{k}}(\bm{r},t)
\tensor{\mu}(\bm{r})\bm{\Phi}^{H}_{n'\lambda'\bm{k}'}(\bm{r},t)
&=& 0.
\nonumber\\
\end{eqnarray}
\end{subequations}
Consequently, the orthogonality, Eq.~(\ref{eq:orthogonality-Phi})
is approved in all cases.

The above relations are derived from 
the eigen equations, Eqs.~(\ref{eq:eigen-PhiE}) and (\ref{eq:eigen-PhiH}).
For example, the relation for $\bm{\Phi}^{E}_{n\lambda\bm{k}}(\bm{r},t)$ 
is proved as,
\begin{eqnarray}
&&E^{2}_{n'\bm{k}'}
\int d\bm{r} \bm{\Phi}^{E}_{n\lambda\bm{k}}(\bm{r},t)
\tensor{\epsilon}(\bm{r})\bm{\Phi}^{E}_{n'\lambda'\bm{k}'}(\bm{r},t)
\nonumber\\
&&=
\int d\bm{r} \bm{\Phi}^{E}_{n\lambda\bm{k}}(\bm{r},t)
\cdot\left[\bm{\nabla}_{\bm{r}}\times
\left[
\tensor{\mu}^{-1}(\bm{r})\bm{\nabla}_{\bm{r}}\times\bm{\Phi}^{E}_{n'\lambda'\bm{k}'}(\bm{r},t)
\right]
\right]
\nonumber\\
&&=
\int d\bm{r} 
\left[\bm{\nabla}_{\bm{r}}\times
\left[\tensor{\mu}^{-1}(\bm{r})
\bm{\nabla}_{\bm{r}}\times
\bm{\Phi}^{E}_{n\lambda\bm{k}}(\bm{r},t)
\right]
\right]
\cdot\bm{\Phi}^{E}_{n'\lambda'\bm{k}'}(\bm{r},t)
\nonumber\\
&&=E^{2}_{n\bm{k}}
\int d\bm{r} \bm{\Phi}^{E}_{n\lambda\bm{k}}(\bm{r},t)
\tensor{\epsilon}(\bm{r})\bm{\Phi}^{E}_{n'\lambda'\bm{k}'}(\bm{r},t),
\end{eqnarray}
where $\tensor{\mu}^{T}(\bm{r})=\tensor{\mu}(\bm{r})$
is used in the transformation from the second line to the third line,
and $\tensor{\epsilon}^{T}(\bm{r})=\tensor{\epsilon}(\bm{r})$
is used in the transformation from the third line to the fourth line.
A similar relation can be derived also for $\bm{\Phi}^{H}_{n\lambda\bm{k}}(\bm{r},t)$.

\section{\label{sec:periodic-system}Wavepacket in a periodic system}
Here we present details
about an optical wavepacket in a periodic system.
Basic features of the wavepacket are discussed
in Appendix~\ref{sec:wavepacket}.
These features are helpful to understand
the effect of an additional modulation 
superimposed onto a periodic structure,
which is discussed in Sec.~\ref{sec:EOM}.
Some comments on a gauge transformation in momentum space
is given in Appendix~\ref{sec:gauge-transformation}.
In Appendix~\ref{sec:expectation-values},
we present detailed procedures to evaluate expectation values
which appear in Appendix~\ref{sec:wavepacket}.

\subsection{\label{sec:wavepacket}wavepacket}
We begin with the wavepacket defined by
\begin{subequations}
\begin{eqnarray}
&&|W\rangle =
\int_{\mathrm{BZ}}d\bm{k}\:
w(\bm{k},\bm{k}_{c},t)
\sum_{\lambda}z_{c\lambda}a^{\dagger}_{n\lambda\bm{k}}|0\rangle
,\label{eq:W-periodic}
\\
&&w(\bm{k},\bm{k}_{c},t)
=
w_{r}(\bm{k}-\bm{k}_{c})e^{-i\vartheta(\bm{k},t)},
\end{eqnarray}
\end{subequations}
where $w_{r}(\bm{k}-\bm{k}_{c})$ is a real function, and 
$w_{r}(\bm{k}-\bm{k}_{c})$ and $z_{c\lambda}$ satisfy
the normalization conditions,
$\int_{\mathrm{BZ}}d\bm{k}\:
w^{2}_{r}(\bm{k}-\bm{k}_{c})=1$
and $\sum_{\lambda}|z_{c\lambda}|^{2}=1$, respectively.
We assume $w_{r}(\bm{k}-\bm{k}_{c})$ has a sharp
peak around 
$\bm{k}_{c} =\int_{\mathrm{BZ}}d\bm{k}\:
w^{2}_{r}(\bm{k}-\bm{k}_{c})\bm{k}$.
It should be noted that, rigorously speaking,
we need to replace this single photon wavepacket
with a coherent (or squeezed) state wavepacket
when we apply the present formalism
to a light beam with macroscopic number of photons.
However, from the linearity of the Maxwell equations, 
the equations of motion for the single photon wavepacket 
is applicable also to the macroscopic coherent beam.

In a fermionic system, we can define the position operator
as the center of the probability density of a fermion,
which is positive-definite both in non-relativistic
and relativistic cases.
However, for a relativistic boson,
the definition of its position is nontrivial. 
In order to find an appropriate definition 
for the position of wavepacket,
we firstly consider the energy and the position weighted
by the energy density evaluated as follows,
\begin{eqnarray}
\langle W|H|W\rangle
&\cong& E_{n\bm{k}_{c}}
,
\label{eq:H-periodic}
\\
\langle W|\bm{\mathcal{R}}|W\rangle
&\cong& E_{n\bm{k}_{c}}\left[
\bm{\nabla}_{\bm{k}_{c}}\vartheta(\bm{k}_{c},t)
-(z_{c}|\bm{\Lambda}_{n\bm{k}_{c}}|z_{c})
\right].
\label{eq:R-periodic}
\end{eqnarray}
It should be noted that $\cong$ in Eqs.~(\ref{eq:H-periodic}) and (\ref{eq:R-periodic}) 
means that the above expectation
values are evaluated
under the assumption that the shape of $w^{2}_{r}(\bm{k}-\bm{k}_{c})$
is sufficiently sharp
compared to the variations of $E_{n\bm{k}}$ and 
$\bm{\Lambda}_{n\bm{k}}(\bm{k})$ around $\bm{k}_{c}$,
and we neglected terms which depend on the shape of $w^{2}_{r}(\bm{k}-\bm{k}_{c})$.

From Eqs.~(\ref{eq:H-periodic}) and (\ref{eq:R-periodic}), the center of gravity is estimated as
\begin{eqnarray}
\frac{\langle W|\bm{\mathcal{R}}|W\rangle}{\langle W|H|W\rangle}
&\cong& \bm{\nabla}_{\bm{k}_{c}}\vartheta(\bm{k}_{c},t)-(z_{c}|\bm{\Lambda}_{n\bm{k}_{c}}|z_{c}).
\end{eqnarray}
Comparing this result with the naive definition
for the position of wavepacket,
$\int_{\mathrm{BZ}}\frac{d\bm{k}}{(2\pi)^3}w^{2}_{r}(\bm{k}-\bm{k}_{c})
\bm{\nabla}_{\bm{k}}\vartheta(\bm{k},t)$,
we can reach the appropriate definition for the position of wavepacket,
\begin{eqnarray}
\bm{r}_{c} &=&
\int_{\mathrm{BZ}}d\bm{k}\:w^{2}_{r}(\bm{k}-\bm{k}_{c})
\left[\bm{\nabla}_{\bm{k}}\vartheta(\bm{k},t)-(z_{c}|\bm{\Lambda}_{n\bm{k}}|z_{c})\right].
\label{eq:rc-periodic}
\nonumber\\
\end{eqnarray}

In order to check the property of the wavepacket,
we consider the expectation values of physical observables.
As shown in Appendix~\ref{sec:expectation-values},
the energy current and the rotation of energy current are
evaluated as
\begin{subequations}
\begin{eqnarray}
\langle W|\bm{\mathcal{P}}|W\rangle
&\cong& E_{n\bm{k}_{c}}\bm{\nabla}_{\bm{k}_{c}}E_{n\bm{k}_{c}}
,\\
\langle W|\bm{\mathcal{J}}|W\rangle
&\cong&
\bm{r}_{c}\times E_{n\bm{k}_{c}}\bm{\nabla}_{\bm{k}_{c}}E_{n\bm{k}_{c}}
+(z_{c}|\bm{\mathcal{S}}_{n\bm{k}_{c}}|z_{c}),
\nonumber\\
\end{eqnarray}
\end{subequations}
where
\begin{widetext}
\begin{subequations}
\begin{eqnarray}
\bm{\mathcal{S}}_{n\bm{k}}
&=&\frac{1}{2}\left[\bm{\mathcal{S}}^{E}_{n\bm{k}}
+\bm{\mathcal{S}}^{H}_{n\bm{k}}\right]
, \label{eq:S}\\
\left[\bm{\mathcal{S}}^{E}_{n\bm{k}}\right]_{\lambda\lambda'}
&=&
-\frac{i}{2}\left[
\langle \bm{\nabla}_{\bm{k}}U^{E}_{n\lambda\bm{k}}|
\times(\tensor{\epsilon} E^{2}_{n\bm{k}}-\Xi^{E}_{\bm{k}})
|\bm{\nabla}_{\bm{k}}U^{E}_{n\lambda'\bm{k}}\rangle
+\langle U^{E}_{n\lambda\bm{k}}|
\bm{S}\times\tensor{\mu}^{-1}\bm{S}
|U^{E}_{n\lambda'\bm{k}}\rangle
\right]
, \label{eq:S-E}\\
\left[\bm{\mathcal{S}}^{H}_{n\bm{k}}\right]_{\lambda\lambda'}
&=&-\frac{i}{2}\left[
\langle \bm{\nabla}_{\bm{k}}U^{H}_{n\lambda\bm{k}}|
\times(\tensor{\mu} E^{2}_{n\bm{k}}-\Xi^{H}_{\bm{k}})
|\bm{\nabla}_{\bm{k}}U^{H}_{n\lambda'\bm{k}}\rangle
+\langle U^{H}_{n\lambda\bm{k}}|
\bm{S}\times\tensor{\epsilon}^{-1}\bm{S}
|U^{H}_{n\lambda'\bm{k}}\rangle
\right]. \label{eq:S-H}
\end{eqnarray}
\end{subequations}
\end{widetext}
It is noted that the first term of $\langle W|\bm{\mathcal{J}}|W\rangle$
is interpreted as the orbital rotational motion, i.e. the rotation of the
center of gravity,
and the second term as the internal one, i.e. the rotation around the center of gravity.
Especially for the locally isotropic system 
in which $\tensor{\epsilon}(\bm{r})$ and $\tensor{\mu}(\bm{r})$ are 
scalar variables,
$\epsilon(\bm{r})$ and $\mu(\bm{r})$ , the contribution from the second terms in
$\bm{\mathcal{S}}^{E}_{n\bm{k}}$ and $\bm{\mathcal{S}}^{H}_{n\bm{k}}$ are rewritten 
by using $\bm{S}\times\bm{S} = i\bm{S}$ as
\begin{eqnarray}
&&-\frac{i}{4}
\bigl[
\langle U^{E}_{n\lambda\bm{k}}|
\bm{S}\times\tensor{\mu}^{-1}\bm{S}
|U^{E}_{n\lambda'\bm{k}}\rangle
\nonumber\\
&&\qquad
+\langle U^{H}_{n\lambda\bm{k}}|
\bm{S}\times\tensor{\epsilon}^{-1}\bm{S}
|U^{H}_{n\lambda'\bm{k}}\rangle
\bigr]
\nonumber\\
&&\to\frac{1}{4}\left[
\langle U^{E}_{n\lambda\bm{k}}|
\mu^{-1}\bm{S}
|U^{E}_{n\lambda'\bm{k}}\rangle
+\langle U^{H}_{n\lambda\bm{k}}|
\epsilon^{-1}\bm{S}
|U^{H}_{n\lambda'\bm{k}}\rangle
\right].
\nonumber\\
\end{eqnarray}
This suggest that the internal rotation correctly includes
the spin of the constituent particle, i.e.,
the polarization of light in the present case.
However, it should be noted that 
the second terms of right-hand side of Eqs.~(\ref{eq:S-E}) and (\ref{eq:S-H})
are not the whole contributions of spin.
Actually, when we consider the circularly polarized light 
in isotropic homogeneous media,
all terms of the internal rotation give the same contribution
and totally represent the rotation originated by the polarization.
In addition, the internal rotation defined above
contains the internal orbital one and the spin one generally.

\subsection{\label{sec:gauge-transformation}Gauge transformation}
When a system has a symmetry 
represented by the unitary matrix $[M_{n\bm{k}}]_{\lambda\lambda'}$,
Maxwell equations are invariant under the transformation,
\begin{eqnarray}
|\tilde{U}^{F}_{n\lambda\bm{k}}\rangle
&=&
\sum_{\lambda'}[M_{n\bm{k}}]_{\lambda'\lambda}|U^{F}_{n\lambda'\bm{k}}\rangle,
\label{eq:GT-U}
\end{eqnarray}
where $F=E$ or $H$.
(Here we consider the case in which
there are degeneracies indexed by the subscript $\lambda$ or $\lambda'$.)
For the sake of convenience.
we call this transformation as 
the gauge transformation in $\bm{k}$-space.
By this gauge transformation, 
$\bm{\Lambda}_{n\bm{k}}$, $\bm{\Omega}_{n\bm{k}}$, 
$\bm{\Delta}_{n\bm{k}}$ and $\bm{\mathcal{S}}_{n\bm{k}}$
are transformed as
\begin{subequations}
\begin{eqnarray}
\tilde{\bm{\Lambda}}_{n\bm{k}}
&=&M^{-1}_{n\bm{k}}\bm{\Lambda}_{n\bm{k}}M_{n\bm{k}}
-iM^{-1}\bm{\nabla}_{\bm{k}}M_{n\bm{k}},
\label{eq:GT-Lambda}\\
\tilde{\bm{\Omega}}_{n\bm{k}}
&=&M^{-1}_{n\bm{k}}\bm{\Omega}_{n\bm{k}} M_{n\bm{k}},
\label{eq:GT-Omega}
\\
\tilde{\bm{\Delta}}_{n\bm{k}}
&=&M^{-1}_{n\bm{k}}\bm{\Delta}_{n\bm{k}}M_{n\bm{k}},
\label{eq:GT-Delta}
\\
\tilde{\bm{\mathcal{S}}}_{n\bm{k}}
&=&M^{-1}_{n\bm{k}} \bm{\mathcal{S}}_{n\bm{k}} M_{n\bm{k}}.
\label{eq:GT-S}
\end{eqnarray}
\end{subequations}
The gauge transformation of Bloch functions in Eq.~(\ref{eq:GT-U}) is equivalent to
that of the corresponding creation operators as
\begin{eqnarray}
\tilde{a}^{\dagger}_{n\lambda\bm{k}}
&=&\sum_{\lambda'}[M_{n\bm{k}}]_{\lambda'\lambda}
a^{\dagger}_{n\lambda'\bm{k}}
\label{eq:GT-a}.
\end{eqnarray}
In terms of this transformed operators,
the wavepacket in Eq.~(\ref{eq:W-periodic})
is represented by
\begin{eqnarray}
&&|W\rangle =
\int_{\mathrm{BZ}}d\bm{k}\:
w(\bm{k},\bm{k}_{c},t)
\sum_{\lambda,\lambda'}[M^{-1}_{n\bm{k}}]_{\lambda'\lambda}
z_{c\lambda}\tilde{a}^{\dagger}_{n\lambda'\bm{k}}|0\rangle.
\label{eq:GT-W}
\nonumber\\
\end{eqnarray}
It should be noted that we have changed only the representation
but not the physical state of wavepacket.
Therefore, the expectation values of physical observables, e.g., 
$H$, $\bm{\mathcal{R}}$, $\bm{\mathcal{P}}$ and $\bm{\mathcal{J}}$,
must be gauge invariant.
Indeed, we can easily show that
the evaluations of $H$ and $\bm{\mathcal{P}}$
in Appendix~\ref{sec:wavepacket}
are gauge invariant because of the invariance of $E_{n\bm{k}}$.
From Eqs.~(\ref{eq:GT-Omega})-(\ref{eq:GT-S})
and Eq.~(\ref{eq:GT-W}),
we can also show the invariance of
$(z_{c}|\bm{\Omega}_{n\bm{k}}|z_{c})$,
$(z_{c}|\bm{\Delta}_{n\bm{k}}|z_{c})$,
and $(z_{c}|\bm{\mathcal{S}}_{n\bm{k}}|z_{c})$.
However, it is not clear whether
the evaluations 
of $\bm{\mathcal{R}}$ and $\bm{\mathcal{J}}$
given in Appendix~\ref{sec:wavepacket}
are also the case.
In order to confirm this point,
it is enough to check whether
the position of wavepacket $\bm{r}_{c}$
in Eq.~(\ref{eq:rc-periodic}) is gauge invariant or not.
In the representation of Eq.~(\ref{eq:GT-W}), 
the derivative of phase factor 
$\bm{\nabla}_{\bm{k}}\vartheta(\bm{k},t)$
and $(z_{c}|\bm{\Lambda}_{n\bm{k}}|z_{c})$
in Appendix~\ref{sec:wavepacket}
are replaced as
\begin{subequations}
\begin{eqnarray}
&&\bm{\nabla}_{\bm{k}}\vartheta(\bm{k},t) = 
i(z_{c}|
e^{i\vartheta(\bm{k},t)}
\bm{\nabla}_{\bm{k}}e^{-i\vartheta(\bm{k},t)}
|z_{c})
\nonumber\\
&&\to i(z_{c}|
\left[e^{i\vartheta(\bm{k},t)}M_{n\bm{k}}\right]
\bm{\nabla}_{\bm{k}}\left[e^{-i\vartheta(\bm{k},t)}M^{-1}_{n\bm{k}}\right]
|z_{c}),
\nonumber\\
\\
&&(z_{c}|\bm{\Lambda}_{n\bm{k}}|z_{c})
\to (z_{c}|M_{n\bm{k}}\tilde{\bm{\Lambda}}_{n\bm{k}}M^{-1}_{n\bm{k}}|z_{c}).
\end{eqnarray}
\end{subequations}
The above formulae and Eq.~(\ref{eq:GT-Lambda})
prove the gauge invariance of $\bm{r}_{c}$ as follows,
\begin{eqnarray}
\tilde{\bm{r}}_{c}
&=& \int_{\mathrm{BZ}}d\bm{k}\:w^{2}_{r}(\bm{k}-\bm{k}_{c})
\nonumber\\
&&\times
\Bigl[
i(z_{c}|
\left[e^{i\vartheta(\bm{k},t)}M_{n\bm{k}}\right]
\bm{\nabla}_{\bm{k}}\left[e^{-i\vartheta(\bm{k},t)}M^{-1}_{n\bm{k}}\right]
|z_{c})
\nonumber\\
&&\qquad
-(z_{c}|M_{n\bm{k}}\tilde{\bm{\Lambda}}_{n\bm{k}}M^{-1}_{n\bm{k}}|z_{c})
\Bigr]
\nonumber\\
&=& \bm{r}_{c}+
\int_{\mathrm{BZ}}d\bm{k}\:w^{2}_{r}(\bm{k}-\bm{k}_{c})
\nonumber\\
&&\times
i(z_{c}|
\left[M_{n\bm{k}}(\bm{\nabla}_{\bm{k}}M^{-1}_{n\bm{k}})
+(\bm{\nabla}_{\bm{k}}M_{n\bm{k}})M^{-1}_{n\bm{k}}\right]
|z_{c})
\Bigr]
\nonumber\\
&=&\bm{r}_{c}.
\end{eqnarray}
Combining this result and the gauge invariance of $E_{n\bm{k}}$
and $(z_{c}|\bm{\mathcal{S}}_{n\bm{k}}|z_{c})$,
we can confirm that
the evaluations 
of $\bm{\mathcal{R}}$ and $\bm{\mathcal{J}}$
in Appendix~\ref{sec:wavepacket}
are also gauge invariant.

\subsection{\label{sec:expectation-values}Expectation values}
Here we presents the detailed evaluations of the expectation values, 
i.e., the Hamiltonian $H$,
the position weighted by the energy density $\bm{\mathcal{R}}$,
the energy current $\bm{\mathcal{P}}$ and 
the rotation of energy current $\bm{\mathcal{J}}$, 
with respect to
the optical wavepacket $|W\rangle$ in a periodic system.
The expectation value of an operator $\mathcal{O}$ is obtained from commutation relations 
between $\mathcal{O}$ and the creation/annihilation operators,
\begin{eqnarray}
\langle W|\mathcal{O}|W\rangle
&=&
\int_{\mathrm{BZ}}d\bm{k}\:\tilde{d\bm{k}'}
\:w^{*}(\bm{k},\bm{k}_{c},t)
w(\bm{k}',\bm{k}_{c},t)
\nonumber\\
&&\times
\langle 0|a_{nz_{c}\bm{k}}
\mathcal{O} a^{\dagger}_{nz_{c}\bm{k}'}|0\rangle
\nonumber\\
&=&
\int_{\mathrm{BZ}}d\bm{k}\:\tilde{d\bm{k}'}
\:w^{*}(\bm{k},\bm{k}_{c},t)
w(\bm{k}',\bm{k}_{c},t)
\nonumber\\
&&\times
\langle 0|
\left[a_{nz_{c}\bm{k}},
\left[\mathcal{O},a^{\dagger}_{nz_{c}\bm{k}'}\right]\right]
|0\rangle,
\end{eqnarray}
where we have introduced the abbreviation,
\begin{eqnarray}
a^{(\dagger)}_{nz_{c}\bm{k}} &=& \sum_{\lambda}z_{c\lambda}a^{(\dagger)}_{n\lambda\bm{k}},
\end{eqnarray} 
and this will be used also for the Bloch functions as
\begin{eqnarray}
|U^{E,H}_{nz_{c}\bm{k}}\rangle &=& \sum_{\lambda}z_{c\lambda}|U^{E,H}_{n\lambda\bm{k}}\rangle.
\end{eqnarray} 
The basic commutation relation between 
$\bm{B}(\bm{r})$ and $\bm{D}(\bm{r})$
can be represented in the following integral form,
\begin{eqnarray}
&&\int d\bm{r}\:d\bm{r}'
\left[\bm{\Phi}^{*}_{1}(\bm{r})\cdot\bm{B}(\bm{r})
,\bm{D}(\bm{r}')\cdot\bm{\Phi}_{2}(\bm{r}')\right]
\nonumber\\
&&=
-i\int d\bm{r}\left[\bm{\nabla}_{\bm{r}}\times\bm{\Phi}_{1}^{*}(\bm{r})\right]
\cdot\bm{\Phi}_{2}(\bm{r})
\nonumber\\
&&=
-i\int d\bm{r}\:
\bm{\Phi}^{*}_{1}(\bm{r})
\cdot\left[\bm{\nabla}_{\bm{r}}\times\bm{\Phi}_{2}(\bm{r})\right].
\label{eq:BD-integral}
\end{eqnarray}
In particular, in a periodic system, 
by the above commutation relation and
Eqs.~(\ref{eq:Maxwell-PhiEH}) and (\ref{eq:Maxwell-PhiHE}),
we can easily show that
\begin{eqnarray}
&&
\left[
a_{n\lambda\bm{k}},
\left[H,a^{\dagger}_{n'\lambda'\bm{k}'}\right]
\right]
\nonumber\\
&&=
E_{n\bm{k}}E_{n'\bm{k}'}\int d\bm{r}
\Bigl[
\bm{\Phi}^{E*}_{n\lambda\bm{k}}(\bm{r})
\tensor{\epsilon}(\bm{r})
\bm{\Phi}^{E}_{n'\lambda'\bm{k}'}(\bm{r})
\nonumber\\
&&\hspace{2.5cm}
+
\bm{\Phi}^{H*}_{n\lambda\bm{k}}(\bm{r})
\tensor{\mu}(\bm{r})
\bm{\Phi}^{E}_{n'\lambda'\bm{k}'}(\bm{r})
\Bigr]
\nonumber\\
&&=E_{n\bm{k}}\delta_{nn'}\delta_{\lambda\lambda'}
\tilde{\delta}(\bm{k}-\bm{k}').
\end{eqnarray}
In the transformation to the last line, we have used
\begin{eqnarray}
&&\int d\bm{r}\:e^{-i(\bm{k}-\bm{k}')\cdot\bm{r}}
F_{\bm{k}\bm{k}'}(\bm{r})
\nonumber\\
&&=
\sum_{\bm{a}}\int_{\mathrm{WS}}d\bm{r}
\:e^{-i(\bm{k}-\bm{k}')\cdot(\bm{a}+\bm{r})}
F_{\bm{k}\bm{k}'}(\bm{a}+\bm{r})
\nonumber\\
&&=
\int_{\mathrm{WS}}\frac{d\bm{r}}{v_{\mathrm{WS}}}
\sum_{\bm{G}}\tilde{\delta}(\bm{k}-\bm{k}'+\bm{G})
e^{-i(\bm{k}-\bm{k}')\cdot\bm{r}}
F_{\bm{k}\bm{k}'}(\bm{r})
,
\nonumber\\
\label{eq:delta}
\end{eqnarray}
where $\bm{a}$ represents an arbitrary lattice vector,
and $F_{\bm{k}\bm{k}'}(\bm{r})$ is a periodic function, i.e.,
$F_{\bm{k}\bm{k}'}(\bm{a}+\bm{r})=F_{\bm{k}\bm{k}'}(\bm{r})$
Since $\bm{k}$ and $\bm{k}'$ are in the first Brillouin zone,
we have also used the following relation implicitly,
\begin{eqnarray}
\sum_{\bm{G}}\tilde{\delta}(\bm{k}-\bm{k}'+\bm{G})e^{-i(\bm{k}-\bm{k}')\cdot\bm{r}}
\Bigr|_{\bm{k},\bm{k}'\in\text{1st BZ}}
= \tilde{\delta}(\bm{k}-\bm{k}').
\nonumber\\
\label{eq:delta-BZ}
\end{eqnarray}
Then we obtain the result
\begin{eqnarray}
\langle W|H|W\rangle
&=&
\int_{\mathrm{BZ}}d\bm{k}
\:w^{2}_{r}(\bm{k}-\bm{k}_{c})E_{n\bm{k}}
\cong E_{n\bm{k}_{c}}.
\nonumber\\
\end{eqnarray}
In the similar manner, 
we obtain the following commutation relation 
which is needed to estimate the expectation value of $\bm{\mathcal{R}}$,
\begin{eqnarray}
&&\left[a_{n\lambda\bm{k}},
\left[
\bm{\mathcal{R}},
a^{\dagger}_{n'\lambda'\bm{k}'}\right]
\right]
\nonumber\\
&&=
E_{n\bm{k}}E_{n'\bm{k}'}
\int d\bm{r}\:\bm{r}
\Bigl[\bm{\Phi}^{E*}_{n\lambda\bm{k}}(\bm{r})
\tensor{\epsilon}(\bm{r})\bm{\Phi}^{E}_{n'\lambda'\bm{k}'}(\bm{r})
\nonumber\\
&&\qquad\qquad\qquad
+\bm{\Phi}^{H*}_{n\lambda\bm{k}}(\bm{r})
\tensor{\mu}(\bm{r})
\bm{\Phi}^{H}_{n'\lambda'\bm{k}'}(\bm{r})
\Bigr]
\nonumber\\
&&=
\frac{i}{4}
\sqrt{E_{n\bm{k}}E_{n'\bm{k}'}}
\left[
(\bm{\nabla}_{\bm{k}}-\bm{\nabla}_{\bm{k}'})
\tilde{\delta}(\bm{k}-\bm{k}')
\right]
\nonumber\\
&&
\quad\times
\left[
\langle U^{E}_{n\lambda\bm{k}}
|\tensor{\epsilon}|
U^{E}_{n'\lambda'\bm{k}'}
\rangle
+
\langle U^{H}_{n\lambda\bm{k}}
|\tensor{\mu}|
U^{H}_{n'\lambda'\bm{k}'}
\rangle
\right].
\end{eqnarray}
In the transformation to the last line,
we have used the relation,
\begin{eqnarray}
&&\int d\bm{r}\:\bm{r}e^{-i(\bm{k}-\bm{k}')\cdot\bm{r}}
F_{\bm{k}\bm{k}'}(\bm{r})
\nonumber\\
&&=
\frac{i}{2}\int d\bm{r}
\left[(\bm{\nabla}_{\bm{k}}-\bm{\nabla}_{\bm{k}'})e^{-i(\bm{k}-\bm{k}')\cdot\bm{r}}\right]
F_{\bm{k}\bm{k}'}(\bm{r})
\nonumber\\
&&=
\frac{i}{2}\sum_{\bm{a}}\int_{\mathrm{WS}}d\bm{r}
\left[(\bm{\nabla}_{\bm{k}}-\bm{\nabla}_{\bm{k}'})e^{-i(\bm{k}-\bm{k}')\cdot(\bm{a}+\bm{r})}\right]
F_{\bm{k}\bm{k}'}(\bm{r})
\nonumber\\
&&=
\frac{i}{2}\int_{\mathrm{WS}}\frac{d\bm{r}}{v_{\mathrm{WS}}}
\:F_{\bm{k}\bm{k}'}(\bm{r})
\nonumber\\
&&\quad\times
\left[(\bm{\nabla}_{\bm{k}}-\bm{\nabla}_{\bm{k}'})
\sum_{\bm{G}}\tilde{\delta}(\bm{k}-\bm{k}'+\bm{G})
e^{-i(\bm{k}-\bm{k}')\cdot\bm{r}}\right],
\nonumber\\
\label{eq:d-delta}
\end{eqnarray}
and Eq.~(\ref{eq:delta-BZ}).
This commutation relation leads to the result,
\begin{eqnarray}
&&\langle W|\bm{\mathcal{R}}|W\rangle
\nonumber\\
&&=
\frac{i}{2}
\int_{\mathrm{BZ}}d\bm{k}\:
E_{n\bm{k}}
\Bigl[
w^{*}(\bm{k},\bm{k}_{c},\bm{r}_{c},t)\bm{\nabla}_{\bm{k}}
w(\bm{k},\bm{k}_{c},\bm{r}_{c},t)
\nonumber\\
&&\hspace{2.5cm}
-[\bm{\nabla}_{\bm{k}}w^{*}(\bm{k},\bm{k}_{c},\bm{r}_{c},t)]
w(\bm{k},\bm{k}_{c},\bm{r}_{c},t)
\Bigr]
\nonumber\\
&&\quad+\frac{i}{4}
\int_{\mathrm{BZ}}d\bm{k}\:
w_{r}^{2}(\bm{k}-\bm{k}_{c})E_{n\bm{k}}
\nonumber\\
&&
\qquad\times
\Bigl[
\langle U^{E}_{nz_{c}\bm{k}}
|\tensor{\epsilon}|\bm{\nabla}_{\bm{k}}U^{E}_{nz_{c}\bm{k}}\rangle
-\langle \bm{\nabla}_{\bm{k}}U^{E}_{nz_{c}\bm{k}}
|\tensor{\epsilon}|U^{E}_{nz_{c}\bm{k}}\rangle
\nonumber\\
&&\qquad\quad
+\langle U^{H}_{nz_{c}\bm{k}}
|\tensor{\mu}|\bm{\nabla}_{\bm{k}}U^{H}_{nz_{c}\bm{k}}\rangle
-\langle \bm{\nabla}_{\bm{k}}U^{H}_{nz_{c}\bm{k}}
|\tensor{\mu}|U^{H}_{nz_{c}\bm{k}}\rangle
\Bigr]
\nonumber\\
&&=
\int_{\mathrm{BZ}}d\bm{k}
\:w^{2}_{r}(\bm{k}-\bm{k}_{c})E_{n\bm{k}}
\left[\bm{\nabla}_{\bm{k}}\vartheta(\bm{k},t)
-(z_{c}|\bm{\Lambda}_{\bm{k}}|z_{c})\right]
\nonumber\\
&&\cong E_{n\bm{k}_{c}}\left[\bm{\nabla}_{\bm{k}_{c}}\vartheta(\bm{k}_{c},t)
-(z_{c}|\bm{\Lambda}_{\bm{k}_{c}}|z_{c})\right].
\end{eqnarray}

The expectation value of $\bm{\mathcal{P}}$
is estimated by using the commutation relation,
\begin{eqnarray}
&&
\left[
a_{n\lambda\bm{k}},
\left[\bm{\mathcal{P}},a^{\dagger}_{n'\lambda'\bm{k}'}\right]
\right]
\nonumber\\
&&=
E_{n\bm{k}}E_{n'\bm{k}'}\int d\bm{r}
\Bigl[
\bm{\Phi}^{E*}_{n\lambda\bm{k}}(\bm{r})
\times
\bm{\Phi}^{H}_{n'\lambda'\bm{k}'}(\bm{r})
\nonumber\\
&&\hspace{2.5cm}
-
\bm{\Phi}^{H*}_{n\lambda\bm{k}}(\bm{r})
\times
\bm{\Phi}^{E}_{n'\lambda'\bm{k}'}(\bm{r})
\Bigr]
\nonumber\\
&&=
\frac{1}{2}
\sqrt{E_{n\bm{k}}E_{n'\bm{k}}}
\:\tilde{\delta}(\bm{k}-\bm{k}')
\nonumber\\
&&\quad\times
\left[
\langle
U^{E}_{n\lambda\bm{k}}
|i\bm{S}|
U^{H}_{n'\lambda'\bm{k}}\rangle
-\langle
U^{H}_{n\lambda\bm{k}}
|i\bm{S}|
U^{E}_{n'\lambda'\bm{k}}
\rangle
\right],
\nonumber\\
\end{eqnarray}
where Eqs.~(\ref{eq:delta}) and (\ref{eq:delta-BZ}) were used.
Combining this commutation relation
and the formula,
\begin{eqnarray}
&&E_{n\bm{k}}\left[
\langle U^{E}_{nz_{c}\bm{k}}|i\bm{S}|U^{H}_{nz_{c}\bm{k}}\rangle
-\langle U^{H}_{nz_{c}\bm{k}}|i\bm{S}|U^{E}_{nz_{c}\bm{k}}\rangle
\right]
\nonumber\\
&&=
\langle U^{E}_{nz_{c}\bm{k}}|
\left[\bm{S}\tensor{\mu}^{-1}\bm{P}_{\bm{k}}\cdot\bm{S}
+\bm{P}_{\bm{k}}\cdot\bm{S}\tensor{\mu}^{-1}\bm{S}\right]
|U^{E}_{nz_{c}\bm{k}}\rangle
\nonumber\\
&&=
\langle U^{E}_{nz_{c}\bm{k}}|
\left[\bm{\nabla}_{\bm{k}}\Xi^{E}_{\bm{k}}\right]
|U^{E}_{nz_{c}\bm{k}}\rangle
\nonumber\\
&&=\bm{\nabla}_{\bm{k}}E^{2}_{n\bm{k}},
\end{eqnarray}
we obtain the result,
\begin{eqnarray}
&&\langle W|\bm{\mathcal{P}}|W\rangle
\nonumber\\
&&=\frac{1}{2}
\int_{\mathrm{BZ}}d\bm{k}
\:w^{2}_{r}(\bm{k}-\bm{k}_{c})
\nonumber\\
&&\quad\times
E_{n\bm{k}}\left[
\langle U^{E}_{nz_{c}\bm{k}}
|i\bm{S}|U^{H}_{nz_{c}\bm{k}}\rangle
-\langle U^{H}_{nz_{c}\bm{k}}
|i\bm{S}|U^{E}_{nz_{c}\bm{k}}\rangle
\right]
\nonumber\\
&&=\frac{1}{2}
\int_{\mathrm{BZ}}d\bm{k}
\:w^{2}_{r}(\bm{k}-\bm{k}_{c})
\bm{\nabla}_{\bm{k}}E^{2}_{n\bm{k}}
\nonumber\\
&&\cong E_{n\bm{k}_{c}}\bm{\nabla}_{\bm{k}}E_{n\bm{k}_{c}}.
\end{eqnarray}

The expectation value of $\bm{\mathcal{J}}$
is derived from the commutation relation,
\begin{eqnarray}
&&\left[
a_{n\lambda\bm{k}},
\left[
\bm{\mathcal{J}},
a^{\dagger}_{n'\lambda'\bm{k}'}\right]
\right]
\nonumber\\
&&=
E_{n\bm{k}}E_{n'\bm{k}'}\int d\bm{r}\:\bm{r}\times
\Bigl[
\bm{\Phi}^{E*}_{n\lambda\bm{k}}(\bm{r})
\times
\bm{\Phi}^{H}_{n'\lambda'\bm{k}'}(\bm{r})
\nonumber\\
&&\hspace{3cm}
-\bm{\Phi}^{H*}_{n\lambda\bm{k}}(\bm{r})
\times
\bm{\Phi}^{E}_{n'\lambda'\bm{k}'}(\bm{r})
\Bigr]
\nonumber\\
&&=
\frac{i}{4}
\sqrt{E_{n\bm{k}}E_{n'\bm{k}'}}
\left[
(\bm{\nabla}_{\bm{k}}-\bm{\nabla}_{\bm{k}'})
\tilde{\delta}(\bm{k}-\bm{k}')\right]
\nonumber\\
&&\quad
\times\Bigl[
\langle U^{E}_{n\lambda\bm{k}}|i\bm{S}|U^{H}_{n'\lambda'\bm{k}'}\rangle
-\langle U^{H}_{n\lambda\bm{k}}|i\bm{S}|U^{E}_{n'\lambda'\bm{k}'}\rangle
\Bigr],
\nonumber\\
\end{eqnarray}
where we have used Eq.~(\ref{eq:d-delta}) 
in the transformation to the last line.
It should be noted that,
as in the previous commutation relations, 
the above commutation relation is also restricted to the case
in which both of $\bm{k}$ and $\bm{k}'$ are 
in the first Brillouin zone.
In addition, our wavepacket is constructed from 
degenerate eigen modes, i.e, eigen modes with the same band index $n$.
Thus the following formula, which will be proved later,
is useful to estimate the expectation value of $\bm{\mathcal{J}}$
with respect to the wavepacket.
\begin{eqnarray}
&&\frac{E_{n\bm{k}}}{4}
\Bigl[
\langle U^{E}_{nz_{c}\bm{k}}|\bm{S}\times|\bm{\nabla}_{\bm{k}}U^{H}_{nz_{c}\bm{k}}\rangle
+\langle\bm{\nabla}_{\bm{k}}U^{E}_{nz_{c}\bm{k}}|\times\bm{S}|U^{H}_{nz_{c}\bm{k}}\rangle
\nonumber\\
&&\quad
-\langle U^{H}_{nz_{c}\bm{k}}|\bm{S}\times|\bm{\nabla}_{\bm{k}}U^{E}_{nz_{c}\bm{k}}\rangle
-\langle\bm{\nabla}_{\bm{k}}U^{H}_{nz_{c}\bm{k}}|\times\bm{S}|U^{E}_{nz_{c}\bm{k}}\rangle
\Bigr]
\nonumber\\
&&=-(z_{c}|\bm{\Lambda}_{n\bm{k}}|z_{c})\times\left(E_{n\bm{k}}\bm{\nabla}_{\bm{k}}E_{n\bm{k}}\right)
+(z_{c}|\bm{\mathcal{S}}_{n\bm{k}}|z_{c}).
\label{eq:formula-for-J}
\end{eqnarray}
Combining the above relation and the commutation relation,
we obtain the result,
\begin{widetext}
\begin{eqnarray}
\langle W|\bm{\mathcal{J}}|W\rangle
&=&
\int_{\mathrm{BZ}}d\bm{k}
\:w^{2}_{r}(\bm{k}-\bm{k}_{c})
\biggl[
\left[\bm{\nabla}_{\bm{k}}\vartheta(\bm{k},t)\right]
\times\left(E_{n\bm{k}}\bm{\nabla}_{\bm{k}}E_{n\bm{k}}\right)
+\frac{E_{n\bm{k}}}{4}\Bigl[
\langle U^{E}_{nz_{c}\bm{k}}|\bm{S}\times|\bm{\nabla}_{\bm{k}}U^{H}_{nz_{c}\bm{k}}\rangle
\nonumber\\
&&
+\langle\bm{\nabla}_{\bm{k}}U^{E}_{nz_{c}\bm{k}}|\times\bm{S}|U^{H}_{nz_{c}\bm{k}}\rangle
-\langle U^{H}_{nz_{c}\bm{k}}|\bm{S}\times|\bm{\nabla}_{\bm{k}}U^{E}_{nz_{c}\bm{k}}\rangle
-\langle\bm{\nabla}_{\bm{k}}U^{H}_{nz_{c}\bm{k}}|\times\bm{S}|U^{E}_{nz_{c}\bm{k}}\rangle
\Bigr]\biggr]
\nonumber\\
&&=
\int_{\mathrm{BZ}}d\bm{k}
\:w^{2}_{r}(\bm{k}-\bm{k}_{c})
\Bigl[
\left[\bm{\nabla}_{\bm{k}}\vartheta(\bm{k},t)
-(z_{c}|\bm{\Lambda}_{n\bm{k}}|z_{c})\right]
\times\left(E_{n\bm{k}}\bm{\nabla}_{\bm{k}}E_{n\bm{k}}\right)
+(z_{c}|\bm{\mathcal{S}}_{n\bm{k}}|z_{c})
\Bigr]
\nonumber\\
&&\cong\bm{r}_{c}\times\left(E_{n\bm{k}}\bm{\nabla}_{\bm{k}}E_{n\bm{k}}\right)
+(z_{c}|\bm{\mathcal{S}}_{n\bm{k}_{c}}|z_{c}).
\end{eqnarray}
\end{widetext}

The proof of Eq.~(\ref{eq:formula-for-J}) needs basic but tedious calculations.
Here we comment that the formula is confirmed 
by using Eqs.~(\ref{eq:Maxwell-UEH})-(\ref{eq:eigen-UH}),
and the $\bm{k}$-derivatives of 
Eqs.~(\ref{eq:eigen-UE}) and (\ref{eq:eigen-UH}).
The outline of the derivation is given as follows,
\begin{widetext}
\begin{eqnarray}
&&(z_{c}|\bm{\Lambda}_{n\bm{k}}|z_{c})
\times\left(E_{n\bm{k}}\bm{\nabla}_{\bm{k}}E_{n\bm{k}}\right)
\nonumber\\
&&\quad
+\frac{E_{n\bm{k}}}{4}
\Bigl[
\langle U^{E}_{nz_{c}\bm{k}}|\bm{S}\times|\bm{\nabla}_{\bm{k}}U^{H}_{nz_{c}\bm{k}}\rangle
+\langle\bm{\nabla}_{\bm{k}}U^{E}_{nz_{c}\bm{k}}|\times\bm{S}|U^{H}_{nz_{c}\bm{k}}\rangle
-\langle U^{H}_{nz_{c}\bm{k}}|\bm{S}\times|\bm{\nabla}_{\bm{k}}U^{E}_{nz_{c}\bm{k}}\rangle
-\langle\bm{\nabla}_{\bm{k}}U^{H}_{nz_{c}\bm{k}}|\times\bm{S}|U^{E}_{nz_{c}\bm{k}}\rangle
\Bigr]
\nonumber\\
&&=
\frac{i}{8}
\Bigl[
\langle\bm{\nabla}_{\bm{k}}U^{E}_{nz_{c}\bm{k}}|\times
(\bm{\nabla}_{\bm{k}}\Xi^{E}_{\bm{k}})|U^{E}_{nz_{c}\bm{k}}\rangle
+
\langle\bm{\nabla}_{\bm{k}}U^{H}_{nz_{c}\bm{k}}|\times
(\bm{\nabla}_{\bm{k}}\Xi^{H}_{\bm{k}})|U^{H}_{nz_{c}\bm{k}}\rangle
\nonumber\\
&&\quad
+
\langle U^{E}_{nz_{c}\bm{k}}|(\bm{\nabla}_{\bm{k}}\Xi^{E}_{\bm{k}})
\times|\bm{\nabla}_{\bm{k}}U^{E}_{nz_{c}\bm{k}}\rangle
+
\langle U^{H}_{nz_{c}\bm{k}}|(\bm{\nabla}_{\bm{k}}\Xi^{H}_{\bm{k}})
\times|\bm{\nabla}_{\bm{k}}U^{H}_{nz_{c}\bm{k}}\rangle
\Bigr]
\nonumber\\
&&\quad
-\frac{i}{4}\Bigl[
\langle\bm{\nabla}_{\bm{k}}U^{E}_{nz_{c}\bm{k}}|\times
(\tensor{\epsilon} E^{2}_{n\bm{k}}-\Xi^{E}_{\bm{k}})
|\bm{\nabla}_{\bm{k}}U^{E}_{nz_{c}\bm{k}}\rangle
+\langle\bm{\nabla}_{\bm{k}}U^{H}_{nz_{c}\bm{k}}|\times
(\tensor{\mu} E^{2}_{n\bm{k}}-\Xi^{H}_{\bm{k}})
|\bm{\nabla}_{\bm{k}}U^{H}_{nz_{c}\bm{k}}\rangle
\Bigr]
\nonumber\\
&&\quad
-\frac{i}{4}\Bigl[
\langle U^{H}_{nz_{c}\bm{k}}|\bm{P}_{\bm{k}}\cdot\bm{S}\tensor{\epsilon}^{-1}\bm{S}
\times|\bm{\nabla}_{\bm{k}}U^{H}_{nz_{c}\bm{k}}\rangle
+\langle\bm{\nabla}_{\bm{k}}U^{E}_{nz_{c}\bm{k}}|
\times\bm{P}_{\bm{k}}\cdot\bm{S}\tensor{\mu}^{-1}\bm{S}|U^{E}_{nz_{c}\bm{k}}\rangle
\nonumber\\
&&\qquad
+\langle U^{E}_{nz_{c}\bm{k}}|\bm{P}_{\bm{k}}\cdot\bm{S}\tensor{\mu}^{-1}\bm{S}
\times|\bm{\nabla}_{\bm{k}}U^{E}_{nz_{c}\bm{k}}\rangle
+\langle\bm{\nabla}_{\bm{k}}U^{H}_{nz_{c}\bm{k}}|
\times\bm{P}_{\bm{k}}\cdot\bm{S}\tensor{\epsilon}^{-1}\bm{S}|U^{H}_{nz_{c}\bm{k}}\rangle
\Bigr]
\nonumber\\
&&=
-\frac{i}{4}\Bigl[
\langle\bm{\nabla}_{\bm{k}}U^{E}_{nz_{c}\bm{k}}|\times
(\tensor{\epsilon} E^{2}_{n\bm{k}}-\Xi^{E}_{\bm{k}})
|\bm{\nabla}_{\bm{k}}U^{E}_{nz_{c}\bm{k}}\rangle
+\langle\bm{\nabla}_{\bm{k}}U^{H}_{nz_{c}\bm{k}}|\times
(\tensor{\mu} E^{2}_{n\bm{k}}-\Xi^{H}_{\bm{k}})
|\bm{\nabla}_{\bm{k}}U^{H}_{nz_{c}\bm{k}}\rangle
\Bigr]
\nonumber\\
&&\quad+
\frac{i}{8}\Bigl[
\langle\bm{\nabla}_{\bm{k}}U^{E}_{nz_{c}\bm{k}}|
\times
(\bm{P}_{\bm{k}}\cdot\bm{S}\tensor{\mu}^{-1}\bm{S}
-\bm{S}\tensor{\mu}^{-1}\bm{P}_{\bm{k}}\cdot\bm{S})|U^{E}_{nz_{c}\bm{k}}\rangle
-\langle U^{E}_{nz_{c}\bm{k}}|
(\bm{P}_{\bm{k}}\cdot\bm{S}\tensor{\mu}^{-1}\bm{S}
-\bm{S}\tensor{\mu}^{-1}\bm{P}_{\bm{k}}\cdot\bm{S})
\times|\bm{\nabla}_{\bm{k}}U^{E}_{nz_{c}\bm{k}}\rangle
\nonumber\\
&&\qquad
+\langle\bm{\nabla}_{\bm{k}}U^{H}_{nz_{c}\bm{k}}|
\times
(\bm{P}_{\bm{k}}\cdot\bm{S}\tensor{\mu}^{-1}\bm{S}
-\bm{S}\tensor{\mu}^{-1}\bm{P}_{\bm{k}}\cdot\bm{S})|U^{H}_{nz_{c}\bm{k}}\rangle
-\langle U^{H}_{nz_{c}\bm{k}}|
(\bm{P}_{\bm{k}}\cdot\bm{S}\tensor{\mu}^{-1}\bm{S}
-\bm{S}\tensor{\mu}^{-1}\bm{P}_{\bm{k}}\cdot\bm{S})
\times|\bm{\nabla}_{\bm{k}}U^{H}_{nz_{c}\bm{k}}\rangle
\Bigr]
\nonumber\\
&&=
(z_{c}|\bm{\mathcal{S}}_{n\bm{k}}|z_{c})
+\frac{i}{8}\bm{\nabla}_{\bm{k}}\times
\Bigl[
\langle U^{E}_{nz_{c}\bm{k}}|
(\bm{P}_{\bm{k}}\cdot\bm{S}\tensor{\mu}^{-1}\bm{S}
-\bm{S}\tensor{\mu}^{-1}\bm{P}_{\bm{k}}\cdot\bm{S})|U^{E}_{nz_{c}\bm{k}}\rangle
+\langle U^{H}_{nz_{c}\bm{k}}|
(\bm{P}_{\bm{k}}\cdot\bm{S}\tensor{\epsilon}^{-1}\bm{S}
-\bm{S}\tensor{\epsilon}^{-1}\bm{P}_{\bm{k}}\cdot\bm{S})|U^{H}_{nz_{c}\bm{k}}\rangle
\Bigr]
\nonumber\\
&&=
(z_{c}|\bm{\mathcal{S}}_{n\bm{k}}|z_{c}).
\end{eqnarray}
\end{widetext}

\section{\label{sec:modulated-system}Expectation values in a modulated system}
When a modulation is introduced into a periodic system, 
the argument given in Appendix \ref{sec:periodic-system} are modified.
Here we consider the modulation represented by Eq.~(\ref{eq:modulation})
which is sufficiently weak and smooth.
It is noted that
the commutation relation Eq.~(\ref{eq:BD-integral}) are not modified
even under any modulation.
From this commutation relation, 
the creation and annihilation operators of approximated eigen modes
satisfy the same commutation relation as that in a periodic system
as shown below.
However, the approximated eigen modes depend on the variable $\bm{r}_{c}$.
Thus we must additionally take into account the operator
$\bm{\nabla}_{\bm{r}_{c}}a^{\dagger}_{n\lambda\bm{k};\bm{r}_{c}}$
for the derivation of the effective Lagrangian.
Fortunately, we can show that the contribution from 
$\bm{\nabla}_{\bm{r}_{c}}a^{\dagger}_{n\lambda\bm{k};\bm{r}_{c}}$
vanishes by the following relation.
\begin{widetext}
\begin{subequations}
\begin{eqnarray}
\left[a_{n\lambda\bm{k};\bm{r}_{c}},
a^{\dagger}_{n'\lambda'\bm{k}';\bm{r}_{c}}\right]
&=&
i\int d\bm{r}
\Bigl[
\bm{\Phi}^{E*}_{n\lambda\bm{k}}(\bm{r})
\cdot
\left[
\bm{\nabla}_{\bm{r}}\times\bm{\Phi}^{H}_{n'\lambda'\bm{k}'}(\bm{r})
\right]
-\left[
\bm{\nabla}_{\bm{r}}\times\bm{\Phi}^{H*}_{n\lambda\bm{k}}(\bm{r})
\right]
\cdot\bm{\Phi}^{E}_{n'\lambda'\bm{k}'}(\bm{r})
\Bigr]
\nonumber\\
&=&
(E_{n\bm{k}}+E_{n'\bm{k}'})
\int d\bm{r}\:
\bm{\Phi}^{E*}_{n\lambda\bm{k}}(\bm{r})
\tensor{\epsilon}(\bm{r})
\bm{\Phi}^{E}_{n'\lambda'\bm{k}'}(\bm{r})
\nonumber\\
&=&
\frac{E_{n\bm{k}}+E_{n'\bm{k}}}
{\sqrt{E_{n\bm{k}}E_{n'\bm{k}}}}
\:
\tilde{\delta}(\bm{k}-\bm{k}')
\langle
U^{E}_{n\lambda\bm{k}}
|\tensor{\epsilon}|
U^{E}_{n'\lambda'\bm{k}}
\rangle
\nonumber\\
&=&\delta_{nn'}\delta_{\lambda\lambda'}\tilde{\delta}(\bm{k}-\bm{k}'),
\\
\left[a_{n\lambda\bm{k};\bm{r}_{c}},
\bm{\nabla}_{\bm{r}_{c}}a^{\dagger}_{n'\lambda'\bm{k}';\bm{r}_{c}}\right]
&=&-\frac{i}{2}\left[
\bm{\nabla}_{\bm{r}_{c}}
\ln\frac{\gamma_{\epsilon}(\bm{r}_{c})}{\gamma_{\mu}(\bm{r}_{c})}
\right]
\int d\bm{r}
\Bigl[
\bm{\Phi}^{E*}_{n\lambda\bm{k}}(\bm{r})
\cdot
\left[
\bm{\nabla}_{\bm{r}}\times\bm{\Phi}^{H}_{n'\lambda'\bm{k}'}(\bm{r})
\right]
+\left[
\bm{\nabla}_{\bm{r}}\times\bm{\Phi}^{H*}_{n\lambda\bm{k}}(\bm{r})
\right]
\cdot\bm{\Phi}^{E}_{n'\lambda'\bm{k}'}(\bm{r})
\Bigr]
\nonumber\\
&=&
\frac{1}{2}\left[\bm{\nabla}_{\bm{r}_{c}}
\ln\frac{\gamma_{\epsilon}(\bm{r}_{c})}{\gamma_{\mu}(\bm{r}_{c})}
\right]
(E_{n\bm{k}}-E_{n'\bm{k}'})
\int d\bm{r}\:
\bm{\Phi}^{E*}_{n\lambda\bm{k}}(\bm{r})
\tensor{\epsilon}(\bm{r})
\bm{\Phi}^{E}_{n'\lambda'\bm{k}'}(\bm{r})
\nonumber\\
&=&
\frac{1}{4}\left[\bm{\nabla}_{\bm{r}_{c}}
\ln\frac{\gamma_{\epsilon}(\bm{r}_{c})}{\gamma_{\mu}(\bm{r}_{c})}
\right]
\frac{E_{n\bm{k}}-E_{n'\bm{k}}}
{\sqrt{E_{n\bm{k}}E_{n'\bm{k}}}}
\:
\tilde{\delta}(\bm{k}-\bm{k}')
\langle
U^{E}_{n\lambda\bm{k}}
|\tensor{\epsilon}|
U^{E}_{n'\lambda'\bm{k}}
\rangle
\nonumber\\
&=&0.
\label{eq:comm-ada}
\end{eqnarray}
\end{subequations}
\end{widetext}
where, in each commutation relation,
we have used Eqs.~(\ref{eq:Maxwell-PhiEH})
from the first to the second expression, 
and Eqs.~(\ref{eq:delta}) and (\ref{eq:delta-BZ})
from the second to the third expression.


Next the expectation values of $H$ and $\bm{\mathcal{R}}$
in modulated system will be estimated.
In the effective Lagrangian, we retain
up to the first order with respect to 
the derivative of $\gamma_{\epsilon}(\bm{r})$ or $\gamma_{\mu}(\bm{r})$.
Therefore we need to estimate the expectation value of $H$ 
up to the first derivatives of the modulation functions.
As for the expectation value of $\bm{\mathcal{R}}$,
we may neglect derivative terms
as was discussed in Sec.~\ref{sec:EOM}.
The commutation relations needed to estimate $\langle W|H|W\rangle$
and $\langle W|\bm{\mathcal{R}}|W\rangle$
are calculated by derivative expansion 
with respect to the modulation functions.
Firstly, the commutation relation for $\langle W|H|W\rangle$
is estimated as follows,
\begin{widetext}
\begin{eqnarray}
&&\left[a_{n\lambda\bm{k};\bm{r}_{c}},
\left[H,a^{\dagger}_{n'\lambda'\bm{k}';\bm{r}_{c}}\right]
\right]
\nonumber\\
&&=
E_{n\bm{k};\bm{r}_{c}}E_{n'\bm{k}';\bm{r}_{c}}
\int d\bm{r}
\left[
\frac{\gamma^{2}_{\epsilon}(\bm{r})}{\gamma^{2}_{\epsilon}(\bm{r}_{c})}
\bm{\Phi}^{E*}_{n\lambda\bm{k}}(\bm{r})
\tensor{\epsilon}(\bm{r})
\bm{\Phi}^{E}_{n'\lambda'\bm{k}'}(\bm{r})
+\frac{\gamma^{2}_{\mu}(\bm{r})}{\gamma^{2}_{\mu}(\bm{r}_{c})}
\bm{\Phi}^{H*}_{n\lambda\bm{k}}(\bm{r})
\tensor{\mu}(\bm{r})
\bm{\Phi}^{H}_{n'\lambda'\bm{k}'}(\bm{r})
\right]
\nonumber\\
&&=
\frac{1}{2}\sqrt{E_{n\bm{k};\bm{r}_{c}}E_{n'\bm{k};\bm{r}_{c}}}
\biggl[
\tilde{\delta}(\bm{k}-\bm{k}')
\Bigl[\langle
U^{E}_{n\lambda\bm{k}}|\tensor{\epsilon}|U^{E}_{n'\lambda'\bm{k}}
\rangle
+\langle
U^{H}_{n\lambda\bm{k}}|\tensor{\mu}|U^{H}_{n'\lambda'\bm{k}}
\rangle
\Bigr]
\nonumber\\
&&\quad
+ie^{-i(\bm{k}-\bm{k}')\cdot\bm{r}_{c}}
\left[
(\bm{\nabla}_{\bm{k}}-\bm{\nabla}_{\bm{k}'})
\tilde{\delta}(\bm{k}-\bm{k}')
\right]
\cdot
\Bigl[
\left[\bm{\nabla}_{\bm{r}_{c}}\ln\gamma_{\epsilon}(\bm{r}_{c})\right]
\langle U^{E}_{n\lambda\bm{k}}
|\tensor{\epsilon}|
U^{E}_{n'\lambda'\bm{k}'}
\rangle
+
\left[\bm{\nabla}_{\bm{r}_{c}}\ln\gamma_{\mu}(\bm{r}_{c})\right]
\langle U^{H}_{n\lambda\bm{k}}
|\tensor{\mu}|
U^{H}_{n'\lambda'\bm{k}'}
\rangle
\Bigr]\biggr]
\nonumber\\
&&\quad+\cdots.
\end{eqnarray}
\end{widetext}
The first and second terms in the second expression
come from the terms of zero-th and first order of $(\bm{r}-\bm{r}_{c})$
respectively. In the transformation to the last expression,
we have used Eqs.~(\ref{eq:delta}), (\ref{eq:delta-BZ}) and (\ref{eq:d-delta}).

In the same manner, 
restricting to the case in which 
both of $\bm{k}$ and $\bm{k}'$ are in the first Brillouin zone
and using Eq.~(\ref{eq:d-delta}),
the commutation relation for $\langle W|\bm{\mathcal{R}}|W\rangle$
is estimated as follows,
\begin{widetext}
\begin{eqnarray}
&&\left[a_{n\lambda\bm{k};\bm{r}_{c}},
\left[
\bm{\mathcal{R}},
a^{\dagger}_{n'\lambda'\bm{k}';\bm{r}_{c}}\right]
\right]
\nonumber\\
&&=E_{n\bm{k};\bm{r}_{c}}E_{n'\bm{k}';\bm{r}_{c}}
\int d\bm{r}\:\bm{r}
\left[
\frac{\gamma^{2}_{\epsilon}(\bm{r})}{\gamma^{2}_{\epsilon}(\bm{r}_{c})}
\bm{\Phi}^{E*}_{n\lambda\bm{k}}(\bm{r})
\cdot
\Bigl[
\tensor{\epsilon}(\bm{r})
\bm{\Phi}^{E}_{n'\lambda'\bm{k}'}(\bm{r})
\Bigr]
+\frac{\gamma^{2}_{\mu}(\bm{r})}{\gamma^{2}_{\mu}(\bm{r}_{c})}
\bm{\Phi}^{H*}_{n\lambda\bm{k}}(\bm{r})
\cdot
\Bigl[
\tensor{\mu}(\bm{r})
\bm{\Phi}^{H}_{n'\lambda'\bm{k}'}(\bm{r})
\Bigr]
\right]
\nonumber\\
&&=
\frac{i}{4}
\sqrt{E_{n\bm{k};\bm{r}_{c}}E_{n'\bm{k}';\bm{r}_{c}}}
\left[
(\bm{\nabla}_{\bm{k}}-\bm{\nabla}_{\bm{k}'})
\tilde{\delta}(\bm{k}-\bm{k}')
\right]
\Bigl[
\langle U^{E}_{n\lambda\bm{k}}
|\tensor{\epsilon}|
U^{E}_{n'\lambda'\bm{k}'}
\rangle
+
\langle U^{H}_{n\lambda\bm{k}}
|\tensor{\mu}|
U^{H}_{n'\lambda'\bm{k}'}
\rangle
\Bigr]
\quad+\cdots.
\end{eqnarray}
\end{widetext}

From these commutation relations, we obtain the following results
which leads to the estimation of the center of gravity
as $\langle W|\bm{\mathcal{R}}|W\rangle/\langle W|H|W\rangle
\cong \nabla_{\bm{k}}\vartheta(\bm{k}_{c},\bm{r}_{c},z_{c},t)
-(z_{c}|\bm{\Lambda}_{n\bm{k}_{c}}|z_{c})$,
\begin{widetext}
\begin{subequations}
\begin{eqnarray}
\langle W|H|W\rangle
&=&
\int_{\mathrm{BZ}}d\bm{k}
\:w^{2}_{r}(\bm{k}-\bm{k}_{c})
E_{n\bm{k};\bm{r}_{c}}
\biggl[
1+
\Bigl[\nabla_{\bm{r}_{c}}\ln\left[
\gamma_{\epsilon}(\bm{r}_{c})\gamma_{\mu}(\bm{r}_{c})
\right]\Bigr]
\cdot
\Bigl[\nabla_{\bm{k}}\vartheta(\bm{k},\bm{r}_{c},z_{c},t)
-\bm{r}_{c}\Bigr]
\nonumber\\
&&\quad
-\left[\nabla_{\bm{r}_{c}}\ln\gamma_{\epsilon}(\bm{r}_{c})\right]
\cdot(z_{c}|\bm{\Lambda}^{E}_{n\bm{k}}|z_{c})
+\left[\nabla_{\bm{r}_{c}}\ln\gamma_{\mu}(\bm{r}_{c})\right]
\cdot(z_{c}|\bm{\Lambda}^{H}_{n\bm{k}}|z_{c})
\biggr]+\cdots
\nonumber\\
&\cong&
\biggl[
1+
\Bigl[\nabla_{\bm{r}_{c}}\ln\left[
\gamma_{\epsilon}(\bm{r}_{c})\gamma_{\mu}(\bm{r}_{c})
\right]\Bigr]
\cdot(z_{c}|\bm{\Lambda}_{n\bm{k}_{c}}|z_{c})
\nonumber\\
&&\quad
-\left[\nabla_{\bm{r}_{c}}\ln\gamma_{\epsilon}(\bm{r}_{c})\right]
\cdot(z_{c}|\bm{\Lambda}^{E}_{n\bm{k}_{c}}|z_{c})
+\left[\nabla_{\bm{r}_{c}}\ln\gamma_{\mu}(\bm{r}_{c})\right]
\cdot(z_{c}|\bm{\Lambda}^{H}_{n\bm{k}_{c}}|z_{c})
\biggr]
E_{n\bm{k}_{c};\bm{r}_{c}}
\nonumber\\
&=&
\left[
1-
\left[\nabla_{\bm{r}_{c}}\ln
\frac{\gamma_{\epsilon}(\bm{r}_{c})}{\gamma_{\mu}(\bm{r}_{c})}
\right]
\cdot(z_{c}|\bm{\Delta}_{n\bm{k}_{c}}|z_{c})
\right]
E_{n\bm{k}_{c};\bm{r}_{c}},
\\
\langle W|\bm{\mathcal{R}}|W\rangle
&=&
\frac{i}{2}
\int_{\mathrm{BZ}}d\bm{k}\:
E_{n\bm{k};\bm{r}_{c}}
\Bigl[
w^{*}(\bm{k},\bm{k}_{c},\bm{r}_{c},z_{c},t)\bm{\nabla}_{\bm{k}}
w(\bm{k},\bm{k}_{c},\bm{r}_{c},t)
-[\bm{\nabla}_{\bm{k}}w^{*}(\bm{k},\bm{k}_{c},\bm{r}_{c},z_{c},t)]
w(\bm{k},\bm{k}_{c},\bm{r}_{c},t)
\Bigr]
\nonumber\\
&&\quad+\frac{i}{4}
\int_{\mathrm{BZ}}d\bm{k}\:
w_{r}^{2}(\bm{k}-\bm{k}_{c})E_{n\bm{k};\bm{r}_{c}}
\Bigl[
\langle U^{E}_{nz_{c}\bm{k}}
|\tensor{\epsilon}|\bm{\nabla}_{\bm{k}}U^{E}_{nz_{c}\bm{k}}\rangle
-\langle \bm{\nabla}_{\bm{k}}U^{E}_{nz_{c}\bm{k}}
|\tensor{\epsilon}|U^{E}_{nz_{c}\bm{k}}\rangle
\nonumber\\
&&\hspace{5cm}
+\langle U^{H}_{nz_{c}\bm{k}}
|\tensor{\mu}|\bm{\nabla}_{\bm{k}}U^{H}_{nz_{c}\bm{k}}\rangle
-\langle \bm{\nabla}_{\bm{k}}U^{H}_{nz_{c}\bm{k}}
|\tensor{\mu}|U^{H}_{nz_{c}\bm{k}}\rangle
\Bigr]+\cdots
\nonumber\\
&=&
\int_{\mathrm{BZ}}d\bm{k}
\:w^{2}_{r}(\bm{k}-\bm{k}_{c})
E_{n\bm{k};\bm{r}_{c}}
\Bigl[\nabla_{\bm{k}}\vartheta(\bm{k},\bm{r}_{c},z_{c},t)
-(z_{c}|\bm{\Lambda}_{n\bm{k}}|z_{c})
\Bigr]+\cdots
\nonumber\\
&\cong&
E_{n\bm{k}_{c};\bm{r}_{c}}
\left[\nabla_{\bm{k}_{c}}\vartheta(\bm{k}_{c},\bm{r}_{c},z_{c},t)
-(z_{c}|\bm{\Lambda}_{n\bm{k}_{c}}|z_{c})
\right].
\end{eqnarray}
\end{subequations}
\end{widetext}

The above estimation for the center of gravity suggests
that the position $\bm{r}_{c}$ defined by Eq.~(\ref{eq:rc-perturbed})
may be regarded as the center of gravity even
in the case with modulation.
In the derivation of the effective Lagrangian, we need to estimated
the inner product between the wavepacket and its time-derivative.
Finally we present the detail for the calculation of this product,
by regarding Eq.~(\ref{eq:rc-perturbed}) as the definition
of the center of wavepacket.
\begin{widetext}
\begin{eqnarray}
\langle W|i\frac{d}{dt}|W\rangle
&=& i\int_{\mathrm{BZ}}d\bm{k}\:
w^{*}(\bm{k},\bm{k}_{c},\bm{r}_{c}, z_{c}, t)
\frac{d}{dt}w(\bm{k},\bm{k}_{c},\bm{r}_{c},z_{c},t)
+i(z_{c}|\dot{z}_{c})
\nonumber\\
&&
+i\int_{\mathrm{BZ}}d\bm{k}\:\tilde{d\bm{k}'}
w^{*}(\bm{k},\bm{k}_{c},\bm{r}_{c}, z_{c}, t)
w(\bm{k}',\bm{k}_{c},\bm{r}_{c}, z_{c}, t)
\langle 0|a_{nz_{c}\bm{k};\bm{r}_{c}}
\left[\dot{\bm{r}}_{c}\cdot\bm{\nabla}_{\bm{r_{c}}}a^{\dagger}_{nz_{c}\bm{k}';\bm{r}_{c}}\right]
|0\rangle
\nonumber\\
&=& \int_{\mathrm{BZ}}d\bm{k}\:
w^{2}_{r}(\bm{k}-\bm{k}_{c})
\frac{d}{dt}\vartheta(\bm{k},\bm{r}_{c},z_{c},t)
+i(z_{c}|\dot{z}_{c})
\nonumber\\
&=& -\dot{\bm{k}}_{c}\cdot
\int_{\mathrm{BZ}}d\bm{k}
\left[\bm{\nabla}_{\bm{k}_{c}}w^{2}_{r}(\bm{k}-\bm{k}_{c})\right]
\vartheta(\bm{k},\bm{r}_{c},z_{c},t)
+i(z_{c}|\dot{z}_{c})
+\frac{d}{dt}\int_{\mathrm{BZ}}d\bm{k}\:
w^{2}_{r}(\bm{k}-\bm{k}_{c})
\vartheta(\bm{k},\bm{r}_{c},z_{c},t)
\nonumber\\
&=& \dot{\bm{k}}_{c}\cdot
\int_{\mathrm{BZ}}d\bm{k}
\left[\bm{\nabla}_{\bm{k}}w^{2}_{r}(\bm{k}-\bm{k}_{c})\right]
\vartheta(\bm{k},\bm{r}_{c},z_{c},t)
+i(z_{c}|\dot{z}_{c})
+\frac{d}{dt}\int_{\mathrm{BZ}}d\bm{k}\:
w^{2}_{r}(\bm{k}-\bm{k}_{c})\vartheta(\bm{k},\bm{r}_{c},z_{c},t)
\nonumber\\
&=& -\dot{\bm{k}}_{c}\cdot
\int_{\mathrm{BZ}}d\bm{k}\:
w^{2}_{r}(\bm{k}-\bm{k}_{c})
\left[\bm{\nabla}_{\bm{k}}\vartheta(\bm{k},\bm{r}_{c},z_{c},t)\right]
+i(z_{c}|\dot{z}_{c})
+\frac{d}{dt}\int_{\mathrm{BZ}}d\bm{k}\:
w^{2}_{r}(\bm{k}-\bm{k}_{c})
\vartheta(\bm{k},\bm{r}_{c},z_{c},t)
\nonumber\\
&=& -\dot{\bm{k}}_{c}\cdot
\left[\bm{r}_{c}+\int_{\mathrm{BZ}}d\bm{k}\:
w^{2}_{r}(\bm{k}-\bm{k}_{c})(z_{c}|\bm{\Lambda}_{n\bm{k}_{c}}|z_{c})\right]
+i(z_{c}|\dot{z}_{c})
+\frac{d}{dt}\int_{\mathrm{BZ}}d\bm{k}\:
w^{2}_{r}(\bm{k}-\bm{k}_{c})\vartheta(\bm{k},\bm{r}_{c},z_{c},t)
\nonumber\\
&\cong&
\bm{k}_{c}\cdot\dot{\bm{r}}_{c}
-\dot{\bm{k}}_{c}\cdot(z_{c}|\bm{\Lambda}_{n\bm{k}_{c}}|z_{c})
+i(z_{c}|\dot{z}_{c})
+\frac{d}{dt}\left[
\int_{\mathrm{BZ}}d\bm{k}\:
w^{2}_{r}(\bm{k}-\bm{k}_{c})
\vartheta(\bm{k},\bm{r}_{c},z_{c},t)-\bm{k}_{c}\cdot\bm{r}_{c}\right],
\end{eqnarray}
\end{widetext}
where Eq.~(\ref{eq:comm-ada}) and $a_{nz_{c}\bm{k};\bm{r}_{c}}|0\rangle=0$
are used in the transformation from the first expression to the second expression.

\section{\label{sec:internal-rotation}Berry curvature and internal rotation}
In a system with generic periodic structure,
it is a tough work to analytically
calculate the Berry curvature and the internal rotation.
However, it is easy to obtain them numerically
by rewriting inner products of Bloch functions 
and their momentum-derivatives 
to the products of conventional expectation values.
Here we present some formulae which are convenient 
for numerical calculations.

For latter convenience, we separate the Berry curvature as, 
\begin{subequations}
\begin{eqnarray}
\bm{\Omega}_{n\bm{k}} 
&=&
\frac{1}{2}\left[
\bm{\Omega}^{E}_{n\bm{k}}+\bm{\Omega}^{H}_{n\bm{k}}
\right]
-i\bm{\Delta}_{n\bm{k}}\times\bm{\Delta}_{n\bm{k}},
\label{eq:OmegaEH}
\\
\bm{\Omega}^{F}_{n\bm{k}}
&=& \bm{\nabla}_{\bm{k}}\times\bm{\Lambda}^{F}_{n\bm{k}}
+i\bm{\Lambda}^{F}_{n\bm{k}}\times\bm{\Lambda}^{F}_{n\bm{k}},
\label{eq:OmegaI}
\end{eqnarray}
\end{subequations}
where $F=E$ or $H$.
In the following, we rewrite the $\bm{k}$-derivative in the above expression
in terms of the Feynman-Hellman relation.
However, in systems with the gauge symmetry,
even if $|U^{E,H}_{n\lambda\bm{k}}\rangle$
is a Bloch function of a physical state,
its derivative may have an unphysical component
proportional to $|K\rangle$.
In other words, Bloch functions and their derivatives should be expanded by
non-orthogonal bases as follows,
\begin{subequations}
\begin{eqnarray}
|V_{\bm{k}}\rangle
&=& \sum_{n,\lambda}
|U^{E}_{n\lambda\bm{k}}\rangle\langle U^{E}_{n\lambda\bm{k}}|\tensor{\epsilon}|V_{\bm{k}}\rangle
\nonumber \\
&&\ \ \ +\sum_{\bm{G},\bm{G}'}|K\rangle[\Gamma^{E}_{\bm{k}}]^{-1}(\bm{G},\bm{G}')
\langle K'|\tensor{\epsilon}|V_{\bm{k}}\rangle
\\
&=& \sum_{n,\lambda}
|U^{H}_{n\lambda\bm{k}}\rangle\langle U^{H}_{n\lambda\bm{k}}|\tensor{\mu}|V_{\bm{k}}\rangle
\nonumber \\
&&\ \ \ +\sum_{\bm{G},\bm{G}'}|K\rangle[\Gamma^{H}_{\bm{k}}]^{-1}(\bm{G},\bm{G}')
\langle K'|\tensor{\mu}|V_{\bm{k}}\rangle,
\end{eqnarray}
\end{subequations}
where
$\Gamma^{E}_{\bm{k}}(\bm{G},\bm{G}') = \bm{K}\tensor{\epsilon}\bm{K}'$ and
$\Gamma^{H}_{\bm{k}}(\bm{G},\bm{G}') = \bm{K}\tensor{\mu}\bm{K}'$.
By using the above expansion and the Feynman-Hellman relation
derived from Eqs.~(\ref{eq:eigen-UE}) and (\ref{eq:eigen-UH}), 
we can rewrite the Berry curvature as
\begin{widetext}
\begin{subequations}
\begin{eqnarray}
\left[\bm{\Omega}^{E}_{n\bm{k}}\right]_{\lambda\lambda'}
&=& 
-i\sum_{m\neq n,\lambda''}
\frac{\langle U^{E}_{n\lambda\bm{k}}|
\left[\bm{\nabla}_{\bm{k}}\Xi^{E}_{\bm{k}}\right]
|U^{E}_{m\lambda''\bm{k}}\rangle
\times\langle U^{E}_{m\lambda''\bm{k}}|
\left[\bm{\nabla}_{\bm{k}}\Xi^{E}_{\bm{k}}\right]
|U^{E}_{n\lambda'\bm{k}}\rangle}
{(E^{2}_{n\bm{k}}-E^{2}_{m\bm{k}})^2}
+\langle U^{E}_{n\lambda\bm{k}}|\tensor{\epsilon}
[\Gamma^{E}_{\bm{k}}]^{-1}\bm{S}\tensor{\epsilon}|U^{E}_{n\lambda'\bm{k}}\rangle
,\label{eq:OmegaE}\\
\left[\bm{\Omega}^{H}_{n\bm{k}}\right]_{\lambda\lambda'}
&=& 
-i\sum_{m\neq n,\lambda''}
\frac{\langle U^{H}_{n\lambda\bm{k}}|
\left[\bm{\nabla}_{\bm{k}}\Xi^{H}_{\bm{k}}\right]
|U^{H}_{m\lambda''\bm{k}}\rangle
\times\langle U^{H}_{m\lambda''\bm{k}}|
\left[\bm{\nabla}_{\bm{k}}\Xi^{H}_{\bm{k}}\right]
|U^{H}_{n\lambda'\bm{k}}\rangle}
{(E^{2}_{n\bm{k}}-E^{2}_{m\bm{k}})^2}
+\langle U^{H}_{n\lambda\bm{k}}|\tensor{\mu}
[\Gamma^{H}_{\bm{k}}]^{-1}\bm{S}\tensor{\mu}|U^{H}_{n\lambda'\bm{k}}\rangle
,\label{eq:OmegaH}\\
\left[\bm{\Delta}_{n\bm{k}}\right]_{\lambda\lambda'}
&=&
\frac{1}{4E_{n\bm{k}}}
\left[
\langle U^{E}_{n\lambda\bm{k}}|\bm{S}|U^{H}_{n\lambda'\bm{k}}\rangle
+\langle U^{H}_{n\lambda\bm{k}}|\bm{S}|U^{E}_{n\lambda'\bm{k}}\rangle
\right].\label{eq:Delta}
\end{eqnarray}
\end{subequations}
Thus $\bm{\Omega}^{i}_{n\bm{k}}$ $(i=E,H)$ is enhanced when the band 
comes close to other bands in energy, with the enhancement being 
inversely proportional to the square of energy difference. 
In contrast, $\bm{\Delta}_{n\bm{k}}$ does not have such an enhancement.
Though Eq.~(\ref{eq:Delta}) seems to diverge at $E_{n\bm{k}}\rightarrow 0$ 
($\bm{k}\rightarrow 0$), it is not the case, as we will see 
in Sec.~\ref{sec:photonic-crystal} for a specific case.
In the long wavelength limit $\bm{k}\rightarrow 0$, the propagating 
light becomes insensitive to spatial 
modulations of $\mu(\bm{r})$ and $\epsilon(\bm{r})$, and the medium
is regarded as uniform. Because 
$\bm{\Delta}_{n\bm{k}}=0$ for a uniform isotropic media, a generic 
periodic medium in a long-wavelength limit also shows 
$\bm{\Delta}_{n\bm{k}}\to 0$.

In the same manner, the internal rotation is also rewritten as follows,
\begin{subequations}
\begin{eqnarray}
\left[\bm{\mathcal{S}}^{E}_{n\bm{k}}\right]_{\lambda\lambda'}
&=&
\frac{1}{2}\Biggl[ 
-i\sum_{m\neq n,\lambda''}
\frac{\langle U^{E}_{n\lambda\bm{k}}|
\left[\bm{\nabla}_{\bm{k}}\Xi^{E}_{\bm{k}}\right]
|U^{E}_{m\lambda''\bm{k}}\rangle
\times\langle U^{E}_{m\lambda''\bm{k}}|
\left[\bm{\nabla}_{\bm{k}}\Xi^{E}_{\bm{k}}\right]
|U^{E}_{n\lambda'\bm{k}}\rangle}
{E^{2}_{n\bm{k}}-E^{2}_{m\bm{k}}}
\nonumber\\
&&\qquad
+E^{2}_{n\bm{k}}\langle U^{E}_{n\lambda\bm{k}}|\tensor{\epsilon}[\Gamma^{E}_{\bm{k}}]^{-1}
\bm{S}\tensor{\epsilon}|U^{E}_{n\lambda'\bm{k}}\rangle
-i\langle U^{E}_{n\lambda\bm{k}}|\bm{S}\times\tensor{\mu}^{-1}\bm{S}|U^{E}_{n\lambda'\bm{k}}\rangle
\Biggr]
\label{eq:S-E-band}
,\\
\left[\bm{\mathcal{S}}^{H}_{n\bm{k}}\right]_{\lambda\lambda'}
&=&
\frac{1}{2}\Biggl[ 
-i\sum_{m\neq n,\lambda''}
\frac{\langle U^{H}_{n\lambda\bm{k}}|
\left[\bm{\nabla}_{\bm{k}}\Xi^{H}_{\bm{k}}\right]
|U^{H}_{m\lambda''\bm{k}}\rangle
\times\langle U^{H}_{m\lambda''\bm{k}}|
\left[\bm{\nabla}_{\bm{k}}\Xi^{H}_{\bm{k}}\right]
|U^{H}_{n\lambda'\bm{k}}\rangle}
{E^{2}_{n\bm{k}}-E^{2}_{m\bm{k}}}
\nonumber\\
&&\qquad
+E^{2}_{n\bm{k}}\langle U^{H}_{n\lambda\bm{k}}|\tensor{\mu}[\Gamma^{H}_{\bm{k}}]^{-1}
\bm{S}\tensor{\mu}|U^{H}_{n\lambda'\bm{k}}\rangle
-i\langle U^{H}_{n\lambda\bm{k}}|\bm{S}\times\tensor{\epsilon}^{-1}\bm{S}|U^{H}_{n\lambda'\bm{k}}\rangle
\Biggr]. \label{eq:S-H-band}
\end{eqnarray}
\end{subequations}
\end{widetext}
It should be noted that $\bm{\Omega}^{E}_{n\bm{k}}$ and $\bm{\Omega}^{H}_{n\bm{k}}$
have very similar expressions
to $\bm{\mathcal{S}}^{E}_{n\bm{k}}$ and $\bm{\mathcal{S}}^{H}_{n\bm{k}}$, 
respectively.
This suggests that there are always some kind of rotation
when the Berry curvatures are nonzero.
In this sense, we have generalized the argument
for the quantum Hall system in Ref.~\cite{Chang-Niu} 
to a photonic system.
In the quantum Hall system, the internal rotation is 
the internal orbital rotation originated by the cyclotron motion
under an external magnetic field.
On the other hand, in the present case, 
the internal rotation are the combination
of the polarization and the internal orbital rotation
originated from a periodic structure.
When there is no periodic structure, anisotropy
nor inhomogeneity 
in $\tensor{\epsilon}(\bm{r})$ and $\tensor{\mu}(\bm{r})$,
Eqs.~(\ref{eq:OmegaE})-(\ref{eq:S-H-band})
are reduced to the Berry curvature, $\frac{\bm{k}}{k^{3}}\sigma_{3}$,
and the spin divided by $\epsilon\mu$, 
$\frac{1}{\epsilon\mu}\cdot\frac{\bm{k}}{k}\sigma_{3}$.
These contributions come only from 
the terms including the spin operator $\bm{S}$,
and $\bm{\Delta}_{n\bm{k}}=0$,
i.e., nonzero $\bm{\Delta}_{n\bm{k}}$ is originated by 
the anisotropy or the periodic structure 
of $\tensor{\epsilon}$ and $\tensor{\mu}$.
Even in generic cases, $\bm{\Delta}_{n\bm{k}}$ has 
a unit of a length, and its magnitude 
is a lattice constant at most.

\section{\label{sec:Maxwell-shift}Transverse shift 
in classical electrodynamics}
Here we prove the consistency between our result
for the transverse shift (Eq.~(\ref{eq:delta_y})),
which is consistent with the TAM conservation
for individual photons (Eq.~(\ref{eq:TAM-photon})),
and the result by Fedoseev \cite{Fedoseev-I,Fedoseev-II},
which is based on classical electrodynamics.
In Refs.~\cite{Fedoseev-I,Fedoseev-II}, each wavepacket is constructed as
a superposition of plane waves with wavevectors $\bm{k}=\bm{k}_{c}+\bm{\kappa}$, 
where $\bm{\kappa}$ distributed around zero vector. 
(In the notation of Refs.~\cite{Fedoseev-I,Fedoseev-II},
$\bm{k}_{c}$ is represented by $\bm{K}$.)
The polarization vector of each constituent plane wave is defined by
Eq.~(23) in Ref.~\cite{Fedoseev-I} with Eq.~(7) in Ref.~\cite{Fedoseev-II},
\begin{subequations}
\begin{eqnarray}
\bm{e}^{(j)}(\bm{\kappa})
&=& z^{(j)}_{s}(\bm{\kappa})\bm{s}^{(j)}(\bm{\kappa})
+z^{(j)}_{p}(\bm{\kappa})\bm{p}^{(j)}(\bm{\kappa}),
\end{eqnarray}
\end{subequations}
where $j=i,\rho,\tau$ for incident, reflected and transmitted beams, respectively,
$z^{(j)}_{s}(\bm{\kappa})$ and $z^{(j)}_{p}(\bm{\kappa})$ 
($|z^{(j)}_{s}(\bm{\kappa})|^{2}+|z^{(j)}_{p}(\bm{\kappa})|^{2}=1$)
represent the polarization state of each plane wave,
$\bm{s}^{(j)}(\bm{\kappa})$ and $\bm{p}^{(j)}(\bm{\kappa})$
are the $s$- and $p$-polarization vectors defined by
\begin{equation}
\bm{s}^{(j)}(\bm{\kappa})=\frac{\bm{n}\times\bm{k}}{|\bm{n}\times\bm{k}|},
\quad
\bm{p}^{(j)}(\bm{\kappa})=\bm{s}(\bm{\kappa})\times \frac{\bm{k}}{|\bm{k}|},
\end{equation}
where $\bm{n}=(0,0,1)$ is normal to the interface,
and we consider the same configuration of the interface
and beams as those in Sec.~\ref{sec:Imbert}.
The relation between the present notation and that in Ref.~\cite{Fedoseev-II}
is represented as $z^{(j)}_{s}(\bm{\kappa}) \leftrightarrow A^{(j)}(\bm{\kappa})
/\sqrt{|A^{(j)}(\bm{\kappa})|^2+|B^{(j)}(\bm{\kappa})|^2}$
and $z^{(j)}_{p}(\bm{\kappa}) \leftrightarrow B^{(j)}(\bm{\kappa})/
\sqrt{|A^{(j)}(\bm{\kappa})|^2+|B^{(j)}(\bm{\kappa})|^2}$,
$\bm{n}\leftrightarrow \bm{N}$ and $\bm{k}/|\bm{k}|\leftrightarrow \bm{m}(\bm{\kappa})$.

By the Maxwell equations,
 $z^{(\rho,\tau)}_{s}(\bm{\kappa})$ and $z^{(\rho,\tau)}_{p}(\bm{\kappa})$ 
are exactly given by
\begin{subequations}
\begin{eqnarray}
&& z^{(j)}_{s}(\bm{\kappa}) = 
\frac{t^{(j)}_{s}(\bm{\kappa})z^{(i)}_{s}(\bm{\kappa})}
{\sqrt{|t^{(j)}_{s}(\bm{\kappa})z^{(i)}_{s}|^{2}+|t^{(j)}_{p}(\bm{\kappa})z^{(i)}_{p}|^{2}}},
\label{eq:z_s}\\
&& z^{(j)}_{p}(\bm{\kappa}) = 
\frac{t^{(j)}_{p}(\bm{\kappa})z^{(i)}_{p}(\bm{\kappa})}
{\sqrt{|t^{(j)}_{s}(\bm{\kappa})z^{(i)}_{s}|^{2}+|t^{(j)}_{p}(\bm{\kappa})z^{(i)}_{p}|^{2}}},
\label{eq:z_p}
\end{eqnarray}
\end{subequations}
where $j=\rho$ or $\tau$,
$t^{(\rho)}_{s}(\bm{\kappa})$ and $t^{(\rho)}_{p}(\bm{\kappa})$
are the amplitude reflection coefficients for 
the $s$- and $p$-polarized plane waves,
$t^{(\tau)}_{s}(\bm{\kappa})$ and $t^{(\tau)}_{p}(\bm{\kappa})$
and the amplitude transmission coefficients for 
the $s$- and $p$-polarized plane waves,
i.e., $t^{(\rho)}_{s} \leftrightarrow R_{s}$, $t^{(\rho)}_{p} \leftrightarrow R_{p}$,
$t^{(\tau)}_{s} \leftrightarrow T_{s}$ and $t^{(\tau)}_{p} \leftrightarrow T_{p}$
in our notation in Sec.~\ref{sec:Imbert}.

In our constitution method for an incident wavepacket,
the polarization state of each constituent plane wave, i.e,
the set of $z^{(i)}_{s}$ and $z^{(i)}_{p}$, is independent of $\bm{\kappa}$.
Otherwise, the concept ``an elliptically-polarized incident wavepacket''
gets fuzzy (see Sec.~\ref{sec:simulation}).
Thus, this is a natural definition for
an elliptically-polarized incident wavepacket.
Its polarization vector is represented also in the following form, 
\begin{equation}
\bm{e}^{(i)}(\bm{\kappa})=
\frac{\bm{p}^{(i)}(\bm{\kappa})+m\bm{s}^{(i)}(\bm{\kappa})}{\sqrt{1+|m|^{2}}}
\label{eq:eI}
\end{equation}
where $m$ is a complex constant, representing the polarization state. 
This $m$ is identical with $m$ defined by Bliokh {\it et al.}~\cite{Bliokh-PRL}, and 
related with our $|z^{I})$ in Sec.~\ref{sec:Imbert} by 
\begin{equation}
|z^{I})=\frac{1}{\sqrt{2(1+|m|^{2})}}
\left(
\begin{array}{c}
1-im \\ 
1+im \end{array}
\right).
\end{equation}
It yields
\begin{equation}
(z^{I}|\bm{\sigma}|z^{I})=\frac{1}{1+|m|^{2}}\left[1-|m|^{2},\ 2\Re (m),\ 2\Im (m)\right].
\label{eq:zm}
\end{equation}
which is used for comparison between the results here and those based on our theory 
of the TAM conservation for individual photons.

We now calculate the transverse shift from Eqs.~(15)-(17) in Ref.~\cite{Fedoseev-II}.
The result is a sum of two terms 
\begin{equation}
\delta y^{(j)}= h^{(j1)}+h^{(j2)},\label{eq:yjh}
\end{equation}
where $j=\rho$ or $\tau$, and
$\delta y^{(\rho)} \leftrightarrow \delta y^{R}$ 
and $\delta y^{(\tau)} \leftrightarrow \delta y^{T}$
in our notation in Sec.~\ref{sec:Imbert}.
From Eqs.~(13a), (13b), (17) and (18) in Ref.~\cite{Fedoseev-II},
the second term of right-hand side, $h^{(j2)}$,
is proportional to the $\kappa_y$-derivative of
$\Im[\ln z^{(j)}_{s}(\bm{\kappa})-\ln z^{(j)}_{p}(\bm{\kappa})]$
at $\bm{\kappa}=0$.
The amplitude reflection/refraction coefficients
depend only on the polar angle, and thus
their derivatives by $\kappa_{y}$ at $\bm{\kappa}=0$ are zero, 
because the $y$-component of $\bm{k}_{c}$ is zero
in the present configuration.
As was mentioned previously,
$z^{(i)}_{s}$ and $z^{(i)}_{p}$
are independent of $\bm{\kappa}$.
Therefore, from Eqs.~(\ref{eq:z_s}) and (\ref{eq:z_p}), $h^{j2}=0$
($j = \rho$, $\tau$),
and we have
\begin{eqnarray}
\delta y^{(j)}
&=&h^{(j1)}
\nonumber\\
&=&
-i\frac{\bm{n}\cdot\left[\bm{e}^{(j)}(0)\times\bm{e}^{(j)*}(0)\right]}
{|\bm{n}\times \bm{k}^{(j)}|}
\nonumber\\
&&
+i\frac{\bm{n}\cdot\left[\bm{e}^{(i)}(0)\times\bm{e}^{(i)*}(0)\right]}
{|\bm{n}\times \bm{k}^{(i)}|},
\label{eq:delta_y-Fedoseev}
\end{eqnarray}
where $\bm{k}^{(j)}$ ($j=\rho$ or $\tau$) are mean wavevectors for 
reflected ($\rho$) and transmitted ($\tau$) wavepackets.
Note that, the correspondence between 
these wavevectors and those in Sec.~\ref{sec:Imbert} are
$\bm{k}^{(i)}\leftrightarrow \bm{k}^{I}$,
$\bm{k}^{(\rho)}\leftrightarrow \bm{k}^{R}$ and 
$\bm{k}^{(\tau)} \leftrightarrow \bm{k}^{T}$.
Eq.~(\ref{eq:delta_y-Fedoseev}) is identical with 
Eq.~(\ref{eq:delta_y}), showing an equivalence 
between Fedoseev's theory based on classical electrodynamics
and ours.

Finally, we rewrite Eq.~(\ref{eq:delta_y-Fedoseev})
in terms of our notation in Sec.~\ref{sec:Imbert}.
For partial reflection,
$A_p$ and $A_s$ are real, and we get 
\begin{equation}
\delta y^{A}
=\frac{2{\Im}(m)}{k^{I}\sin\theta_{I}}
\left[
\frac{(A_s/A_p)\cos\theta_{A}}{1+(A_s/A_p)^{2}|m|^{2}}
-\frac{\cos\theta_{I}}{1+|m|^{2}}
\right],
\label{eq:yAF-partial}
\end{equation}
where $A=T$ or $R$.
By rewriting Eq.~(\ref{eq:yAF-partial}) in terms 
of $|z^{I})$, the shift is equal to our result in Eq.~(\ref{eq:Imbert-partial}) 
but not to Eq.~(5) in Ref.~\cite{Bliokh-PRL}.
For total reflection, $R_p$ and $R_s$ are complex numbers with $|R_p|=|R_s|=1$, and we get
\begin{equation}
\delta y^{A}=\frac{-2\cos\theta_{I}}{k^{I}\sin\theta_{I}}
\frac{\Im (m) [\Re (R_p^{*}R_s)+1]+ \Re (m) \Im (R_p^{*}R_s) }{1+|m|^{2}}. 
\label{eq:hAF-total}
\end{equation}
This is exactly the same as ours in Eq.~(\ref{eq:Imbert-total}). 
To summarize, for every case,
the calculation based on classical electrodynamics
gives the identical result with ours based on our quantum-mechanical formalism,
and this result is consistent with the TAM conservation for individual photons.

\section{\label{sec:Omega-2D}Berry curvature in a two-dimensional photonic crystal}
In order to discuss the TM and TE modes,
it is convenient to introduce the following unit vectors,
\begin{eqnarray}
\bm{e}_{K} &=& \frac{\bm{K}}{K}
,\ 
\bm{e}_{\mathrm{I}} = \frac{\bm{e}_{z}\times\bm{e}_{K}}{|\bm{e}_{z}\times\bm{e}_{K}|},
\nonumber\\
\end{eqnarray} 
and the Bloch functions are represented by
\begin{subequations}
\begin{eqnarray}
\epsilon|U^{E}_{\mathrm{TE}\:m\bm{k}} \rangle&=& 
\bm{e}_{\mathrm{I}}\otimes |U^{D}_{\mathrm{TE}\:m\bm{k}}\rangle
,\\
\mu|U^{H}_{\mathrm{TE}\:m\bm{k}} \rangle&=& 
\bm{e}_{z}\otimes |U^{B}_{\mathrm{TE}\:m\bm{k}}\rangle,
\end{eqnarray}
\end{subequations}
for the TE modes and 
\begin{subequations}
\begin{eqnarray}
\epsilon|U^{E}_{\mathrm{TM}\:m\bm{k}} \rangle&=& 
\bm{e}_{z}\otimes |U^{D}_{\mathrm{TM}\:m\bm{k}}\rangle
,\\
\mu|U^{H}_{\mathrm{TM}\:m\bm{k}} \rangle&=& 
\bm{e}_{\mathrm{I}}\otimes |U^{B}_{\mathrm{TM}\:m\bm{k}}\rangle,
\end{eqnarray}
\end{subequations}
for the TM modes.
The superscripts, $D$ and $B$, mean that they correspond to 
the electric and magnetic flux densities, respectively, satisfying
the transversality condition, i.e. being perpendicular to $\bm{K}$.

The matrices for the eigen equations given in Eqs.~(\ref{eq:XiE}) and (\ref{eq:XiH}) 
are simplified in 
the case with scalar $\epsilon(\bm{r})$ and $\mu(\bm{r})$
as follows,
\begin{subequations}
\begin{eqnarray}
\Xi^{E}_{\bm{k}}(\bm{G},\bm{G}')
&=&\Theta(\bm{G},\bm{G}')\mu^{-1}(\bm{G},\bm{G}')
,\\
\Xi^{H}_{\bm{k}}(\bm{G},\bm{G}')
&=&\Theta(\bm{G},\bm{G}')\epsilon^{-1}(\bm{G},\bm{G}')
\end{eqnarray}
\end{subequations}
where
$
\Theta_{\bm{k}}(\bm{G},\bm{G}')
=(\bm{K}\cdot\bm{K}'I-\bm{K}'\otimes\bm{K})
$,
and their derivatives are represented by
\begin{subequations}
\begin{eqnarray}
\nabla_{k_{x}}\Theta_{\bm{k}}(\bm{G},\bm{G}')
&=&\left(
\begin{array}{ccc}
0 & -K_{y} & -k_{z} \\
-K_{y}' & K_{x}+K_{x}' & 0 \\
-k_{z} & 0 & K_{x}+K_{x}'
\end{array}
\right)
,\nonumber\\
\\
\nabla_{k_{y}}\Theta_{\bm{k}}(\bm{G},\bm{G}')
&=&\left(
\begin{array}{ccc}
K_{y}+K_{y}' & -K_{x}' & 0 \\
-K_{x} & 0 & -k_{z} \\
0 & -k_{z} & K_{y}+K_{y}'
\end{array}
\right)
,\nonumber\\
\\
\nabla_{k_{z}}\Theta_{\bm{k}}(\bm{G},\bm{G}')
&=&\left(
\begin{array}{ccc}
2k_{z} & 0 & -K_{x}' \\
0 & 2k_{z} & -K_{y}'\\
-K_{x} & -K_{y} & 0
\end{array}
\right).
\end{eqnarray}
\end{subequations}
because $\bm{G}$ and $\bm{G}'$ have no $z$-components.

From the above formula together with Eqs.(\ref{eq:OmegaE}), 
(\ref{eq:OmegaH}) and by setting $k_z=0$,
we can easily show that the Berry curvature 
of a non-degenerate TM (TE) mode
has only $z$-component,
\begin{widetext}
\begin{subequations}
\begin{eqnarray}
\Omega^{E,z}_{\mathrm{TM}\:n\bm{k}}
&=& 
2\sum_{m\neq n}
\frac{\Im\left[\langle U^{E}_{\mathrm{TM}\:n\bm{k}}|
\left[\nabla_{k_{x}}\Xi^{E}_{\bm{k}}\right]
|U^{E}_{\mathrm{TM}\:m\bm{k}}\rangle
\langle U^{E}_{\mathrm{TM}\:m\bm{k}}|
\left[\nabla_{k_{y}}\Xi^{E}_{\bm{k}}\right]
|U^{E}_{\mathrm{TM}\:n\bm{k}}\rangle
\right]}
{(E^{2}_{\mathrm{TM}\:n\bm{k}}-E^{2}_{\mathrm{TM}\:m\bm{k}})^2}
,\\
\Omega^{H,z}_{\mathrm{TM}\:n\bm{k}}
&=& 
2\sum_{m\neq n}
\frac{\Im\left[\langle U^{H}_{\mathrm{TM}\:n\bm{k}}|
\left[\nabla_{k_{x}}\Xi^{H}_{\bm{k}}\right]
|U^{H}_{\mathrm{TM}\:m\bm{k}}\rangle
\langle U^{H}_{\mathrm{TM}\:m\bm{k}}|
\left[\nabla_{k_{y}}\Xi^{H}_{\bm{k}}\right]
|U^{H}_{\mathrm{TM}\:n\bm{k}}\rangle
\right]}
{(E^{2}_{\mathrm{TM}\:n\bm{k}}-E^{2}_{m\mathrm{TM}\:\bm{k}})^2}
+\langle U^{H}_{\mathrm{TM}\:n\bm{k}}|\mu
[\Gamma^{H}_{\bm{k}}]^{-1}S^{z}\mu|U^{H}_{\mathrm{TM}\:n\bm{k}}\rangle
,\nonumber\\
\\
\Omega^{E,z}_{\mathrm{TE}\:n\bm{k}}
&=& 
2\sum_{m\neq n}
\frac{\Im\left[\langle U^{E}_{\mathrm{TE}\:n\bm{k}}|
\left[\nabla_{k_{x}}\Xi^{E}_{\bm{k}}\right]
|U^{E}_{\mathrm{TE}\:m\bm{k}}\rangle
\langle U^{E}_{\mathrm{TE}\:m\bm{k}}|
\left[\nabla_{k_{y}}\Xi^{E}_{\bm{k}}\right]
|U^{E}_{\mathrm{TE}\:n\bm{k}}\rangle
\right]}
{(E^{2}_{\mathrm{TE}\:n\bm{k}}-E^{2}_{\mathrm{TE}\:m\bm{k}})^2}
+\langle U^{E}_{\mathrm{TE}\:n\bm{k}}|\epsilon
[\Gamma^{E}_{\bm{k}}]^{-1}S^{z}\epsilon|U^{E}_{\mathrm{TE}\:n\bm{k}}\rangle
,\\
\Omega^{H,z}_{\mathrm{TE}\:n\bm{k}}
&=& 
2\sum_{m\neq n}
\frac{\Im\left[\langle U^{H}_{\mathrm{TE}\:n\bm{k}}|
\left[\nabla_{k_{x}}\Xi^{H}_{\bm{k}}\right]
|U^{H}_{\mathrm{TE}\:m\bm{k}}\rangle
\langle U^{H}_{\mathrm{TE}\:m\bm{k}}|
\left[\nabla_{k_{y}}\Xi^{H}_{\bm{k}}\right]
|U^{H}_{\mathrm{TE}\:n\bm{k}}\rangle
\right]}
{(E^{2}_{\mathrm{TE}\:n\bm{k}}-E^{2}_{\mathrm{TE}\:m\bm{k}})^2}.
\end{eqnarray}
\end{subequations}
\end{widetext}
Note that 
the Berry curvature does not necessarily 
decrease with the energy increases
in contrast to the system without periodic structure,
because nearly degenerate points due to the band structure
enhance the magnitude of the Berry curvature.

Following the same argument as the Berry curvature,
the internal rotation also has only the $z$-component
for non-degenerate bands.
\begin{widetext}
\begin{subequations}
\begin{eqnarray}
\mathcal{S}^{E,z}_{\mathrm{TM}\:n\bm{k}}
&=&
\sum_{m\neq n}
\frac{\Im\left[\langle U^{E}_{\mathrm{TM}\:n\bm{k}}|
\left[\nabla_{k_{x}}\Xi^{E}_{\bm{k}}\right]
|U^{E}_{\mathrm{TM}\:m\bm{k}}\rangle
\langle U^{E}_{\mathrm{TM}\:m\bm{k}}|
\left[\nabla_{k_{y}}\Xi^{E}_{\bm{k}}\right]
|U^{E}_{\mathrm{TM}\:n\bm{k}}\rangle
\right]}
{E^{2}_{\mathrm{TM}\:n\bm{k}}-E^{2}_{\mathrm{TM}\:m\bm{k}}}
,\\
\mathcal{S}^{H,z}_{\mathrm{TM}\:n\bm{k}}
&=&
\sum_{m\neq n}
\frac{\Im\left[\langle U^{H}_{\mathrm{TM}\:n\bm{k}}|
\left[\nabla_{k_{x}}\Xi^{H}_{\bm{k}}\right]
|U^{H}_{\mathrm{TM}\:m\bm{k}}\rangle
\langle U^{H}_{\mathrm{TM}\:m\bm{k}}|
\left[\nabla_{k_{y}}\Xi^{H}_{\bm{k}}\right]
|U^{H}_{\mathrm{TM}\:n\bm{k}}\rangle
\right]}
{E^{2}_{\mathrm{TM}\:n\bm{k}}-E^{2}_{\mathrm{TM}\:m\bm{k}}}
\nonumber\\
&&\quad
+\frac{1}{2}\left[
E^{2}_{\mathrm{TM}\:n\bm{k}}\langle U^{H}_{\mathrm{TM}\:n\bm{k}}|\mu[\Gamma^{H}_{\bm{k}}]^{-1}
S^{z}\mu|U^{H}_{\mathrm{TM}\:n\bm{k}}\rangle
+\langle U^{H}_{\mathrm{TM}\:n\bm{k}}|\epsilon^{-1}S^{z}|U^{H}_{\mathrm{TM}\:n\bm{k}}\rangle
\right]
,\\
\mathcal{S}^{E,z}_{\mathrm{TE}\:n\bm{k}}
&=&
\sum_{m\neq n}
\frac{\Im\left[\langle U^{E}_{\mathrm{TE}\:n\bm{k}}|
\left[\nabla_{k_{x}}\Xi^{E}_{\bm{k}}\right]
|U^{E}_{\mathrm{TE}\:m\bm{k}}\rangle
\langle U^{E}_{\mathrm{TE}\:m\bm{k}}|
\left[\nabla_{k_{y}}\Xi^{E}_{\bm{k}}\right]
|U^{E}_{\mathrm{TE}\:n\bm{k}}\rangle
\right]}
{E^{2}_{\mathrm{TE}\:n\bm{k}}-E^{2}_{\mathrm{TE}\:m\bm{k}}}
\nonumber\\
&&\quad
+\frac{1}{2}\left[
E^{2}_{\mathrm{TE}\:n\bm{k}}\langle U^{E}_{\mathrm{TE}\:n\lambda\bm{k}}|\epsilon[\Gamma^{E}_{\bm{k}}]^{-1}
S^{z}\epsilon|U^{E}_{\mathrm{TE}\:n\lambda'\bm{k}}\rangle
+\langle U^{E}_{\mathrm{TE}\:n\bm{k}}|\mu^{-1}S^{z}|U^{E}_{\mathrm{TE}\:n\bm{k}}\rangle
\right]
,\\
\mathcal{S}^{H,z}_{\mathrm{TE}\:n\bm{k}}
&=&
\sum_{m\neq n}
\frac{\Im\left[\langle U^{H}_{\mathrm{TE}\:n\bm{k}}|
\left[\nabla_{k_{x}}\Xi^{H}_{\bm{k}}\right]
|U^{H}_{\mathrm{TE}\:m\bm{k}}\rangle
\langle U^{H}_{\mathrm{TE}\:m\bm{k}}|
\left[\nabla_{k_{y}}\Xi^{H}_{\bm{k}}\right]
|U^{H}_{\mathrm{TE}\:n\bm{k}}\rangle
\right]}
{E^{2}_{\mathrm{TE}\:n\bm{k}}-E^{2}_{\mathrm{TE}\:m\bm{k}}}.
\end{eqnarray}
\end{subequations}
\end{widetext}
It is interesting that 
the internal rotation of a photon can be
perpendicular to the propagating direction
in a two-dimensional photonic crystal.

We also calculate $\bm{\Delta}_{n\bm{k}}$ from
Eq.~(\ref{eq:Delta}).
For non-degenerate bands, there is no contribution to the Berry 
curvature $\bm{\Omega}_{n\bm{k}}$
from the vector product of $\bm{\Delta}_{\mathrm{TM(TE)}\:n\bm{k}}$
in Eq.~(\ref{eq:OmegaEH}),
because $\bm{\Delta}_{\mathrm{TM(TE)}\:n\bm{k}}$
is a simple vector variable, not a set of matrices.
Meanwhile, $\bm{\Delta}_{\mathrm{TM(TE)}\:n\bm{k}}$ may modify 
the energy spectrum when a modulation is applied.
$\bm{\Delta}_{n\bm{k}}$ is given as follows
\begin{subequations}
\begin{eqnarray}
\bm{\Delta}_{\mathrm{TM}\:n\bm{k}}
&=&
\frac{1}{2E^2_{\mathrm{TM}\:n\bm{k}}}
\Im\left[
\langle U^{D}_{\mathrm{TM}\:n\bm{k}}|
\epsilon^{-1}\mu^{-1}\bm{P}_{\bm{k}}\epsilon^{-1}
|U^{D}_{\mathrm{TM}\:n\bm{k}}\rangle
\right],\nonumber \\
\label{eq:DeltaTM}
\\
\bm{\Delta}_{\mathrm{TE}\:n\bm{k}}
&=&
\frac{1}{2E^2_{\mathrm{TE}\:n\bm{k}}}
\Im\left[
\langle U^{B}_{\mathrm{TE}\:n\bm{k}}|
\mu^{-1}\bm{P}_{\bm{k}}\epsilon^{-1}\mu^{-1}
|U^{B}_{\mathrm{TE}\:n\bm{k}}\rangle
\right]
.\nonumber \\ \label{eq:DeltaTE}
\end{eqnarray}
\end{subequations}
In many cases, we can approximately regard
the magnetic permeability $\mu$ to be constant.
Then $\bm{\Delta}_{\mathrm{TM}\:n\bm{k}}$ vanishes
from Eq.~(\ref{eq:DeltaTE}), whereas $\bm{\Delta}_{\mathrm{TE}\:n\bm{k}}$
does not in general.

\section{\label{sec:Delta} Remarks on $\bm{\Delta}_{\mathrm{TE}\:n\bm{k}}$}
Here we evaluate 
$\bm{\Delta}_{\mathrm{TE}\:n\bm{k}}$ for the two-dimensional photonic crystal
discussed in Sec.~\ref{sec:photonic-crystal}
and see its effect on the energy dispersion 
and group velocity for each of the first and second bands
of TE mode.
From Eq.~(\ref{eq:modulated-E}), an additional correction 
appears in the energy of each TE mode as
\begin{eqnarray}
\frac{\mathcal{E}_{\mathrm{TE}\:n\bm{k}_{c};\bm{r}_{c}}}
{E_{\mathrm{TE}\:n\bm{k}_{c};\bm{r}_{c}}}
&=&1-\left[\bm{\nabla}_{\bm{r}_{c}}\ln\gamma_{\epsilon}(\bm{r}_{c})\right]
\cdot \bm{\Delta}_{\mathrm{TE}\:n\bm{k}_{c}}
\end{eqnarray}
where $E_{\mathrm{TE}\:n\bm{k}_{c};\bm{r}_{c}}=
\gamma_{\epsilon}(\bm{r}_{c})E_{\mathrm{TE}\:n\bm{k}_{c}}$.
Figure \ref{fig:Delta} shows $\bm{\Delta}_{\mathrm{TE}\:n\bm{k}}$
for the first and second bands of TE mode
and we can see $\bm{\Delta}_{\mathrm{TE}\:n\bm{k}}\lesssim 0.1 a$.
Therefore, the correction is at most a few percent as long as 
the modulation is sufficiently weak, i.e., 
$|a\bm{\nabla}_{\bm{r}_{c}}\ln\gamma_{\epsilon}(\bm{r}_{c})|\ll 1$.
In order to make the argument complete,
we also calculate a correction to the group velocity of a TE mode,
\begin{subequations}
\begin{eqnarray}
&&\bm{\nabla}_{\bm{k}_{c}}\mathcal{E}_{\mathrm{TE}\:n\bm{k}_{c};\bm{r}_{c}}
\nonumber\\
&&=\bm{\nabla}_{\bm{k}_{c}}E_{\mathrm{TE}\:n\bm{k}_{c};\bm{r}_{c}}
-
\bm{\nabla}_{\bm{k}_{c}}
\Bigl[  
\left[\bm{\nabla}_{\bm{r}_{c}}\gamma_{\epsilon}(\bm{r}_{c})\right]
\cdot 
\bm{\Delta}_{\mathrm{TE}\:n\bm{k}_{c}}E_{\mathrm{TE}\:n\bm{k}_{c}}
\Bigr]
\nonumber\\
&&\cong
\bm{\nabla}_{\bm{k}_{c}}E_{\mathrm{TE}\:n\bm{k}_{c};\bm{r}_{c}}
+\tensor{\Pi}_{\mathrm{TE}\:n\bm{k}_{c}}\dot{\bm{k}}_{c},
\label{eq:DeltaEpsilon}
\\
&&\tensor{\Pi}^{ij}_{\mathrm{TE}\:n\bm{k}}
=\nabla^{i}_{\bm{k}}\Delta^{j}_{\mathrm{TE}\:n\bm{k}}+
\left[
\nabla^{i}_{\bm{k}}\ln E_{\mathrm{TE}\:n\bm{k}}
\right]
\Delta^{j}_{\mathrm{TE}\:n\bm{k}}.
\end{eqnarray}
\end{subequations}
Here we used the relation $\dot{\bm{k}_{c}}\cong 
-[\bm{\nabla}_{\bm{r}_{c}}\gamma_{\epsilon}(\bm{r}_{c})]E_{\mathrm{TE}\:n\bm{k}_{c}}$
for smooth and weak modulation.
By plugging Eq.~(\ref{eq:DeltaEpsilon}) to the equation of motion for $\bm{r}_{c}$
in Eq.~(\ref{eq:EOM-r}), 
$\tensor{\Pi}_{\mathrm{TE}\:n\bm{k}}$ is a variable to be compared
with the Berry curvature.
Figure~\ref{fig:Pi} shows that 
the effect of $\tensor{\Pi}_{\mathrm{TE}\:n\bm{k}}$
is negligibly small compared to the effect of the Berry curvature
in the present case.
However, it is noted that,
even when a Berry connection is nonzero,
the corresponding Berry curvature can vanishes.
(This is easily understood by the analogy
of a Berry connection and a Berry curvature to
a vector potential and a magnetic field.)
In such a case, $\bm{\Delta}_{\mathrm{TE}\:n\bm{k}}$
and $\tensor{\Pi}_{\mathrm{TE}\:n\bm{k}}$
are not necessarily minor corrections.
These corrections may become enough measurable
for a generic modulation additional to a periodic structure,
while, for assuring the validity of our argument,
we mainly consider a slowly-varying modulation
in this paper.

\begin{figure}[hbt]
$\begin{array}{cc}
\includegraphics[scale=0.18]{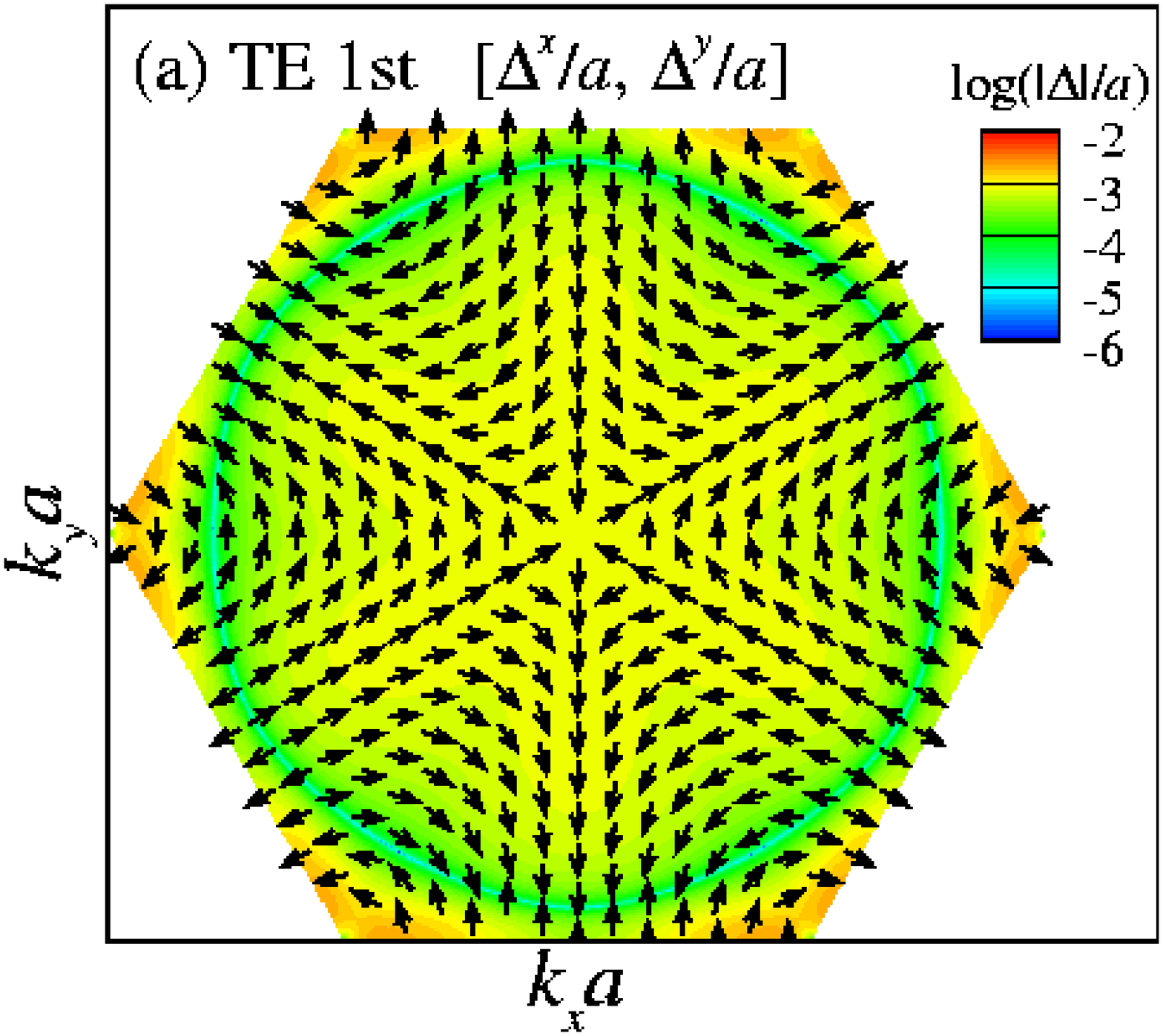}
& \includegraphics[scale=0.18]{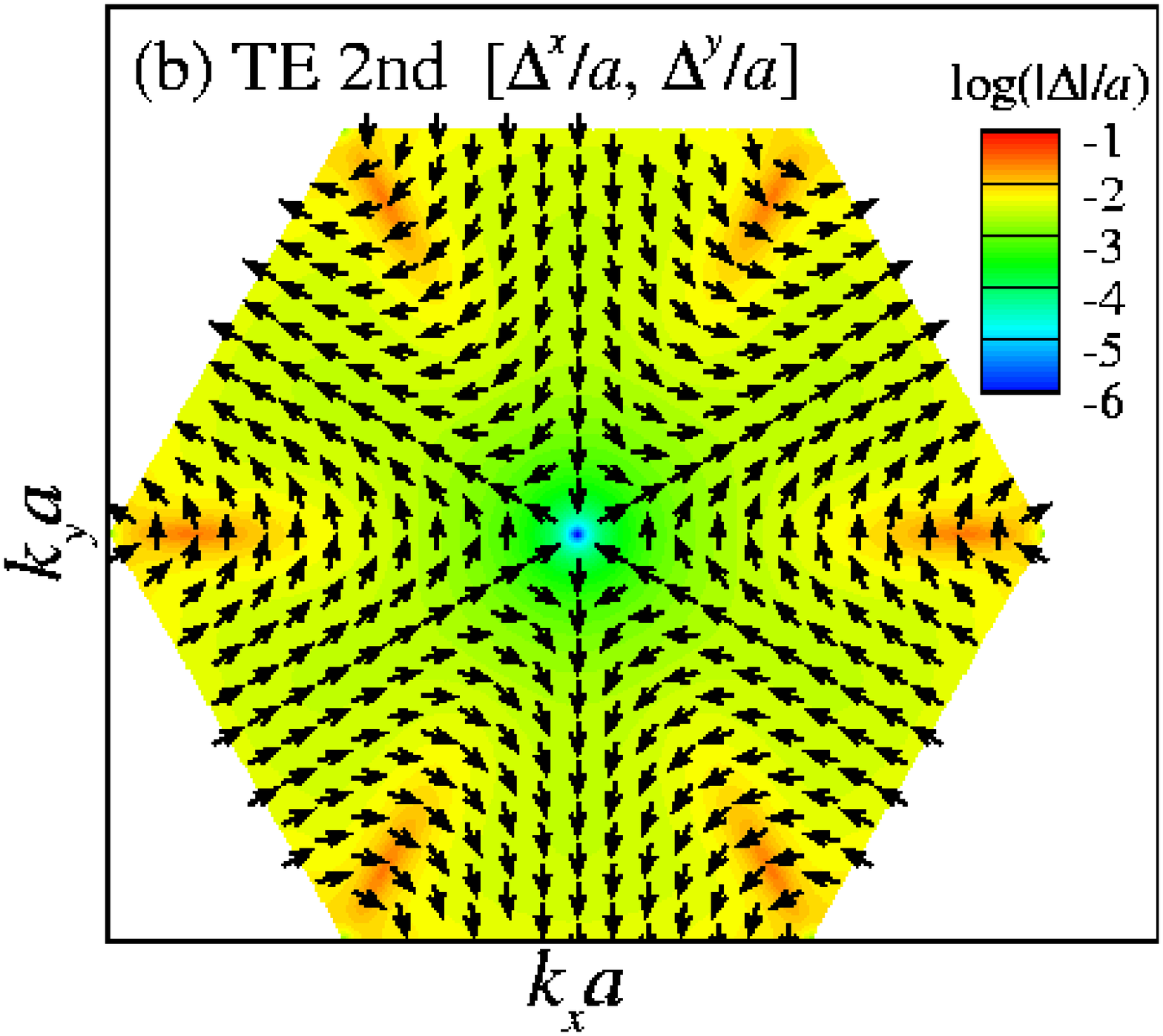}
\end{array}$
\caption{
Difference between the electric and magnetic parts
of the Berry connection
for each of (a) the TE first band and (b) the TE second band.
i.e., $\bm{\Delta}_{\mathrm{TE}\:n\bm{k}}
=(\bm{\Lambda}^{E}_{\mathrm{TE}\:n\bm{k}}
-\bm{\Lambda}^{H}_{\mathrm{TE}\:n\bm{k}})/2$.
}
\label{fig:Delta}
\end{figure}
\begin{figure}[hbt]
\includegraphics[scale=0.18]{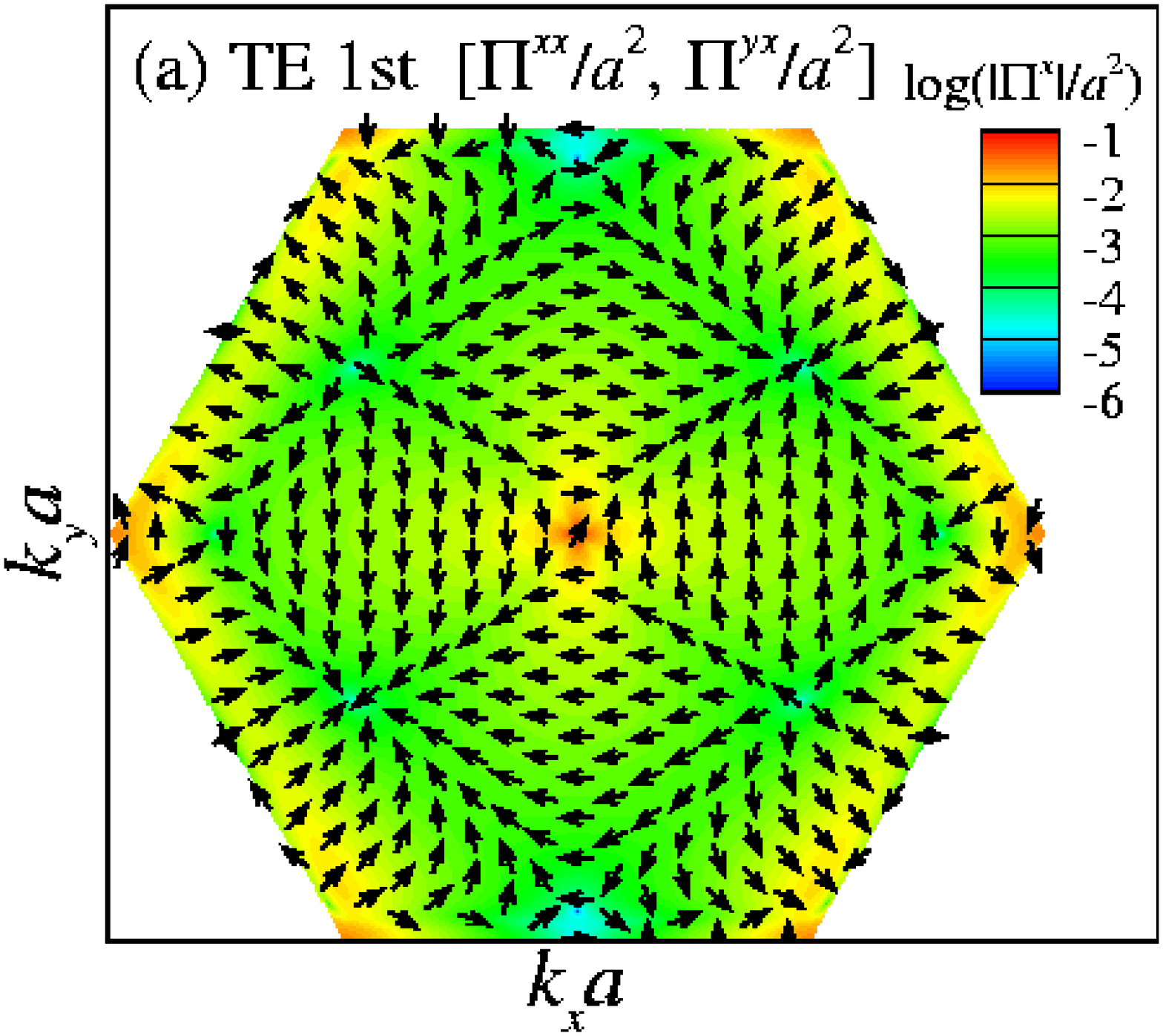}
\includegraphics[scale=0.18]{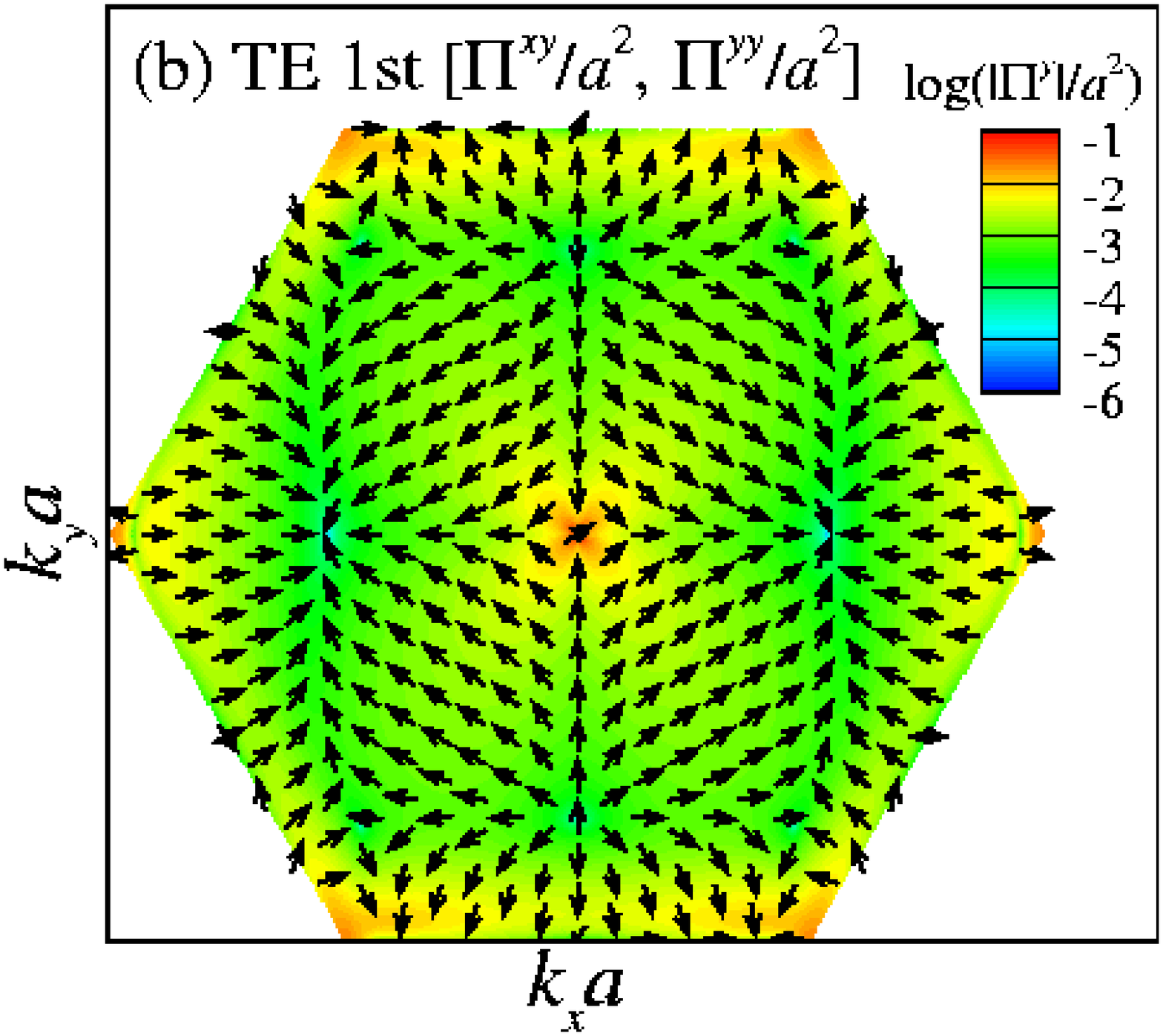}
\includegraphics[scale=0.18]{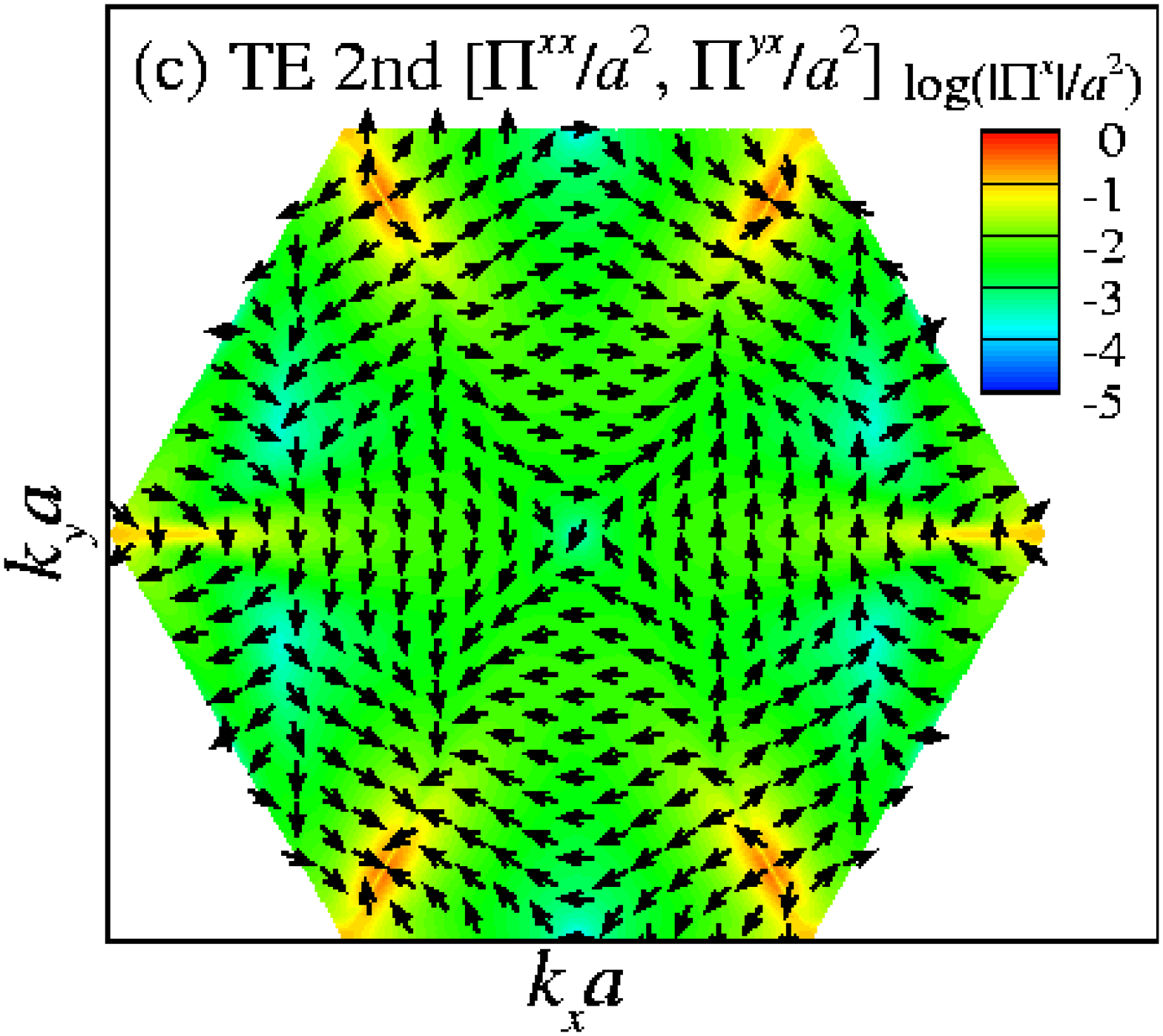}
\includegraphics[scale=0.18]{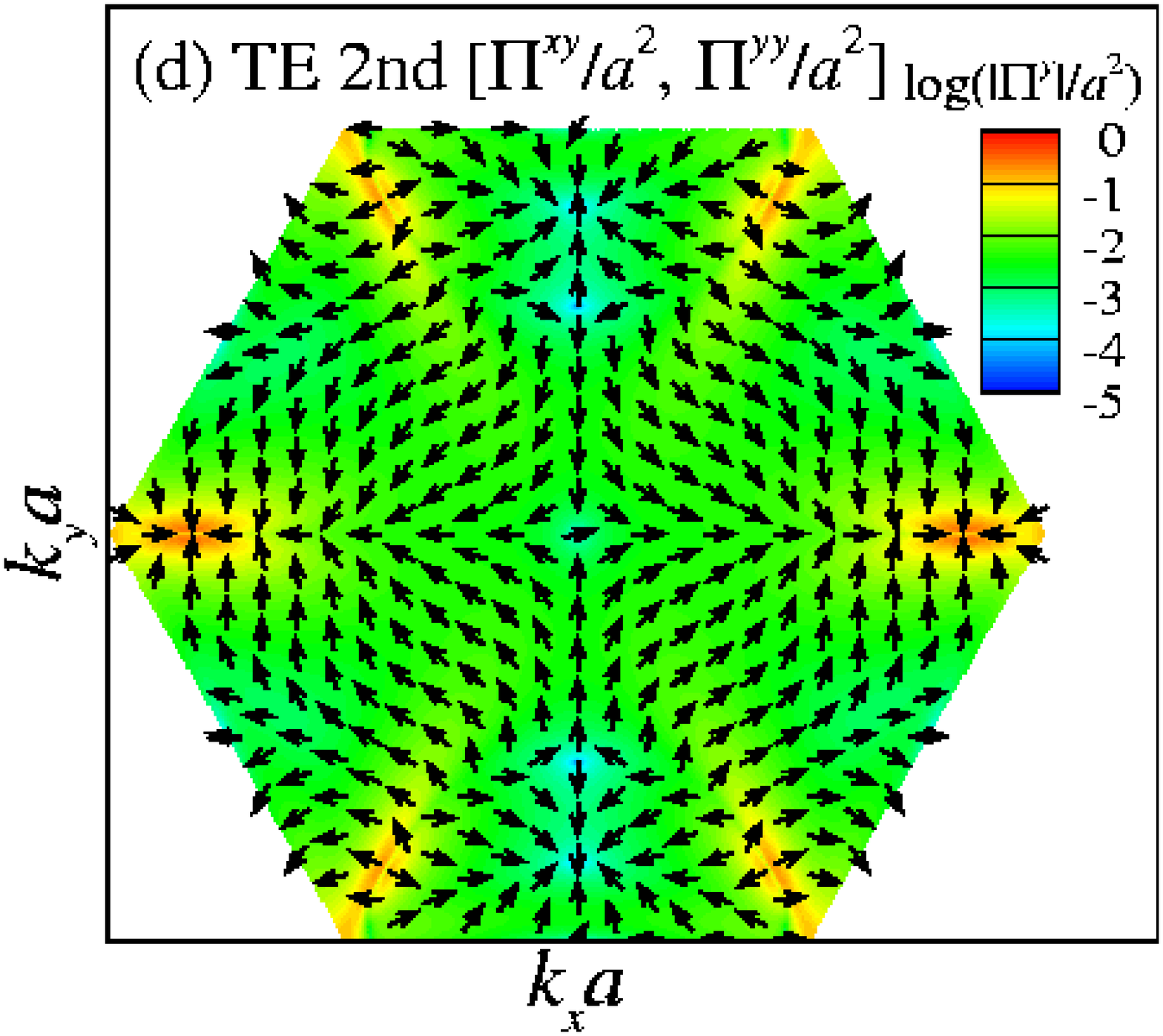}
\caption{
$\tensor{\Pi}_{\mathrm{TE}\:n\bm{k}}$,
which is related to the correction of group velocity
for each of
(a) the TE first band with a $x$-directional modulation,
(b) the TE first band with a $y$-directional modulation,
(c) the TE second band with a $x$-directional modulation
and (d) the TE second band with a $y$-directional modulation.
}
\label{fig:Pi}
\end{figure}

\section{\label{sec:magneto-deflection}Difference from 
magnetically induced deflections}
It was proposed theoretically \cite{photonic-Hall-th}
and observed experimentally \cite{photonic-Hall-ex}
that, in a Faraday-active random medium subject to
a magnetic field perpendicular to an incident beam,
the diffusion flow of light is deflected
in a direction perpendicular
to both the incident light beam and
the externally applied magnetic field.
This effect seems to be more similar to the conventional electrical Hall effect
than the optical Hall effect is,
because the effect is caused by the external magnetic field
and the direction of deflection is perpendicular to it.
However, it should be noted that, unlike electrons,
photons are not charged,
and their orbital motions do not directly couple 
to an external magnetic field.
This effect is theoretically interpreted by
the magnetically induced off-diagonal components
of a diffusion tensor and 
experimentally proved to be due to 
the magnetically induced changes in the optical 
properties of scatterers
\cite{photonic-Hall-th,photonic-Hall-ex}.
In this sense, this effect is similar to
the anomalous Hall effect due to 
the skew scattering mechanism,
rather than to the conventional Hall effect.
On the other hand, the optical Hall effect 
is originated by the anomalous velocity 
of an optical wavepacket which
appears without external magnetic field
nor scatterers.

This kind of phenomena, i.e.,
magnetically induced deflection,
is not restricted to random media.
The deflection of light by a magnetic field
in a nonscattering homogeneous medium
has also been discussed theoretically \cite{Landau}
and observed experimentally \cite{Rikken-Tiggelen}.
When the effect of absorption 
in a Faraday-active medium is negligible,
the linear effect of external magnetic field $\bm{B}$
on this medium is described by the dielectric tensor,
\begin{subequations}
\begin{eqnarray}
\tensor{\epsilon}_{ij}
&=& n^{2}(\delta_{ij}+2i\epsilon_{ijk}\Delta_{k}),
\\
\bm{\Delta}
&=& \frac{\gamma}{2n^{2}}\bm{B},
\end{eqnarray}
\end{subequations}
where $n$ is the refractive index of the medium in the case of $\bm{B}=0$,
$\Re\gamma$ and $\Im\gamma$
represent the strength of the magnetic circular birefringence
and that of magnetic circular dichroism respectively,
while we set $\Im\gamma=0$.
The eigen modes of the dielectric displacement $\bm{D} =\tensor{\epsilon}\bm{E}$
in such a medium are explicitly given in Ref.~\cite{comment-I},
and they are represented in terms of the orthogonal unit vectors
$\bm{e}_{k}$, $\bm{e}_{\theta}$, and $\bm{e}_{\phi}$
in the spherical coordinate of the $\bm{k}$-space as
\begin{subequations}
\begin{eqnarray}
&&\bm{D}_{+} \propto (\bm{e}_{B}\cdot\bm{e}_{k})\bm{e}_{\theta}
+i(C_{B}+\Delta|\bm{e}_{B}\times\bm{e}_{k}|^{2})\bm{e}_{\phi},
\\
&&\bm{D}_{-} \propto 
(C_{B}+\Delta|\bm{e}_{B}\times\bm{e}_{k}|^{2})\bm{e}_{\theta}
-i(\bm{e}_{B}\cdot\bm{e}_{k})\bm{e}_{\phi},
\\
&&C_{B} =\sqrt{(\bm{e}_{B}\cdot\bm{e}_{k})^2
+\Delta^{2}|\bm{e}_{B}\times\bm{e}_{k}|^{4}}.
\end{eqnarray}
\end{subequations}
Here $\bm{e}_{B}$ is a unit vector defined by 
$\bm{\Delta}=\Delta\:\bm{e}_{B}$
with the condition $\bm{e}_{B}\cdot\bm{e}_{k}\ge 0$,
and $\bm{e}_{\phi}\parallel \bm{e}_{B}\times\bm{e}_{k}$,
$\bm{e}_{\theta} = \bm{e}_{\phi}\times\bm{e}_{k}$.
(When $\Im\gamma\neq 0$, $\Delta$ is a complex-valued parameter.)
These eigen modes have the dispersion relations and the group velocities,
\begin{subequations}
\begin{eqnarray}
E_{\pm,\bm{k}} &=&
\frac{v k}{\sqrt{1-2\Delta^2|\bm{e}_{B}\times\bm{e}_{k}|^{2}\mp 2\Delta C_{B}}},
\\
\bm{v}_{\pm,\bm{k}}&=&
\frac{E_{\pm,\bm{k}}}{k}\left[
\bm{e}_{k}\mp \frac{\Delta}{C_{B}}
|\bm{e}_{B}\times\bm{e}_{k}|(\bm{e}_{B}\cdot\bm{e}_{k})\bm{e}_{\theta}
\right],
\end{eqnarray}
\end{subequations}
where $v= 1/n$. The direction of Poynting vector of each mode
coincide with $\bm{v}_{\bm{k}\pm}$ as long as $\Im\gamma=0$.
It should be noted that 
the deflection occurs within the plane determined by $\bm{k}$ and $\bm{B}$.
The angle $\delta\theta$ between the propagating directions of two eigen modes
with the same $\bm{k}$
is given in Ref.\cite{comment-I} and represented in the present notation as
\begin{eqnarray}
\delta\theta &=& 
2\arctan\frac{\Delta|\bm{e}_{B}\times\bm{e}_{k}|(\bm{e}_{B}\cdot\bm{e}_{k})}
{\sqrt{(\bm{e}_{B}\cdot\bm{e}_{k})^2
+\Delta^{2}|\bm{e}_{B}\times\bm{e}_{k}|^{4}}}.
\end{eqnarray}
For the exact Voigt geometry ($\bm{e}_{B}\cdot\bm{e}_{k}=0$),
there appears no deflection \cite{comment-I,reply-I}.
The physic of this phenomenon is intuitively interpreted
by considering the first order perturbation with respect to $\Delta$ 
and the situation in which the angle between $\bm{e}_{B}$ and $\bm{e}_{k}$ 
are not close to the Voigt geometry, i.e.,
$\bm{e}_{B}\cdot\bm{e}_{k}\gg |\Delta||\bm{e}_{B}\times\bm{e}_{k}|^{2}$.
The approximated eigenvalues and group velocities are represented as follows,
\begin{subequations}
\begin{eqnarray}
E_{\pm,\bm{k}} &\cong& v (k\pm\bm{\Delta}\cdot\bm{k}),
\\
\bm{v}_{\pm,\bm{k}}  &\cong& v (\bm{e}_{k}\pm\bm{\Delta}).
\end{eqnarray}
\end{subequations}
This effect comes from the magnetically induced change 
in the dispersion relation of each mode
due to the Pitaevskii magnetization, $\pm v\bm{\Delta}\cdot\bm{k}$ \cite{Landau}.
On the other hand, the optical Hall effect 
is caused by the anomalous velocity due to the geometrical propriety of a wavepacket.

In the above perturbative picture,
$\bm{D}_{\pm}$ are approximately equivalent to
right/left circularly polarized modes which have
the spin angular momenta, $\pm\bm{e}_{k}$.
Therefore, the above interpretation based on the Pitaevskii magnetization
means that an external magnetic field couples to 
the spin of photon
through a Faraday-active medium.
From this consideration, we reasonably expect that an external magnetic field
couples not only to the spin 
but also to a generic internal rotation of photon
in the form of dipole coupling.
(Consequently, this effect is expected for 
Laguerre-Gauss beams \cite{OAM}
which have internal orbital angular momenta.)
As shown in Sec.~\ref{sec:photonic-crystal},
there appear eigen modes with large internal rotations
in a two-dimensional photonic crystal
without inversion symmetry.
Here we take the configuration in which 
the photonic crystal is periodic in the $xy$-plane
and uniform along the $z$-direction.
Considering an eigen mode with $k_{z}=0$,
its internal rotation
is oriented in the $z$-direction, i.e.,
perpendicular to its propagating direction.
Thus, when the photonic crystal is composed of 
Faraday-active media and subject to an external magnetic field,
it is expected that the magnetically induced deflection can be enhanced.
In addition, this effect would be observed even in the Voigt geometry.
The details of this problem is beyond the scope of the present study
and we will discuss it elsewhere.
Here we just note that this effect in a Faraday-active photonic crystal 
is due to the magnetically induced change of dispersion relation
as well as that in a homogeneous Faraday-active medium,
and is different from the optical Hall effect in a photonic crystal
discussed in Sec.~\ref{sec:photonic-crystal}.



\begin{thebibliography}{99}
\bibitem{Berry}
M.~V.~Berry, Proc. R. Soc. A \textbf{392}, 45 (1984).
;J. Mod. Opt. \textbf{34}, 1401 (1987).

\bibitem{GPP}
\textit{Geometrical Phases in Physics},
edited by A.~Shapere and F.~Wilczek 
(World Scientific, Singapore, 1989).
\bibitem{GPQS}
\textit{The Geometrical Phase in Quantum Systems},
edited by A. Bohm, A. Mostafazadeh, H. Koizumi, Q. Niu, and J. Zwanziger
(Springer-Verlag, Berlin, 2003). 

\bibitem{Karplus-Luttinger}
R.~Karplus and J.~M.~Luttinger,
Phys. Rev. \textbf{95}, 1154 (1954).
\bibitem{Luttinger}
J.~M.~Luttinger,  Phys. Rev. \textbf{112}, 739 (1958).

\bibitem{TKNN}
D.~J.~Thouless, M.~Kohmoto, M.~P.~Nightingale, and M.~den~Nijs,
 Phys. Rev. Lett. \textbf{49}, 405 (1982).
\bibitem{Kohmoto}
M.~Kohmoto, Ann. Phys. (N.Y.) \textbf{160}, 343 (1985).
\bibitem{Aoki-Ando}
H.~Aoki and T.~Ando,
Phys. Rev. Lett. \textbf{57}, 3093 (1986).

\bibitem{MN}
M.~Onoda and N.~Nagaosa,
J. Phys. Soc. Jpn. \textbf{71}, 19 (2002).
\bibitem{Jungwirth}
T.~Jungwirth, Q.~Niu, and A.~H.~MacDonald,
Phys. Rev. Lett. 88, 207208 (2002).

\bibitem{Fang}
Z.~Fang, N.~Nagaosa, K.~S.~Takahashi, A.~Asamitsu, R.~Mathieu, 
T.~Ogasawara, H.~Yamada, M.~Kawasaki, Y.~Tokura, and K.~Terakura,
Science \textbf{302}, 92 (2003).

\bibitem{MNZ}
S.~Murakami, N.~Nagaosa, and S.-C.~Zhang,
Science \textbf{301}, 1348 (2003).
\bibitem{Sinova}
J.~Sinova,
D.~Culcer, Q.~Niu, N.~A.~Sinitsyn, 
T.~Jungwirth, and A.~H.~MacDonald,
Phys. Rev. Lett. \textbf{92}, 126603 (2004).

\bibitem{Rytov}
S.~M.~Rytov, Dokl. Akad. Nauk SSSR \textbf{18}, 263 (1938).
\bibitem{Vladimirski}
V.~V.~Vladimirski, Dokl. Akad. Nauk SSSR \textbf{31}, 222 (1941). 
\bibitem{Pancharatnam}
S.~Pancharatnam,
The Proceedings of the Indian Academy of Sciences Vol.~XLIV, No.~5, Sec.~A,
247 (1956).

\bibitem{Chiao-Wu}
R.~Y.~Chiao and Y.~S.~Wu,
Phys. Rev. Lett. \textbf{57}, 933 (1986).
\bibitem{Tomita-Chiao}
A.~Tomita and R.~Y.~Chiao,
Phys. Rev. Lett. \textbf{57}, 937 (1986).
\bibitem{Berry-II}
M.~V.~Berry, Nature \textbf{326}, 277 (1987).

\bibitem{MSN}
M.~Onoda, S.~Murakami and N.~Nagaosa,
Phys. Rev. Lett. \textbf{93}, 083901 (2004).

\bibitem{Fedorov}
F.~I.~Fedorov, Dokl. Akad. Nauk SSSR \textbf{105}, 465 (1955).

\bibitem{Imbert} 
C.~Imbert, 
Phys. Rev. D \textbf{5}, 787 (1972).

\bibitem{Boulware}
D.~G.~Boulware,
Phys. Rev. D \textbf{7}, 2375 (1973).
\bibitem{Ashby-Miller}
N.~Ashby and S.~C.~Miller Jr., 
Phys. Rev. D \textbf{7}, 2383 (1973).

\bibitem{Schilling}
H.~Schilling, Ann. Phys. (Leipzig) \textbf{16}, 122 (1965). 

\bibitem{Fedoseev-I}
V.~G.~Fedoseev,
Opt. Spektrosk. {\bf 71}, 829 (1991) [
Opt. Spectrosc. (USSR) {\bf 71}, 483 (1991)].

\bibitem{Fedoseev-II}
V.~G.~Fedoseev,
Opt. Spektrosk. {\bf 71}, 992 (1991) [
Opt. Spectrosc. (USSR) {\bf 71}, 570 (1991)].

\bibitem{Pillon}
F.~Pillon, H.~Gilles and S.~Girard, 
Appl. Opt. \textbf{43}, 1863 (2004).

\bibitem{Dooghin}
A.~V.~Dooghin , N.~D.~Kundikova, V.~S.~Liberman, and B.~Ya.~Zel'dovich,
Phys. Rev. A \textbf{45}, 8204 (1992).

\bibitem{Liberman-Zeldovich}
V.~S.~Liberman and B.~Ya.~Zel'dovich,
Phys. Rev. A \textbf{46}, 5199 (1992).

\bibitem{Bliokh}
K.~Yu.~Bliokh and Yu.~P.~Bliokh,
Phys. Rev. E \textbf{70}, 026605 (2004).

\bibitem{Jackiw}
R.~Jackiw and A.~Kerman,
Phys. Lett. \textbf{71A}, 158 (1979).

\bibitem{Pattanayak}
A.~K.~Pattanayak and W.~C.~Schieve,
Phys. Rev. E \textbf{50}, 3601 (1994).

\bibitem{Chang-Niu}
M.-C.~Chang and Q.~Niu,
Phys. Rev. B \textbf{53}, 7010 (1996).

\bibitem{Sundaram-Niu}
G.~Sundaram and Q.~Niu, 
Phys. Rev. B \textbf{59}, 14915 (1999).

\bibitem{JMW}
J.~D.~Joannopoulos, R.~D.~Meade, and J.~N.~Winn,
\textit{Photonic Crystals} (Princeton University Press, Princeton, 1995).
\bibitem{Sakoda}
K.~Sakoda,
\textit{Optical Properties of Photonic Crystals}
(Springer, Berlin, 2005).

\bibitem{Born}
M.~Born and E.~Wolf, \textit{Principles of Optics, 7th edition} 
(Cambridge University Press, Cambridge, 1999).

\bibitem{photonic-Hall-th}
B.~A.~van~Tiggelen,
Phys. Rev. Lett. \textbf{75}, 422 (1995).

\bibitem{photonic-Hall-ex}
G.~L.~J.~A.~Rikken and B.~A.~van~Tiggelen,
Nature \textbf{381}, 54 (1996).

\bibitem{Landau}
L.~D.~Landau, E.~M.~Lifshitz, and
L.~P.~Pitaevskii,
\textit{Electrodynamics of Continuous Media}
(Pergamon, Oxford, 1984).

\bibitem{Rikken-Tiggelen}
G.~L.~J.~A.~Rikken and B.~A.~van~Tiggelen,
Phys. Rev. Lett. \textbf{78}, 847 (1997).

\bibitem{comment-I}
G.~W.~'t~Hooft, G.~Nienhuis, and
J.~C.~J. Paasschens,
Phys. Rev. Lett. \textbf{80}, 1114 (1998).

\bibitem{reply-I}
G.~L.~J.~A.~Rikken and B.~A.~van~Tiggelen,
Phys. Rev. Lett. \textbf{80}, 1115 (1998).

\bibitem{comment-II}
J.~Yimin and M.~Liu,
Phys. Rev. Lett. \textbf{90}, 099401 (2003).

\bibitem{reply-II}
G.~L.~J.~A.~Rikken and B.~A.~van~Tiggelen,
Phys. Rev. Lett. \textbf{90}, 099402 (2003).

\bibitem{Haldane-Raghu}
F.~D.~M.~Haldane and S.~Raghu, cond-mat/0503588.
\bibitem{Raghu-Haldane}
S.~Raghu and F.~D.~M.~Haldane, cond-mat/0602501.

\bibitem{Sawada-Nagaosa}
K.~Sawada and N.~Nagaosa, Phys. Rev. Lett. \textbf{95}, 237402 (2005).

\bibitem{Bliokh-PRL}
K.~Yu.~Bliokh and Yu.~P.~Bliokh,
Phys. Rev. Lett. \textbf{96}, 073903 (2006).

\bibitem{Dirac}
P.~A.~M.~Dirac, \textit{Lectures on Quantum Mechanics}
(Yeshiva University, New York, 1964).

\bibitem{Kristensen}
M.~Kristensen and J.~P.~Woerdman,
Phys. Rev. Lett. \textbf{72}, 2171 (1994).

\bibitem{note-force}
Except for a simple system with a quadratic dispersion,
the concept of ``force'' often becomes ambiguous,
while the concept of ``acceleration'' is still well-defined.
This issue is rather crucial for systems with spin-orbit interactions,
i.e., systems which are relativistic in nature.
Here, for the sake of convenience,
we refer to the time derivative of (lattice) momentum
as the driving force.

\bibitem{note-second-quantization}
Although the arguments in Refs.\cite{Chang-Niu, Sundaram-Niu}
based on the first quantized formalism,
the reformulation in the second quantized one is straightforward.
Then the similar argument can be applicable to photonic systems.
In a relativistic boson system like a photonic system,
we cannot construct a positive definite provability density. 
This is why we have formulated our theory
in the second quantized formalism.

\bibitem{Goos-Hanchen}
F.~Goos and M.~H{\" a}nchen,
Ann. Phys. (Leipzig) \textbf{1}, 333 (1947).

\bibitem{Jackson}
J.~D.~Jackson, \textit{Classical Electrodynamics, 3rd edition} 
(John Wiley \& Sons, Inc., New York, 1999).

\bibitem{note-Omega}
This form of the Berry curvature, 
$\bm{\Omega}_{\bm{k}}=
\frac{\bm{k}}{k^{3}}\sigma_{3}$, is
specific to the massless particle with spin-1.
For the relativistic fermion with mass $m$ and spin-$\frac{1}{2}$,
the Berry curvature in the helicity basis is expressed by
\begin{eqnarray*}
\bm{\Omega}_{\bm{k}} &=& 
\frac{1}{2E^{2}_{\bm{k}}}
\left[
\frac{m}{E_{\bm{k}}}
\left(
\sigma_{1}\bm{e}_{\theta}
+\sigma_{2}\bm{e}_{\phi}
\right)+
\bm{e}_{k}\sigma_{3}
\right],
\end{eqnarray*}
where $E_{\bm{k}} =\sqrt{k^2+m^2}$.
In the massless limit, 
this coincides with the Berry curvature of photon
except for the overall coefficient due to different magnitude of spin.
In the non-relativistic limit, i.e, increasing $m$ with fixing $k$,
the Berry curvature decrease as $1/m^{2}$
because the spin-orbit interaction also scales in the same manner.

\bibitem{Duval}
C.~Duval, Z.~Horv{\' a}th, and P.~A.~Horv{\' a}thy,
Phys. Rev. D \textbf{74}, 021701(R) (2006).

\bibitem{OAM}
L.~Allen, S.~M.~Barnett,
and M.~.J.~ Padgett,
\textit{Optical Angular Momentum} 
(Institute of Physics Publishing, Bristol and Philadelphia, 2003).
\bibitem{Fedoseev-LG}
V.~G.~Fedoseyev, Opt. Commun. \textbf{193}, 9 (2001).
\bibitem{Dasgupta}
R.~Dasgupta and P.~K.~Gupta, Opt. Commun. \textbf{257}, 91 (2006).
\bibitem{Sasada}
H.~Okuda and H.~Sasada,
Opt. Exp. \textbf{14}, 8393 (2006).


\bibitem{phononic-crystal-I}
M.~S.~Kushwaha, P.~Halevi, L.~Dobrzynski, and B.~Djafari-Rouhani,
Phys. Rev. Lett. \textbf{71}, 2022 (1993).
\bibitem{phononic-crystal-II}
M.~S.~Kushwaha, P.~Halevi, G.~Mart{\' i}nez,
L.~Dobrzynski and B.~Djafari-Rouhani,
Phys. Rev. B. \textbf{49}, 2313 (1994).

\end{thebibliography}
\end{document}